\documentclass[12pt]{PhDthesisPSnPDF}
\usepackage{amsmath, amsthm, amssymb, amsfonts,graphicx}

\title{Infinite Derivative Gravity:\\ A finite number of predictions}

\author{James Edholm}

\collegeordept{Physics \\ Department of Physics}
\university{Lancaster University}
\crest{\includegraphics[width=4cm]{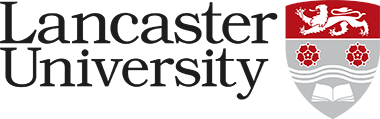}}

\renewcommand{\submittedtext}{A thesis submitted to Lancaster University for the degree of }
\degree{Doctor of Philosophy in the Faculty of Science and Technology}
\degreedate{February 2019}
\supervisor{\emph{\\Supervised by Dr David Burton and Dr Anupam Mazumdar}}

\begin{document}

\newcommand{\non}{\nonumber}
\newcommand{\D}{\nabla}
\newcommand{\p}{\partial}
\newcommand{\nn}{\nonumber}
\newcommand{\bg}{\begin{equation}}
\newcommand{\en}{\end{equation}}
\renewcommand\[{\begin{equation}}
\renewcommand\]{\end{equation}}
\newcommand{\al}{\alpha}
\newcommand{\bt}{\beta}
\newcommand{\ga}{\gamma}
\newcommand{\da}{\delta}
\newcommand{\n}{\nabla}
\newcommand{\ba}{\begin{eqnarray}}
\newcommand{\ea}{\end{eqnarray}}
\newcommand{\LF}{\left(}
\newcommand{\RF}{\right)}
\newcommand{\LT}{\left[}
\newcommand{\RT}{\right]}
\newcommand{\Ld}{\left.}
\newcommand{\Rd}{\right.}
\newcommand{\cO}{{\cal O}}
\newcommand{\cF}{{\cal F}}
\newcommand{\Ra}{\Rightarrow}
\newcommand{\bl}{\begin{align}}
\newcommand{\el}{\end{align}}
\newcommand{\fl}{}
\newtheorem{theorem}{Theorem}[section]
\theoremstyle{definition}
\newtheorem{definition}{Definition}[section]

\renewcommand\baselinestretch{1.5}
\baselineskip=18pt plus1pt


 \maketitle  








\begin{abstracts}        
Ghost-free Infinite Derivative Gravity (IDG) is a modified gravity theory which 
can avoid the singularities predicted by General Relativity. 
This thesis examines the effect of IDG on four areas of importance 
for theoretical cosmologists and experimentalists.
First, the gravitational potential produced by a point source is derived and compared to experimental evidence, around both Minkowski and (Anti) de Sitter backgrounds. 
Second, the conditions necessary for avoidance of singularities for perturbations around Minkowski 
and (Anti) de Sitter  spacetimes are found, as well as for background Friedmann-Robertson-Walker spacetimes. 
Third, the modification to perturbations during primordial inflation is derived and shown to give a constraint on the mass scale of IDG, and to allow further tests of the theory. Finally, the effect of IDG on the production and propagation of gravitational waves is derived and it is shown that IDG gives almost precisely the same predictions as General Relativity  for the power emitted by a binary system. 

\end{abstracts}




\frontmatter
\begin{dedication} 

For my parents: Margaret and Peter

\end{dedication}



\begin{acknowledgements}      
\newenvironment{changemargin}[2]{%
\begin{list}{}{%
\setlength{\topsep}{0pt}%
\setlength{\leftmargin}{#1}%
\setlength{\rightmargin}{#2}%
\setlength{\listparindent}{\parindent}%
\setlength{\itemindent}{\parindent}%
\setlength{\parsep}{\parskip}%
}%
\item[]}{\end{list}}
\begin{changemargin}{-1cm}{-1cm}
Special thanks go to my mother Margaret, for her encouragement, and to Bonnie, 
for always being there to brighten my day.

I would like to particularly thank Dr David Burton, who kindly agreed to 
supervise me when my previous supervisor moved 
to a new job. David has been a tremendous help and has always 
been willing to guide me whenever I was stuck.

Aindri\'u Conroy also deserves a special mention for
informally supervising me and collaborating on several papers. 
Aindri\'u was instrumental in helping my understanding of the field
as well as being a true friend.

I would also like to thank the students in the Theoretical 
Particle Cosmology group at Lancaster who have helped me along 
the way: Charlotte Owen (despite her perpetual insistence on being 
fashionably late!), Ilia Teimouri, Spyridon Talaganis, Saleh Qutub, 
Leonora Donaldson-Wood \& Amy Lloyd-Stubbs who all contributed greatly to 
my time at Lancaster. 

I am also very grateful for all the friends inside and outside 
Lancaster who kept me motivated and supported, in particular 
my flatmates who (eventually!) became actual mates: Anuja, Gustaf, Sash \& Craig; 
the other physicists: Eva, Denise, James, Marjan, Simon \& Kunal;
my school friends: both Ben, Chad \& Will who were always ready to 
answer any mathematical queries with their usual attempts at humour
and Dan, GB, Adam \& Will who weren't as helpful with my maths problems but always made
me laugh;
as well as James, Will, Sam, Tash, Robbie, Sarah, Harriet \& Cat who 
reminded me that there was life outside the Lancaster bubble during my regular
excursions to London \& Bristol. 

Finally my thanks go to Lancaster University for funding my research and to my 
viva examiners Carsten van de Bruck and John McDonald for taking the time 
to read my thesis and for fruitful discussions.
\end{changemargin}
\end{acknowledgements}




\begin{declaration}        

This thesis is my own work and no portion of the work referred to in this thesis has been submitted in support of an application for another degree or qualification at this or any other institute of learning.

\end{declaration}


\vspace{20mm}
\begin{quote}
``\textit{If I have seen further than others, it is by standing on the shoulders of giants}''
Isaac Newton

\end{quote}


\setcounter{secnumdepth}{2} 
\setcounter{tocdepth}{2}    
\vspace{10mm}
\tableofcontents            


\listoffigures  



\begin{pubs}        
\newif\ifshowcitations\showcitationsfalse%
\newif\ifshowlinks\showlinksfalse%

%
%
%

\ifshowlinks%
  \usepackage[
         colorlinks=true,
         urlcolor=blue,       
         ]{hyperref}
  \newcommand*{\inspireurl}[1]{\\\href{#1}{INSPIRE-HEP entry}}
\else
  \makeatletter
  \newcommand*{\inspireurl}[1]{\@bsphack\@esphack}
  \makeatother
\fi
\ifshowcitations%
  \newcommand*{\citations}[1]{\\* #1}
\else
  \makeatletter
  \newcommand*{\citations}[1]{\@bsphack\@esphack}
  \makeatother
\fi
\renewcommand{\labelenumii}{\arabic{enumi}.\arabic{enumii}}


\underline{Chapter 3}
\begin{itemize}
\item
\textbf{``Behavior of the Newtonian potential for ghost-free gravity and singularity-free gravity''}\\
J. Edholm, A. Koshelev, A. Mazumdar, arXiv:1604.01989, \textit{Physical Review D}, 10.1103/PhysRevD.94.104033
\item
\textbf{``Newtonian potential and geodesic completeness in infinite derivative gravity''}\\
J. Edholm, A. Conroy, arXiv:1705.02382, \textit{Physical Review D}, 10.1103/PhysRevD.96.044012
\item
\textbf{``Revealing Infinite Derivative Gravity’s true potential:
The weak-field limit around de Sitter backgrounds''}\\
    J. Edholm, arXiv:1801.00834 \textit{Physical Review D} 10.1103/PhysRevD.97.064011
\end{itemize}
\underline{Chapter 4}
\begin{itemize}
\item
\textbf{``Newtonian potential and geodesic completeness in infinite derivative gravity''}\\
J. Edholm, A. Conroy, arXiv:1705.02382, \textit{Physical Review D}, 10.1103/PhysRevD.96.044012
\item
\textbf{``Criteria for resolving the cosmological singularity in Infinite
Derivative Gravity around expanding backgrounds''} \\
 J. Edholm, A. Conroy,  arXiv:1710.01366, \textit{Physical Review D}, 10.1103/PhysRevD.96.124040
 \item
 \textbf{``Conditions for defocusing around more general metrics in Infinite Derivative Gravity''}\\ 
    J. Edholm, arXiv:1802.09063, \textit{Physical Review D}, 10.1103/PhysRevD.97.084046
\end{itemize}
\underline{Chapter 5}
\begin{itemize}
\item
\textbf{``UV completion of the Starobinsky model, tensor-to-scalar ratio, and constraints on nonlocality''}\\
J. Edholm, arXiv:1611.05062, \textit{Physical Review D} 10.1103/PhysRevD.97.084046
\end{itemize}
\underline{Chapter 6}
\begin{itemize}
\item
\textbf{``Gravitational radiation in Infinite Derivative Gravity and connections to Effective Quantum Gravity''}\\ 
    J. Edholm, arXiv:1806.00845, \textit{Physical Review D} 10.1103/PhysRevD.98.044049
\end{itemize}

\end{pubs}



\mainmatter




\chapter{Introduction}

Einstein's theory of General Relativity (GR) has been extraordinarily successful at describing 
gravity in the hundred years since it was proposed. 
It predicts gravitational lensing \cite{Einstein:1911vc},
gravitational redshift \cite{Chernoff:1993th} and 
gravitational waves  \cite{Abbott:2016blz}.
It explains the precession of the perihelion of
Mercury \cite{Will:2005va}, a problem which had been unanswered for decades.

However, we do not have a full quantum completion of gravity. Gravity is not renormalizable, 
i.e. we cannot remove unwanted loop divergences in Feynmann diagrams, and generates singularities, 
or places where the spacetime curvature becomes infinite. 
 Candidate theories such as string theory \cite{Calcagni:2017sdq,Blumenhagen:2013fgp},
loop quantum gravity \cite{Nicolai:2005mc,Ashtekar:2012np} and emergent gravity \cite{Verlinde:2016toy} have not provided an overarching description 
of black holes or the Big Bang. 

Previous classical modifications to GR include $f(R)$ gravity and finite higher-derivative terms. 
$f(R)$ gravity does not help with the unwanted behaviour in the 
ultraviolet (UV), i.e. at short distances (without fine-tuning) \cite{Lee:2012dk}. 
On the other hand, finite higher derivative theories can help with the UV behaviour,
but result in terms with negative kinetic energy, known as ghosts. 
Ostrogradsky's theorem of Hamiltonian analysis  \cite{Woodard:2015zca}
tells us why  ghosts are produced for these theories, but  Infinite Derivative Gravity 
 (IDG)
is constructed specifically to avoid ghosts, 
by ensuring that the propagator has at most one extra pole compared to GR \cite{Barnaby:2007ve}. 
We will later show explicitly that IDG can be constrained so that ghosts are avoided.
\section{What's wrong with General Relativity?}

Einstein's great insight was to see that gravity is caused by the curvature of spacetime \cite{Einstein:1916vd}. 
The simplest possible action which describes this is the Einstein-Hilbert action
 \cite{Einstein:1916vd}
 \ba \label{eq:EHaction}
        S_{\text{EH}}=\frac{1}{16\pi G}\int d^{4}x \, \sqrt{-g}R,
\ea
where $G$ is the gravitational constant and $R$
 is the Ricci curvature scalar. $g$ is the determinant
of the spacetime metric $g_{\mu\nu}$, which we take to have the signature ($-$,+,+,+) . General Relativity has 
successfully passed all experimental tests to date, notably predicting gravitational 
lensing \cite{Einstein:1911vc} and gravitational waves \cite{Abbott:2016blz}, 
as well as explaining the longstanding issue of the precession 
of the perihelion of Mercury  \cite{Will:2005va}.
It is used
to launch spacecraft and to enable GPS satellites to correct for 
gravitational redshift \cite{Chernoff:1993th} in order to locate ourselves to an astounding level of accuracy \cite{Will:2014kxa}.

Varying the Einstein-Hilbert action \eqref{eq:EHaction} with respect to the metric leads to the GR equations of motion
\ba \label{eq:einsteinequation}
        8\pi G\hspace{0.5mm} T_{\mu\nu} = R_{\mu\nu} - \frac{1}{2}g_{\mu\nu} R,
\ea        
where $R_{\mu\nu}$ is the Ricci curvature tensor and $T_{\mu\nu}$ is the stress-energy tensor
which describes the distribution of matter and energy. 

Despite its many successes, it was soon realised that GR predicts singularities, or divergences in the spacetime curvature.
These were first thought to be anomalies or to only happen 
in very specific cases. It was eventually found that 
GR will always produce singularities due to the Penrose-Hawking singularity theorems and the Null Energy Condition (NEC) \cite{hawking_ellis_1973}.
 The NEC can be thought of as a covariant description of the idea that
 the combination of the density and the pressure of matter  is a positive quantity, which 
 applies to all matter that we have seen so far \cite{Elder:2013gya}.

\subsection{Previous modifications to General Relativity}
The simplest modification to the GR action \eqref{eq:EHaction} is to keep the dependence on the Ricci scalar $R$, 
but replace the linear dependence with a function $f(R)$. The so-called $f(R)$ theories involve the action 
\ba \label{eq:fofraction}
        S=\frac{1}{16\pi G}\int d^{4}x \, \sqrt{-g}f(R).
\ea
A notable 
example is Starobinsky gravity, where $f(R)=R+\alpha R^2$, where $\alpha$ is 
a dimensionless constant,
which can be used to generate cosmological inflation \cite{Starobinsky:1980te}.

Unfortunately, more general modifications are plagued by the fact that ghosts are generically produced by actions which lead to equations of motion
with terms with more than 2 derivatives. For example, fourth order
equations of motion are produced by 
Stelle's extension of GR, which is therefore commonly known as
fourth order gravity and has the action~\cite{stelle:1977}

\ba \label{eq:fourthderivaction}
        S = \int d^4x \sqrt{-g} \left(M^2_P R -\alpha R^2 + \beta R_{\mu\nu} R^{\mu\nu}  \right),
\ea
where $M_P=\sqrt{\frac{1}{8\pi G}}$ is the reduced Planck mass and $\alpha$, $\beta$ and $\gamma$ are 
constants. Although the action \eqref{eq:fourthderivaction} allows renormalisation, or the removal of unwanted divergences in 
Feynmann diagrams \cite{stelle:1977,Veltman:1975vx}, 
the UV modified propagator 
 has an additional  pole compared to General Relativity when the momentum $k_\mu$ satisfies $k^\mu k_\mu=\frac{M^2_P}{2\left(\beta-3\alpha\right)}$ \cite{stelle:1977}.
This extra pole has a negative residue and therefore is an excitation with negative kinetic energy,
known as a ghost \cite{Conroy:2017uds}.

Notably $f(R)$ gravity escapes this fate because such theories can be described 
by a single extra scalar degree of freedom added 
to\ GR, which does not 
necessarily lead to a ghost.

Classically speaking, tachyons are another cause for concern in modified gravity theories.
Particles with imaginary mass will travel faster than light \cite{Feinberg:1967zza},
so to avoid this fate we can only introduce poles at $k^2+m^2$ with real mass 
$m^2\geq 0$.
\subsection{Other modified gravity theories}
In four spacetime dimensions, 
the Gauss-Bonnet term  $G=
R^2-4R^{\mu\nu}R_{\mu\nu} + R^{\mu\nu\rho\sigma}
R_{\mu\nu\rho\sigma}$ is  a topological surface term and so adding it to the Einstein-Hilbert
action has no effect on the equations of motion. \eqref{eq:fourthderivaction}
is sometimes written with an extra term $\gamma R^{\mu\nu\rho\sigma}
R_{\mu\nu\rho\sigma}$, but the action can be separated into  \eqref{eq:fourthderivaction}
and the surface term. 
The $d$-dimensional action
\ba \label{eq:gbaction}
        S = \int d^dx \sqrt{-g} \left(M^2_P R +R^2-4R^{\mu\nu}R_{\mu\nu} + R^{\mu\nu\rho\sigma}
R_{\mu\nu\rho\sigma} \right),
\ea
is known as Gauss-Bonnet gravity, which only modifies gravity in 5 or more spacetime 
dimensions.

\section{Motivation for infinite derivatives}    
The Ostrogradsky instability produces a ghost for a generic theory where a finite number 
of higher derivatives (more than 2) act on terms 
in the action \cite{Woodard:2015zca}. This however does not apply when there are an infinite number of derivatives. These
infinite derivatives were first used in string theory in order to
avoid singularities \cite{Tseytlin:1995uq}
before being applied to gravity \cite{Tomboulis:1997gg,Biswas:2005qr}.

The most general  quadratic\footnote{We look here only at quadratic curvatures,
but future work could look at the effect of cubic or higher order curvatures.} covariant, parity-invariant, metric-compatible, torsion-free action is \cite{Biswas:2005qr}
\begin{eqnarray}\label{fullaction}
S_{\text{IDG}}&=&\int d^{4}x \, \frac{\sqrt{-g}}{2}\Big[M^2_PR+RF_{1}(\Box)R
+R_{\mu\nu}F_{2}(\Box)R^{\mu\nu}+R_{\mu\nu\rho\sigma}F_{3}(\Box)R^{\mu\nu\rho\sigma}\Big],\non\\
&&
\text{with} \quad F_{i}(\Box)=\sum^{\infty}_{n=0}f_{i_{n}}\frac{\Box^{n}}{M^{2n}} \,,
\end{eqnarray}
where \{$f_{i_0},f_{i_1},f_{i_2},...,f_{i_n}$\} are dimensionless 
coefficients and $\Box=g^{\mu\nu} \D_\mu \D_\nu$ is the d'Alembertian 
operator.\footnote{Each $\Box$ term comes with an associated mass scale $M^2$, 
where $M < M_p =
(16\pi G)^{-1/2}\sim 2.4 \times 10^{18}$ GeV. The physical significance of $M$ could be any
scale beyond $10^{-2}$ eV, which arises from constraints on studying the $1/r$-fall of the Newtonian
potential~\cite{Edholm:2016hbt}. When writing the $F_i(\Box)$, we suppress the scale $M$ in order to simplify our formulae for the rest of
this thesis. However, for any physical comparison one has to bring in the scale $M$ along with
$M_p$.} $M$ is a new mass scale, which together with the $f_{i_n}$s regulates the length scale at which the higher derivative 
terms  start to have a significant effect. 

The IDG action has been the subject of many papers in the past decade, 
with researchers looking at the bending of light round the sun \cite{Feng:2017vqd,Xu:2017luq},
exact solutions of the theory \cite{Li2015}, the diffusion equation \cite{Calcagni:2010ab,Calcagni2018ta},
black hole stability \cite{Calcagni:2018pro}, causality \cite{Giaccari:2018nzr},
and time reflection positivity \cite{Christodoulou:2018jbn}.

It was shown in \cite{Gorka:2012hs} that the initial value problem is well-posed
even when there are infinitely many derivatives and that only a finite number of initial conditions are required. Exact plane wave solutions to IDG were studied 
in \cite{Kilicarslan:2019njc}. In the time since this thesis was submitted for viva voce examination, more papers have been released looking at the stability of the Minkowski spacetime \cite{Briscese:2018bny} and the Casimir effect \cite{Buoninfante:2018bkc}.

\newpage
\section{Summary of previous results}
In this section we give a brief summary of the aspects of Infinite Derivative Gravity
which are not covered in the rest of this thesis
but are nevertheless important in their own right. 

\subsection{Entropy}
The entropy $S_{\text{BH}}$ of an axisymmetric black hole was famously found by Bekenstein and
 Hawking to depend on the area of the black hole event horizon \cite{Bekenstein:1973ur,Hawking:1974sw}, with 
 the exact form
\ba
        S_{\text{BH}}=\frac{\text{Area}}{4G}.
\ea
It was found in \cite{Conroy:2015wfa} that in the linear regime, 
this formula was not modified by the addition of the
extra terms in the IDG action,
given the necessary constraints to avoid introducing any extra 
degrees of freedom into the propagator, although
there are corrections at the non-linear level.
It was further found in \cite{Talaganis:2017evj} that the entropy of a rotating Kerr black hole was not modified
by the Ricci scalar or Ricci tensor terms in \eqref{fullaction}, but there would be corrections
from the Riemann tensor term.

In \cite{Conroy:2015nva}, the entropy around an (Anti) de Sitter, or (A)dS, metric in D-dimensions was calculated, giving us
constraints on the coefficients in the action \eqref{fullaction}, which later help us when
calculating the conditions to avoid singularities.
   
\subsection{Boundary terms of the theory}   
The Einstein-Hilbert action contains second derivatives of the metric, so in theory one would need to 
fix the metric and its derivatives on the boundary of any spacetime manifold. However,  the Gibbons-Hawking-York (GHY) boundary terms for GR were found in 1977.
The GHY boundary cancels out the terms generated by the EH action on the boundary \cite{Gibbons:1976ue,Chakraborty:2016yna}. The IDG boundary terms were found 
in \cite{Teimouri:2016ulk} using the Arnowitt-Deser-Miser (ADM) decomposition \cite{poisson_2004}, which splits the spacetime
metric into spatial slices of constant time, followed by a coframe slicing, which leads to a simpler line element. 
This was first done for the example of an infinite derivative scalar field, before extending the method
to the gravitational case. To use the d'Alembertian operator in the coframe,
it was necessary to split it into a spatial derivative operator and a time derivative operator.

It was shown in \cite{Deruelle:2009zk} that for a general action depending on the Riemann tensor 
$R_{\mu\nu\rho\sigma}$ or
its contractions 
\begin{eqnarray}\label{5.1}S=\frac{1}{16\pi G}\int_{\mathcal{M}}d^{4}x \, \sqrt{-g}f(R_{\mu\nu\rho\sigma})\,,
\end{eqnarray}
it is possible to add the auxiliary fields $\varrho_{\mu\nu\rho\sigma}$ and $\varphi^{\mu\nu\rho\sigma}$. 
These fields are independent of the metric and of each other, 
but retain the symmetries of the Riemann tensor. This allows us to write the action \eqref{fullaction} as
\begin{eqnarray}\label{5.2}
S=\frac{1}{16\pi G}\int_{\mathcal{M}}d^{4}x \,\sqrt{-g}\left[f(\varrho_{\mu\nu\rho\sigma})
+\varphi^{\mu\nu\rho\sigma}\left(R_{\mu\nu\rho\sigma}-\varrho_{\mu\nu\rho\sigma}\right)\right]\,.
\end{eqnarray} 
After a long and technical calculation laid out in \cite{Teimouri:2016ulk}, we are able to find the total derivative term in the action by using the Gauss, Codazzi and Ricci equations as well as the 
equations of motion for the auxiliary fields. 
As the boundary term is defined as the term needed to cancel out the 
total derivative term in the action,
therefore
we arrive at the IDG boundary terms 
\ba\label{40}
S_{tot}&=&S_{gravity}+S_{boundary}\nonumber \\
&=&\frac{1}{16\pi G}\int_{\mathcal{M}}
d^{4}x \, \sqrt{-g}\Big[\varrho+\alpha\big(\varrho F_{1}(\Box)\varrho
+\varrho_{\mu\nu}F_{2}(\Box)\varrho^{\mu\nu}\nonumber \\
&&+\varrho_{\mu\nu\rho\sigma}F_{3}(\Box)
\varrho^{\mu\nu\rho\sigma}\big)+\varphi^{\mu\nu\rho\sigma}
\left(R_{\mu\nu\rho\sigma}-\varrho_{\mu\nu\rho\sigma}\right)\Big]\nonumber \\
&& +\frac{1}{8\pi
G}\oint_{\partial\mathcal{M}}d\Sigma_{\mu} \, n^{\mu}
\Big[K+\alpha\big(2KF_{1}(\Box)\rho
-4KF_{1}(\Box)\Omega-KF_{2}(\Box)\Omega \nonumber \\
&&-\left.\left. K_{ij}F_{2}(\Box)\Omega^{ij}+K_{ij}
F_{2}(\Box)\rho^{ij}-4K_{ij}F_{3}(\Box)\Omega^{ij}
-2\Omega_{ij}X_{1}^{ij}-\frac{1}{2}\Omega_{ij}X_{2}^{ij}\right)\right],~~~~~~~~~
\ea
where $\Omega_{ij}= n^\gamma n^\delta\varrho_{\gamma i\delta j}$
is the contraction of the auxiliary field with $n^\mu$, the normal vector to the 
ADM hypersurface and $K_{ij}= - \D_i n_j$ is the extrinsic curvature of the hypersurface.
The $X_{ij}$ terms are additional terms which appear due to the existence of the
d'Alembertian in the action and depend on the choice of coframe slicing.
It is notable that the $F(\Box)$ terms feature in the boundary - the infinite
derivative terms filter through from the bulk.

The fact that the boundary terms can be found  is a useful check
on the theory. The method of finding the 
boundary terms can also be used when calculating the Hamiltonian of the theory. 

\subsection{Hamiltonian analysis of the theory}
By examining the propagator of the theory, we will later show using Lagrangian analysis
that there will be no ghost degrees of freedom. Intuitively, one would expect the same
thing from a Hamiltonian analysis, but it is reassuring to study it explicitly. 

By starting with a toy model of an infinite derivative scalar field and using the 
ADM decomposition, it is
possible to find the Hamiltonian and the constraints that arise \cite{Mazumdar:2017kxr}.
It is pleasing to see that one can show that by choosing the coefficients in the action
correctly, we can make sure there are no extra poles in the theory compared to 
GR.

\subsection{Quantum aspects of the theory}
 The unitarity of the theory, i.e. the requirement that the sum of all possible outcomes
 is equal to unity, was examined in 
\cite{Talaganis:2017ycf} and the Slavnov identities found in \cite{Talaganis:2017dwo}. It was shown that
the mass scale of the theory could vary depending on the number of particles in 
the system \cite{Talaganis:2017dqy}, so that
the effective mass scale would be inversely proportional 
to the square root of the number of particles.

By looking at an infinite derivative scalar
toy model, it was shown that one could cancel the divergences 
in one-loop scattering diagrams and that the
exponential suppression of the propagator could aid convergence 
\cite{Talaganis:2014ida,Talaganis:2016ovm}. 
The behaviour of the 2-loop, 2-point function can also be 
controlled \cite{Talaganis:2016ovm}.
It was shown that one-particle irreducible (1PI) 
Feynmann diagrams could be made UV finite 
(i.e. does not diverge in the ultraviolet, or high energy, regime) by using an infinite derivative scalar toy model \cite{Talaganis:2017tnr}. 
For those wanting a more detailed understanding of the quantum aspects of 
IDG, the review paper \cite{Biswas:2014tua} may be helpful.

Renormalisibility of the theory was also 
studied in more general dimensions (i.e. a number spacetime dimensions not equal to four) \cite{Modesto:2014lga,Modesto:2015lna},
which found that IDG could be described as super-renormalisable.

\newpage
\subsection{Using IDG as a source for the accelerated expansion of the universe}
\begin{figure}[!htbp]
\centering
\includegraphics[width=127mm]{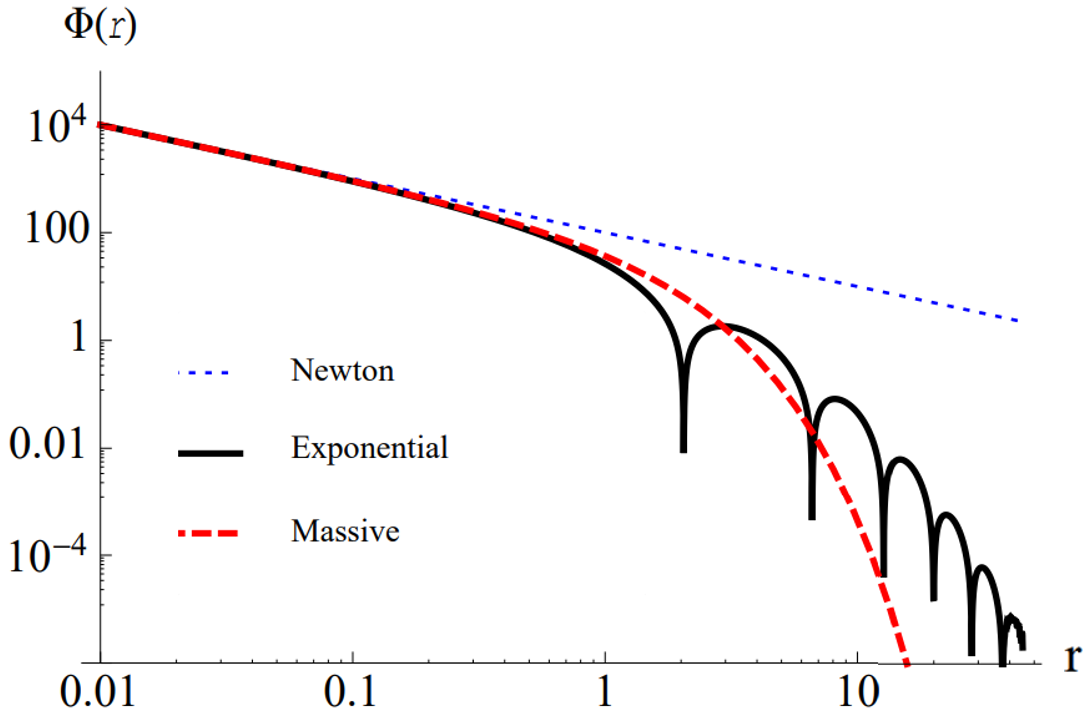}
\caption{A plot of the gravitational potential against the distance from a source in different models. 
The different curves represent the
standard GR prediction, the exponential choice of form factors 
\eqref{eq:exponentialform} \cite{Conroy:2014eja} and the 
pure Yukawa suppression of massive gravity for comparison.
The authors chose a mass of 1m$^{-1}$ for the massive gravity and exponential curves. }
\label{irplot}
\end{figure}
In this thesis, we use IDG to suppress the strength of gravity at small scales in
order to avoid singularities. However, they can also be used to produce effects at 
large distances.

Infinite derivatives, in the form of a infinite series of the inverse d'Alembertian $1/\Box$, have been been used to
explain the accelerated expansion of the universe. This acceleration, first observed in the 1990s, 
requires either some unknown ``dark energy'' (normally assumed to have the form of a cosmological constant $\Lambda$) to drive expansion or modifications to the
gravitational action \cite{Perlmutter:1998np,Riess:1998cb,Heymans:2012gg}.

\textit{Massive gravity} actions extend the Einstein-Hilbert action \eqref{eq:EHaction} with  
terms of the form $\frac{m_{gr}^2}{\Box^2} R$, where $m_{gr}$ is the
non-zero mass of the graviton 
\cite{Rham:2015mxa,Maggiore:2013mea,Maggiore:2014sia,Dirian:2014xoa,Foffa:2013vma,Modesto:2013jea}. 

An infinite series of inverse d'Alembertian operators 
were used 
to generate the accelerated expansion in \cite{Conroy:2014eja}.
The authors used inverse d'Alembertian operators instead of
d'Alembertians, so that the action took the form of \eqref{fullaction}
with the $F_i(\Box)$ replaced with
\ba \label{eq:inverseboxchoice}
        \bar{F}_i(\Box)= \sum_{n=1}^\infty f_{i_{-n}} (M^2/\Box)^n,
\ea
where the $f_{i_{-n}}$ are dimensionless coefficients.
The non-linear field equations were found, 
allowing the
authors to derive the Newtonian potential at large distances shown in Fig. \ref{irplot}.
For this plot, the form factors (terms which appear in the equations of motion)
\ba
        \bar{a}(\Box)&\equiv& 1 +M^{-2}_P\left(\frac{1}{2}\bar{F}_2(\Box)
        + \bar{F}_3(\Box)\right)\Box, \non\\
        \bar{c}(\Box)&\equiv& 1+M^{-2}_P\left(-2\bar{F}_1(\Box)
         -\frac{1}{2}\bar{F}_2(\Box)
        + \frac{1}{3}\bar{F}_3(\Box) \right)\Box,
\ea 
were chosen so that
\ba \label{eq:exponentialform}
        \bar{a}(\Box)=\bar{c}(\Box)=\exp(-M^2/\Box).
\ea        

The authors of \cite{Conroy:2014eja} found that the choice \eqref{eq:exponentialform} gave two modifications to the Newtonian potential.
Firstly, similar to other massive gravity theories, the potential was exponentially suppressed 
at large distances.
 Secondly, the potential was modified by an oscillating factor $\cos(Mr)$. This would allow IDG to be differentiated from other massive gravity theories by observations.
\vspace{0.9cm}

\newpage   
\section{Organisation of thesis}   
The content of this thesis is organised as follows:
\vspace{5mm}

\hspace{-6.1mm}\textbf{Chapter 2}:   
We show that the IDG action can be derived from the most general metric-based,
 torsion-free, 
quadratic in curvature action. We give the full non-linear equations 
of motion and use these to derive the equations of motion around a flat 
background and an (Anti) de Sitter, or (A)dS background. 
Finally we find the propagator around a Minkowski background, 
giving the necessary conditions to avoid ghosts, and give the 
quadratic variation of the action around maximally symmetric backgrounds 
(Minkowski and (A)dS), which we later use to investigate perturbations
in the Cosmic Microwave Background.
 
\vspace{5mm}\hspace{-6.1mm}\textbf{Chapter 3}: We find the Newtonian potential, 
which is the perturbation to a background metric caused by a static, spherically
symmetric point source. We first look at a perturbation around a flat background
and use this to constrain 
our mass scale as well as show that we return to the GR prediction at large distances. 
We later look at a de Sitter background and show that using the Minkowski metric as a background is
an acceptable approximation for a de Sitter background with the current Hubble 
parameter.
 
\vspace{5mm}\hspace{-6.1mm}\textbf{Chapter 4}:
We examine the concept of a singularity and how GR will always generate 
geometric singularities due to the Hawking-Penrose singularity theorems. 
This is because GR requires null rays to \textit{focus},
which means that one cannot trace the path of light to past infinity,
known as geodesic incompleteness. 
We show how IDG can avoid this fate by fulfilling the defocusing conditions, 
both for perturbations around Minkowski \& (A)dS backgrounds, 
and for a background Friedmann-Robertson-Walker (FRW) solution, thus solving the Big Bang Singularity Problem.  

\newpage
\vspace{5mm}\hspace{-6.1mm}\textbf{Chapter 5}: IDG can be thought of as an 
extension to Starobinsky inflation and we show how this changes the inflationary observables
in the Cosmic Microwave Background. This allows us to use data from the Planck satellite 
to place further constraints on our mass scale.

\vspace{5mm}\hspace{-6.1mm}\textbf{Chapter 6}: We examine how IDG affects 
the production and propagation of gravitational radiation.
 First we derive the modified quadrupole formula, 
 then look at the backreaction produced by the emission of gravitational waves  
 and give a prediction for the power produced by a system. 
 We take the example of a binary system in both a circular and an 
 elliptical orbit and use the Hulse-Taylor binary to constrain our mass scale.

\chapter{Infinite Derivative Gravity}
\label{chap:idg}
\section{Infinite Derivative Gravity}
We begin by deriving the action \eqref{fullaction}.
Here we look at the derivation around Minkowski spacetime \cite{Biswas:2011ar}, but 
the calculation can also be carried out in a maximally symmetric spacetime
such as (Anti) de Sitter \cite{Biswas:2016egy}.
Our starting point is the most general, quadratic in curvature, covariant, metric-tensor based modification 
to the Einstein-Hilbert action 
in four dimensions, which is
given by  \cite{Biswas:2011ar}
\ba \label{eq:mostgeneralaction}
        S= \int d^4x \sqrt{-g} \left(R_{\mu_1\nu_1\lambda_1\sigma_1} 
        \mathcal{O}^{\mu_1\nu_1\lambda_1\sigma_1}_{\mu_2\nu_2\lambda_2\sigma_2}
        R^{\mu_2\nu_2\lambda_2\sigma_2}\right),
\ea        
where the operator $\mathcal{O}^{\mu_1\nu_1\lambda_1\sigma_1}_{\mu_2\nu_2\lambda_2\sigma_2}$
is a general covariant operator  and all the possible 
Riemannian curvatures (Riemann, Weyl and Ricci tensors and Ricci scalar)
are represented by $R_{\mu_1\nu_1\lambda_1\sigma_1}$. 

In full, \eqref{eq:mostgeneralaction} is given by\footnote{There is no term
of the form $F(\Box)R$ as this would be
a total derivative and therefore equivalent to
a boundary term.} \cite{Biswas:2011ar} 
\begin{eqnarray} \label{eq:generalactionaroundminko}
S&=&\int d^4x\frac{\sqrt{-g}}{2}\Big[M_P^2R+R F_1(\Box)R+R F_2(\Box)\nabla_{\nu}\nabla_{\mu}R^{\mu\nu}+R_{\mu\nu} F_3(\Box)R^{\mu\nu}\nonumber \\
&+&R_{\mu}^{\nu} F_4(\Box)\nabla_{\nu}\nabla_{\lambda}R^{\mu\lambda}+R^{\lambda\sigma} F_5(\Box)\nabla_{\mu}\nabla_{\sigma}\nabla_{\nu}\nabla_{\lambda}R^{\mu\nu}+R F_6(\Box)\nabla_{\mu}\nabla_{\nu}\nabla_{\lambda}\nabla_{\sigma}R^{\mu\nu\lambda\sigma}\nonumber \\
&+&R_{\mu\lambda} F_7(\Box)\nabla_{\nu}\nabla_{\sigma}R^{\mu\nu\lambda\sigma}+R_{\lambda}^{\rho} F_8(\Box)\nabla_{\mu}\nabla_{\sigma}\nabla_{\nu}\nabla_{\rho}R^{\mu\nu\lambda\sigma} \nonumber \\
& + &R^{\mu_1\nu_1} F_9(\Box)\nabla_{\mu_1}\nabla_{\nu_1}\nabla_{\mu}\nabla_{\nu}\nabla_{\lambda}\nabla_{\sigma}R^{\mu\nu\lambda\sigma}+ R_{\mu\nu\lambda\sigma} F_{10}(\Box)R^{\mu\nu\lambda\sigma} \nonumber \\
& + & R^{\rho}{ }_{\mu\nu\lambda} F_{11}(\Box)\nabla_{\rho}\nabla_{\sigma}R^{\mu\nu\lambda\sigma}+ R_{\mu\rho_1 \nu\sigma_1} F_{12}(\Box)\nabla^{\rho_1}\nabla^{\sigma_1}\nabla_{\rho}\nabla_{\sigma}R^{\mu\rho\nu\sigma}
\nonumber\\ &+&
R_{\mu}^{\;\nu_1\rho_1\sigma_1}F_{13}(\Box)\nabla_{\rho_1}\nabla_{\sigma_1}\nabla_{\nu_1}\nabla_{\nu}\nabla_{\lambda}\nabla_{\sigma}R^{\mu\nu\lambda\sigma} \nonumber \\
& + & R^{\mu_1\nu_1\rho_1\sigma_1} F_{14}(\Box)\nabla_{\rho_1}\nabla_{\sigma_1}
\nabla_{\nu_1}\nabla_{\mu_1}\nabla_{\mu}\nabla_{\nu}\nabla_{\lambda}\nabla_{\sigma}R^{\mu\nu\lambda\sigma}\Big]\,.
\label{d4}
\end{eqnarray}

We can simplify \eqref{eq:generalactionaroundminko} by  
commuting derivatives, recalling that we are examining fluctuations around a Minkowski 
background. 
Additionally, the antisymmetric nature of the Riemann tensor \eqref{eq:riemannantisymm}
and the (second) Bianchi identity \eqref{eq:secondbianchi} allow us to write the action in a simplified form.
For example, we can commute the $\mu$ and $\nu$ indices in the $F_6(\Box)$ term as follows
\ba
         \D_\mu \D_\nu \D_\lambda \D_\sigma 
        R^{\mu\nu\lambda\sigma} &=& 
       \frac{1}{2} \left[\D_\mu \D_\nu \D_\lambda \D_\sigma 
        R^{\mu\nu\lambda\sigma}+\D_\nu \D_\mu \D_\lambda \D_\sigma 
        R^{\mu\nu\lambda\sigma}\right]\non\\
        &=& 
       \frac{1}{2} \left[\D_\mu \D_\nu \D_\lambda \D_\sigma 
        R^{\mu\nu\lambda\sigma}+\D_\mu \D_\nu \D_\lambda \D_\sigma 
        R^{\nu\mu\lambda\sigma}\right]\non\\
        &=& 0.
\ea        
where we relabeled the indices on the second term to get to the second line
and used the antisymmetry of the Riemann tensor to get to the third line.
By repeating this procedure for the other terms, we can write
\eqref{eq:generalactionaroundminko} as \eqref{fullaction}, i.e.
 \cite{Biswas:2011ar}\begin{eqnarray}\label{fullaction2}
S&=&\int d^{4}x \, \frac{\sqrt{-g}}{2}\Big[M^2_PR+RF_{1}(\Box)R
+R_{\mu\nu}F_{2}(\Box)R^{\mu\nu}+R_{\mu\nu\rho\sigma}F_{3}(\Box)R^{\mu\nu\rho\sigma}\Big],\non\\
&&
\text{with} \quad F_{i}(\Box)=\sum^{\infty}_{n=0}f_{i_{n}}\frac{\Box^{n}}{M^{2n}}
\,.
\end{eqnarray}

\subsection{Equations of motion}
We are now ready to derive the equations of motion for the Infinite Derivative Gravity action \eqref{fullaction2}, following the method of \cite{Biswas:2013cha}. 
Using the identities in Appendix \ref{sec:variationcurvature}
and Appendix \ref{sec:variationdalembertianactingcurvature}, 
one can proceed by varying the action to find the gravitational
stress energy tensor $T^{\mu\nu}$ 
\ba
        T^{\mu\nu} \equiv - \frac{2}{\sqrt{-g}} \frac{\delta S}{\delta g_{\mu\nu}}.
\ea
 The full equations of motion for the action (\ref{fullaction}) are given by \cite{Biswas:2013cha}
\ba \label{eq:fullequationofmotion}
                T^{\alpha\beta} &=& M^2_P G^{\alpha\beta}  +2 G^{\alpha\beta} F_1(\Box) R 
                +2 g^{\alpha\beta} R F_1 (\Box)R 
                - 2 \left(\nabla^\alpha \nabla^\beta - g^{\alpha\beta} \Box\right) F_1 (\Box)R \nonumber\\
                &&-  \Omega^{\alpha\beta}_1 + \frac{1}{2}g^{\alpha\beta} \left(\Omega^\sigma_{1\sigma}+\bar{\Omega}_1\right) 
                + 2 {R^{(\beta}}_\mu F_2(\Box)R^{\mu\alpha)}
                -\frac{1}{2}g^{\alpha\beta} R^{\mu\nu} F_2 (\Box)R_{\mu\nu}\nonumber\\
                && - 2 \D^{(\alpha} \D_\mu F_2 (\Box)R^{\mu\beta)} + \Box F_2 (\Box)R^{\alpha\beta}  
                +  g^{\alpha\beta} \D_\mu \D_\nu F_2 (\Box)R^{\mu\nu}  -  \Omega^{\alpha\beta}_2 \nonumber\\
                &&     + \frac{1}{2}g^{\alpha\beta} \left( \Omega^\sigma_{2\sigma} 
                + \bar{\Omega}_2 \right) - 2 \Delta^{\alpha\beta}_2  -\frac{1}{2}g^{\alpha\beta} C^{\mu\nu\lambda\sigma} F_3 (\Box)C_{\mu\nu\lambda\sigma} 
                + 2 {C^\alpha}_{\rho\theta\psi} F_3 (\Box)C^{\beta\rho\theta\psi}\nonumber\\
                && - 2 \left[ 2 \nabla_\mu \nabla_\nu + R_{\mu\nu} \right] F_3 (\Box)C^{\beta\mu\nu\alpha} -  \Omega^{\alpha\beta}_3 +\frac{1}{2} g^{\alpha\beta} \left(\Omega^\gamma_{3\gamma} + \bar{\Omega}_3 \right) - 4 \Delta^{\alpha\beta}_3, 
\ea
where
the $C_{\mu\nu\rho\sigma}$ is the Weyl curvature tensor and we have defined 
\ba
        \Omega^{\alpha\beta}_1 &=& \sum^\infty_{n=1} f_{1_n} \sum^{n-1}_{l=0} 
        \nabla^\alpha \Box^l R \nabla^\beta \Box^{(n-l-1)}R, 
        \quad \bar{\Omega}_1 = \sum^\infty_{n=1} f_{1_n} \sum^{n-1}_{l=0} \Box^{l}R  \Box^{(n-l)} R, \nonumber\\
        \Omega_2^{\alpha\beta} &=& \sum^\infty_{n=1} f_{2_n} \sum^{n-1}_{l=0} 
        \D_\nu \left[\Box^l  R^{\nu}_\sigma \Box^{(n-l-1)} R^{(\beta|\sigma|;\alpha)} 
        - \Box^l R^{\nu;(\alpha}_\sigma \Box^{(n-l-1)} R^{\beta)\sigma}\right], \nonumber\\
        \Delta_2^{\alpha\beta} &=& \sum_{n=1}^\infty f_{2_n} \sum_{l=0}^{n-1} 
       \D_\nu \left[\Box^l R^\nu_\sigma \D^{(\alpha}\Box^{(n-l-1)} R^{\beta)\sigma}
        -\D^{(\alpha} \Box^l R^\nu_\sigma \Box^{(n-l-1)} R^{\nu\sigma}\right], \non\\
        \Omega^{\alpha\beta}_3 &=& \sum^\infty_{n=1} f_{3_n} \sum^{n-1}_{l=0}  \D^\alpha \Box^l
        C^{\mu }{}_{\nu\lambda\sigma}\D^\beta \Box^{(n-l-1)} {C_\mu}^{\nu\lambda\sigma},
        \quad \bar{\Omega}_3 =  \sum_{n=1}^\infty f_{3_n} \sum^{n-1}_{l=0} \Box^l 
        C^{\mu}{}_{\nu\lambda\sigma}\Box^{n-1} C_\mu{}^{\nu\lambda\sigma}, \nonumber\\
        \Delta^{\alpha\beta}_3 &=& \frac{1}{2} \sum_{n=1}^\infty f_{3_n} \sum^{n-1}_{l=0} \D_\nu \left[\Box^l C^{\lambda\nu }{}_{\sigma\mu} \D^{(\alpha} \Box^{(n-l-1)}{C_\lambda}^{\beta)\sigma\mu} 
        -\D^{(\alpha} \Box^l C^{\lambda\nu}{}_{\sigma\mu} C_\lambda{}^{\beta)\sigma\mu(n-l-1)} \right].~~~~~~~~
\ea
\newpage
As expected, these equations satisfy the generalised Bianchi identities \\
$\nabla^\alpha T_{\alpha\beta}=0$. This is because the action \eqref{fullaction2} is metric-based, and any metric-based Lagrangian will be diffeomorphism invariant
and therefore satisfy the generalised Bianchi identities
 \cite{Carroll:2004st}.

By contracting \eqref{eq:fullequationofmotion} with $g_{\alpha\beta}$,
we find that the trace equation is 
\ba \label{eq:tracefulleqns}
                T &=& g_{\alpha\beta} T^{\alpha\beta} 
                = -M^2_P R + 6 \Box F_1 (\Box)R  + \Omega^\sigma_{1\sigma} +2 \bar{\Omega}_1
                  -2 \nabla_\mu \nabla_\nu \left(F_2 (\Box)R^{\mu\nu} \right) +\Omega^\sigma_{2\sigma} \nonumber\\
                && +2 \bar{\sigma}_2  - 2 \Delta^\alpha_{2\alpha} -2 C^{\mu\nu\lambda\sigma} F_3 (\Box)C_{\mu\nu\lambda\sigma} + 2 C_{\beta\mu\nu\sigma} F_3 (\Box)C^{\beta\mu\nu\sigma}
                 + \Omega^\gamma_{3\gamma} + 2\bar{\Omega}_3   - 4 \Delta^\alpha_{3\alpha.}~~~~~~~~
\ea
\section{Equations around various backgrounds}
We now simplify the equations by linearising around specific backgrounds.
We assume a background solution $\bar{g}_{\mu\nu}$ 
and then make a small variation $h_{\mu\nu}$, 
so that\footnote{We define $h^{\mu\nu}=- \delta g^{\mu\nu}$, which follows from the requirement that the Kronecker delta is invariant under 
a linear variation to the metric, up to linear order in the variation \cite{Conroy:2017uds}.} $g_{\mu\nu}\to \bar{g}_{\mu\nu} +h_{\mu\nu}$,
$g^{\mu\nu}\to \bar{g}^{\mu\nu} - h^{\mu\nu}$. 
We later use the linear equations of motion to see what happens for perturbations
caused by a small amount of matter, or random fluctuations caused by quantum effects.

\subsection{Equations around a flat background}
For a flat background, we can make two simplifications. 
Firstly, the background curvatures are zero,
so any terms in \eqref{eq:fullequationofmotion} which are quadratic in the curvatures will 
not contribute to the linearised equations of motion. 
Secondly, for linearised perturbations around a flat background, 
covariant derivatives commute.
 
For the perturbed metric $g_{\mu\nu}\to \eta_{\mu\nu} +h_{\mu\nu}$,
$g^{\mu\nu}\to \eta^{\mu\nu} - h^{\mu\nu}$, we can write the linearised 
equations of motion as~\cite{Biswas:2011ar}
\ba \label{eq:minkofulleqns}
        T^{\alpha\beta} &=& M^2_P \left(r^{\alpha\beta}-\frac{1}{2}g^{\alpha\beta}r\right) 
        - 2 \left(\nabla^\alpha \nabla^\beta - g^{\alpha\beta} \bar{\Box}\right) F_1 (\bar{\Box})r \nonumber\\
                && - 2 \D^{(\alpha} \D_\mu F_2 (\bar{\Box})r^{\mu\beta)} + \bar{\Box} F_2 (\bar{\Box})r^{\alpha\beta}  
                +  g^{\alpha\beta} \D_\mu \D_\nu F_2 (\bar{\Box})r^{\mu\nu}   \nonumber\\
                &&     -4 \nabla_\mu \nabla_\nu  F_3 (\bar{\Box})c^{\beta\mu\nu\alpha},  
\ea
where the background d'Alembertian is denoted by $\bar{\Box}$, $c_{\mu\nu\rho\sigma}$ is the linearised
Weyl tensor, and the linearised Riemann and Ricci curvatures around a flat background are 
\ba \label{eq:linaerisedcurvmink}
        &&r_{\rho\mu\sigma\nu} = \frac{1}{2} \left( \partial_\sigma \partial_\mu h_{\rho\nu} 
        + \partial_\nu \partial_\rho h_{\mu\sigma} - \partial_\nu \partial_\mu h_{\rho\sigma} 
        - \partial_\sigma \partial_\rho h_{\mu\nu} \right), \non\\
        &&r_{\mu\nu} = \frac{1}{2} \left(\partial_\sigma \partial_\mu  h^\sigma_\nu + \partial_\nu \partial_\sigma h^\sigma_\mu 
        - \partial_\nu \partial_\mu h - \bar{\Box} h_{\mu\nu} \right),\non\\
        &&r = \partial_\mu \partial_\nu h^{\mu\nu} - \bar{\Box} h.
\ea
We can write \eqref{eq:minkofulleqns} in the simplified form
\ba \label{eq:minkowskieqns}       
        \frac{1}{M^2_P} T_{\mu\nu} &=& a(\bar{\Box}) r_{\mu\nu} -\frac{1}{2}g_{\mu\nu}
c(\bar{\Box})r -\frac{1}{2}\p_\mu \p_\nu f(\bar{\Box})r, 
\ea   
where the functions $a(\Box)$, $c(\Box)$ and $f(\Box)$ are defined as
\ba \label{eq:acfdefinitionsmink}
        a(\bar{\Box}) &=& 1 + M^{-2}_P \bigg( F_2(\bar{\Box}) + 2 F_3(\bar{\Box}) \bigg)
\bar{\Box}, \nonumber\\
        c(\bar{\Box}) &=& 1 + M^{-2}_P \left( -4 F_1(\bar{\Box}) - F_2(\bar{\Box}) + \frac{2}{3}
F_3(\bar{\Box})\right) \bar{\Box}, \nonumber\\
        f(\bar{\Box}) &=& M^{-2}_P \left( 4 F_1(\bar{\Box}) + 2 F_2(\bar{\Box}) + \frac{4}{3}
F_3(\bar{\Box}) \right).    
\ea            
Note that $\bar{\Box} f(\bar{\Box}) = a(\bar{\Box})-c(\bar{\Box})$ and that contracting \eqref{eq:minkowskieqns} with $g^{\mu\nu}$ leads to the trace equation of motion around Minkowski spacetime
\ba \label{eq:traceeqnaroundminkowski}
      \frac{1}{M^2_P} T &=&\frac{1}{2}\left( a(\bar{\Box})  -3
c(\bar{\Box})\right)r .  
\ea         
In terms of the perturbation to the metric $h_{\mu\nu}$, 
\eqref{eq:minkowskieqns} can be written as \cite{Conroy:2017uds}
\ba \label{eq:lineareomsintermsofh}
        -2\kappa T_{\mu\nu} \hspace{-2mm}&=&\hspace{-2mm} a(\bar{\Box}) \left(\bar{\Box} h_{\mu\nu}
         \hspace{-0.8mm}-\hspace{-0.8mm} \partial_\sigma  \left( \partial_\mu h^\sigma_\nu + \partial_\nu h^\sigma_\mu\right) \right)
        + c(\bar{\Box})\left( \partial_\mu \partial_\nu h +\eta_{\mu\nu} 
        \partial_\sigma \partial_\tau h^{\sigma\tau} - \eta_{\mu\nu} \bar{\Box} h \right) \non\\
        &&+ f(\bar{\Box}) \hspace{1mm}\partial_\mu 
        \partial_\nu \partial_\sigma \partial_\tau h^{\sigma\tau},
\ea    
where $\kappa=M^{-2}_P$.
\subsection{Equations around a de Sitter background}
The de Sitter (dS) metric is a subset of the general FRW metric
\eqref{eq:FRWmetric} with the scale factor $a(t)=e^{Ht}$ where 
$H$, the Hubble parameter $H=\frac{\dot{a}}{a}$, is a constant here. 
The scale factor is a dimensionless quantity which 
describes the evolution of proper distance between two locally stationary objects.
The line element can be written as
\ba \label{eq:desittermetricwithexp}
        ds^2= -dt^2 + e^{2Ht}\left(dx^2+dy^2+dz^2\right).
\ea        
The dS metric has the property of being maximally symmetric, which for 4 spacetime dimensions means that 
there are 10 Killing vectors \cite{Carroll:2004st}. 
We therefore have the constant background curvatures
\ba
        \bar{R}^\mu{}_{\nu\rho\sigma}= \frac{\bar{R}}{12}\left(\delta^\mu_\rho g_{\nu\sigma} - \delta^\mu_\sigma g_{\nu\rho}\right), 
        \quad \bar{R}^\mu_\nu = \frac{\bar{R}}{4} \delta^\mu_\nu,\quad \bar{R} =12H^2.
\ea 
It can be seen that dS is a vacuum solution, i.e. a solution which fulfils the equations of motion with a zero stress-energy tensor, to IDG by examining the trace equation \eqref{eq:tracefulleqns}. The background
curvatures are constants and the Weyl tensor is zero, so the only terms which contribute to the equations of motion are
the GR terms, which for $T=0$ gives us $\bar{R}=4M^{-2}_P\Lambda$. The linearised Ricci curvatures around a de Sitter background are given by \cite{Biswas:2016etb}
\ba \label{eq:linearisedcurvds}
        r^\mu_\nu &=& \frac{1}{2} \left( \bar{\D}_\sigma \bar{\D}^\mu h^\sigma_\nu 
        + \bar{\D}_\sigma \bar{\D}_\nu h^{\sigma\mu} - \bar{\D}_\nu \bar{\D}^\mu h - \bar{\Box} h_\nu^\mu \right) - \frac{\bar{R}}{4} h_\nu^\mu, \nonumber\\
        r &=& \bar{\D}_\mu \bar{\D}_\nu h^{\mu\nu} - \bar{\Box} h -\frac{\bar{R}}{4} h.
\ea   
We will use the traceless Einstein tensor \cite{Biswas:2016etb}
\ba \label{eq:tracelesseinstein}
        S^\mu_\nu \equiv R^\mu_\nu -\frac{1}{4}\delta^\mu_\nu R,
\ea        
and define
\ba
        \tilde{F}_1(\Box) \equiv F_1(\Box) + \frac{1}{4} F_2(\Box),
\ea
so that we can recast the action \eqref{fullaction2} as 
\ba \label{eq:actionwithtraceless}
        S= \int d^4 x \frac{\sqrt{-g}}{2}\left[ M^2_P +R \tilde{F}_1(\Box) R + S^\mu_\nu F_2(\Box) S^\nu_\mu + C_{\mu\nu\rho\sigma} F_3(\Box) C^{\mu\nu\rho\sigma}-2\Lambda\right],~~~~
\ea         
where we have included a cosmological constant $\Lambda$. Around a de Sitter background, we can write the linearised equations of motion for the 
action \eqref{eq:actionwithtraceless} as\footnote{The modified Gauss-Bonnet 
term $GB_n \equiv R \Box^n R\ -4 R_{\mu\nu} \Box^n R^{\mu\nu}
+ R_{\mu\nu\alpha\beta} \Box^n R^{\mu\nu\alpha\beta}$ does not contribute to 
the equations of motion \cite{Li:2015bqa}. We have used this fact  to discard 
the $F_3(\Box)$ term in the action.} 
\ba \label{eq:dsfieldeqns} 
        && \kappa T^\mu_\nu = \left(1 + 24  M^{-2}_p H^2  \tilde{f}_{1_0}+(\bar{\Box}+ 6 H^2) M^{-2}_p F_2 (\bar{\Box}) \right)  
        r^\mu_\nu - \delta^\mu_\nu \Lambda \nonumber\\
       && -\frac{1}{2}  \left( 1+ 24  M^{-2}_p H^2 \tilde{f}_{1_0}
       -4(\bar{\Box}+3H^2)M^{-2}_p \tilde{F}_1 (\bar{\Box})
        \right)\delta^\mu_\nu r  -2 M^{-2}_p\bar{\D}^\mu \bar{\D}_\nu  \tilde{F}_1 (\bar{\Box}) r\nonumber\\
       &&+ M^{-2}_p \delta^\mu_\nu \bar{\D}_\sigma \bar{\D}^\tau F_2 (\bar{\Box}) r^\sigma_\tau
       - 2  M^{-2}_p \bar{\D}^\sigma \bar{\D}_{(\nu} F_2 (\bar{\Box}) r^{\mu)}_\sigma. 
\ea
where $\tilde{f}_{1_0}$ is the zeroth order coefficient of $\tilde{F}_1(\Box)$. We can simplify \eqref{eq:dsfieldeqns}  using the commutation relation given in~\cite{Biswas:2016etb} for a generic scalar $S$,
\ba
        \D_\mu F(\Box) S = F(\Box-3H^2) \D_\mu S,
\ea          
and also the following relations, where covariant derivatives and d'Alembertians with bars use the background dS metric.
For a generic scalar $S$, one can verify order by order that 
\ba 
        \bar{\D}^\mu F(\bar{\Box}) \bar{\D}_\nu S &=& F(\bar{\Box}-5H^2)\bar{\D}^\mu \bar{\D}_\nu       
          S\non\\
           &&+ \frac{1}{4} \left(F(\bar{\Box}+3H^2) -F(\bar{\Box}-5H^2)\right) \delta^\mu_\nu \bar{\Box} S,~~~~
\ea
\newpage
and for a symmetric tensor $t_{\mu\nu}=t_{(\mu\nu)}$
\ba \label{eq:commutatorofdwithfboxforranktwotensor}
         \bar{\D}_\mu F(\bar{\Box}) t^{\mu\nu} 
        =F\left(\bar{\Box}+5H^2\right) \bar{\D}_\mu t^{\mu\nu} - 2 H^2 X(\bar{\Box})\bar{\D}_\nu
        t^\mu_\mu,~~~~~~
\ea      
where $X(\bar{\Box})$ is the infinite sum
\ba \label{eq:definitionofxofboxcommutor}
        X(\bar{\Box}) \equiv \sum^\infty_{n=1}
        f_n \sum_{m=0}^{n-1}
        \left[\left(\bar{\Box}+5H^2\right)^m \left(\bar{\Box}-3H^2\right)^{n-1-m}\right],~~~~~~
\ea
which results from commuting the covariant derivative from the left hand side of $F(\bar{\Box})$
to the right hand side.
 These relations allows us to write the equations of motion in the same simplified form we used around a flat background,
\ba \label{eq:dsfieldeqnssimplified}
        \frac{1}{M^2_P} T^\mu_\nu &=& a(\bar{\Box}) r^\mu_\nu -\frac{1}{2}\delta^\mu_\nu
c(\bar{\Box})r -\frac{1}{2}\bar{\D}^\mu \p_\nu f(\bar{\Box})r, 
\ea   
but where $a$, $c$ and $f$ are now 
\ba 
        &&a(\bar{\Box})\equiv 1 + 24  M^{-2}_p  H^2  \tilde{f}_{1_0}+(\bar{\Box}-2H^2)
 M^{-2}_p F_2 (\bar{\Box}),\nonumber\\
        &&c(\bar{\Box})\equiv1+   M^{-2}_p \Big\{24  H^2 \tilde{f}_{1_0}-4(\bar{\Box}+3H^2)\tilde{F}_1
(\bar{\Box})\nonumber\\
       &&- \frac{1}{2} F_2(\bar{\Box}+8H^2)\bar{\Box}- \frac{1}{2} F_2(\bar{\Box})(\bar{\Box}+8H^2)+4H^2
\bar{\Box} F_2'(\bar{\Box})\Big\}, \nonumber\\
        &&   f(\Box)\equiv  M^{-2}_p
  \left(4\tilde{F}_1
(\bar{\Box})+ 2F_2(\bar{\Box}) -8H^2 X(\bar{\Box}-5H^2) \right),~~~~~~
\ea        
where $X(\bar{\Box})$ is defined in \eqref{eq:definitionofxofboxcommutor}, and we have defined
\ba
        F'_2(\Box)=\sum_{n=1}^\infty f_{2_n} n \Box^{n-1}.
\ea        
Note that unlike the Minkowski case, $a(\Box)-c(\Box)\neq f(\Box)\Box$
in general, and that these equations do not reduce to the Minkowski case
in the limit $H\to 0$ because of the way we recast the action.

\subsection{Propagator}
\subsubsection{Propagator around a flat background}\label{sec:propflat}
We will find the propagator 
by splitting up the perturbed metric into the 
various spin projection operators. Any massive or massless symmetric tensor field can be split up into 
these operators, which represent different degrees of freedom. In particular,
$P^2$ and $P^1$ represent transverse and traceless spin-2 and spin-1 degrees
of freedom,  while $P^0_s$ and $P^0_s$ represent the spin-0 scalar multiplets \cite{Biswas:2013kla}.

These four 
operators form a complete set of projection 
operators, i.e. \cite{VanNieuwenhuizen:1973fi}
\ba
        P^i_a P^j_b = \delta^{ij} \delta_{ab} P^i_a \quad \text{ and } \quad
        P^2 +P^1 +P^0_s+ P^0_w =1.
\ea        
The operators $P^0_{sw}$ and $P^0_{ws}$ represent a mixing of the 
two scalar multiplets, such that \cite{Biswas:2013kla}
\ba
        P^0_{ij} P^0_k = \delta_{jk} P^0_{ij}, \quad \quad P^0_{ij} P^0_{kl}
        = \delta_{il} \delta_{jk} P^0_k, \quad \quad P^0_k P^0_{ij} =\delta_{ik} P^0_{ij}.
\ea         In 4-dimensional spacetime, the operators are defined as \cite{VanNieuwenhuizen:1973fi} 
\begin{subequations}\label{eq:definitionspinprojops}
        \ba  
        P^{(2)}{}_{\mu\nu\rho\sigma} &=& \frac{1}{2} \left( \theta_{\mu\rho} \theta_{\nu\sigma} + \theta_{\mu\sigma} \theta_{\nu\rho} \right) - \frac{1}{3} \theta_{\mu\nu} \theta_{\rho\sigma}, \\
        P^{(1)}{}_{\mu\nu\rho\sigma} &=& \frac{1}{2} \left( \theta_{\mu\rho} \omega_{\nu\sigma} + \theta_{\mu\sigma} \omega_{\nu\rho} + \theta_{\nu\rho} \omega_{\mu\sigma} + \theta_{\nu\sigma} \omega_{\mu\rho} \right),      \\
        P^{(0)}_s{}_{\mu\nu\rho\sigma} &=& \frac{1}{3} \theta_{\mu\nu} \theta_{\rho\sigma},\quad \quad
       ~~\hspace{1.2mm} P^{(0)}_w{}_{\mu\nu\rho\sigma} = \omega_{\mu\nu} \omega_{\rho\sigma},\\
        P^{(0)}_{sw}{}_{\mu\nu\rho\sigma} &=&  \frac{1}{\sqrt{3}} \theta_{\mu\nu} \omega_{\rho\sigma},\quad \quad
        P^{(0)}_{ws}{}_{\mu\nu\rho\sigma} = \frac{1}{\sqrt{3}} \omega_{\mu\nu} \theta_{\rho\sigma},
\ea
\end{subequations}
where the transverse and longitudinal projectors in  momentum space are respectively 
\ba \label{eq:defnofthetaomega}
        \theta_{\mu\nu} = \eta_{\mu\nu} + \frac{1}{k^2}k_\mu k_\nu,
        \quad  \quad\omega_{\mu\nu} = - \frac{1}{k^2} k_\mu k_\nu,
\ea
where $k^2=k^\alpha k_\alpha$, so that the flat spacetime metric is split up as $\eta_{\mu\nu}=\theta_{\mu\nu}+\omega_{\mu\nu}$.
 One can use the relation
\ba
        \Pi^{-1}_{\mu\nu}{}^{\rho\sigma} h_{\rho\sigma}=M^{-2}_P T_{\mu\nu},
\ea
where $\Pi^{-1}$ is the inverse propagator, to split the individual components of  the equations of motion (in the
form \eqref{eq:lineareomsintermsofh})  into the spin projection operators. 
For example, by transforming into momentum space using $\p_\mu \to i k_\mu$, one finds that the $c(\Box)$ term becomes
\ba
        c(\Box) \left(\eta_{\mu\nu} \p_\rho\p_\sigma h^{\rho\sigma} + 
        \p_\mu \p_\nu h -\eta_{\mu\nu} \Box h\right)
        \to - c(-k^2)k^2 \big[  P^0_w + 3 P^0_s  \big]_{\mu\nu}{}^{\rho\sigma} h_{\rho\sigma}.
\ea     
 
By repeating this calculation for each term and merging the spin projection operators, one finds that around a flat background, the propagator is given by
\cite{Biswas:2011ar,Biswas:2013kla,Buoninfante:2016iuf} 
\begin{eqnarray}\label{piforidg}
        \Pi(k^2)_{\text{IDG}}&=& \frac{P^{(2)}}{k^2a(-k^2)}+\frac{P_s^{(0)}}{k^2\left(a(-k^2)-3c(-k^2)\right)},
\end{eqnarray}
where $a$ and $c$ are defined in \eqref{eq:acfdefinitionsmink} and $P^{(2)}$ 
and $P^{(0)}_s$ are the spin-2 and spin-0 parts of the propagator respectively. 
We note that $\nabla_\mu \to i k_\mu$ in momentum space, so $\Box\to -k^2=-k_\mu k^\mu$. 

We want to show that there are no ghosts (excitations with negative kinetic energy, which lead to instabilities \cite{Abramo:2009qk,Conroy:2017uds}), which are generated when there 
are negative residues in the propagator.
The simplest choice which would not generate any negative residues is to have no poles in the propagator. 
We therefore set 
\ba \label{eq:choiceofaequalsc}
        a(-k^2)=c(-k^2)=e^{\gamma(-k^2)},
\ea        
where $\gamma$ is an entire function which is defined below. In order to
remove singularities, we need to suppress gravity at high energy scales,
so we require that $\gamma>0$ as $k\to \infty$.

\textbf{Definition 2.2.1}
A function that is analytic at each point on the ``entire'' (finite)
complex plane is known as an \textbf{entire function}. 

This definition depends on the notion of a function being analytic, which is defined as \cite{ablowitz2003}:

\textbf{Definition 2.2.2}
A function $f(z)$, where $z\in \mathbb{C}$, is said to be \textbf{analytic} at
a point $z_0$ if it is differentiable in a \textit{neighbourhood} of $z_0$.
Similarly, a function $f(z)$ is said to be \textbf{analytic in a region} if it 
is analytic in every point in that region.

By definition, the 
exponential of an entire function has no zeroes and therefore there are no poles in the propagator. 
While this choice might seem overly restrictive, it transpires that all possible functions with no zeroes are encapsulated by the choice
\eqref{eq:choiceofaequalsc} \cite{Conroy:2017uds}.

Inserting \eqref{eq:choiceofaequalsc} into \eqref{piforidg} means that our propagator then simplifies to 
\begin{eqnarray}\label{piforidgwithaequalc}
        \Pi(k^2)_{\text{IDG}}&=& \frac{1}{a(-k^2)}\left(\frac{P^{(2)}}{k^2}-\frac{P^{(0)}_s}{2k^2}\right)\,,
\end{eqnarray}
where $a(-k^2)=\exp[\gamma(-k^2)]$. \eqref{piforidgwithaequalc} is simply the GR propagator with the additional factor $1/a(-k^2)$. Hence
we have a clear path to return to GR: $a(-k^2)\to 1$.

\subsubsection{More complicated choices}
While $a=c$ is the most natural choice, we show in Section \ref{sec:defocusing} that it does not avoid the
Hawking-Penrose singularity for perturbations around a flat background. 
We therefore look at a more complicated case. We know that if there are multiple extra poles 
in the propagator compared to GR 
then a ghost is generated \cite{Conroy:2017uds,Biswas:2005qr}, a claim that we will now prove.
A propagator $\Pi$ with two adjacent poles $m_0$ and $m_1$, with $m_0^2<m_1^2$ takes the form
\ba
        \Pi(-k^2) = \left(k^2 +m^2_0\right)\left(k^2+m_1^2\right)\bar{a}(-k^2)
\ea         
where $\bar{a}(-k^2)$ does not contain any roots 
in the range $-m_1^2<k^2<-m^2_0$, and therefore
does not change sign in this range. Decomposing the inverse
propagator into partial fractions yields 
\ba \label{eq:inversepropwithtwopoles}
        \Pi^{-1}(-k^2) = \frac{1}{\bar{a}(-k^2)}
        \left(\frac{1}{m^2_1-m^2_0}\right) \left(\frac{1}{k^2+m_0^2}
        -\frac{1}{k^2+m_1^2} \right)
\ea        
The residues of  \eqref{eq:inversepropwithtwopoles} at $k^2=-m_1^2$
and $k^2=-m_0^2$ have different signs.
This means that one of these poles must be ghost-like \cite{Biswas:2005qr}.

However, a single extra pole will not necessarily produce a ghost, and we can explicitly avoid the production of a ghost by requiring that the residue of the pole is positive. 

We make the following choice which will give us a single extra pole\footnote{More
generally we can make other choices, for example by introducing the entire functions
$\alpha(z)$ and $\beta(z)$ so that the scalar part of the propagator 
becomes $\frac{\beta(-k^2)P^{(0)}_s}
        {e^{\alpha\left(-k^2\right)}(m^2-k^2)}$ \cite{Koshelev:2017ebj}. }  \cite{Conroy:2016sac}
\ba \label{eq:choiceofawithonepole}
        c(\Box) = \frac{a(\Box)}{3}\left[1+2 
        \left( 1 - \alpha M^{-2}_P\Box\right)\tilde{a}(\Box)\right],
\ea 
where $\tilde{a}(\Box)$ is also the exponent of an entire function. By Taylor expanding the trace equation \eqref{eq:traceeqnaroundminkowski} and using \eqref{eq:choiceofawithonepole} to rewrite $a(\Box)-3c(\Box)$,
one finds that the dimensionless coefficient $\alpha$ is given by
\ba
        \alpha=6f_{1_0}+2f_{2_0}-\frac{M^2_P}{M^2}.
\ea        
We therefore produce the propagator 
\begin{eqnarray}\label{piforidgonepole}
        \Pi(k^2)_{\text{IDG}}&=& \frac{1}{a(-k^2)} 
        \left[\frac{P^{(2)}}{k^2}-\frac{m^2}{2\tilde{a}(-k^2)}\frac{P^{(0)}_s}
        {k^2(k^2+m^2)}\right]\,,
\end{eqnarray}
where we have defined the mass of the spin-0 particle as $m^2=M^2_P/\alpha$
and the spin projection operators $P^{(2)}$ and $P^{(0)}$ are defined in \eqref{eq:definitionspinprojops} . 
Excluding the benign GR pole at $k^2=0$, 
\eqref{piforidgonepole} has a single pole
at $k^2=-m^2$. We require $m^2>0$ to avoid an imaginary mass which would produce a tachyon.

\subsubsection{Comparison to the 4th derivative propagator}
We can compare \eqref{piforidgonepole} to the propagator for the 4th derivative action \eqref{eq:fourthderivaction}
\ba
        \Pi_{\text{4th deriv}} = \Pi_{\text{GR}} -\frac{P^{(2)}}{k^2+m^2}
\ea        
where $m^2=M^2_P/\left[2(3\alpha-\beta)\right]$. We can see that at the extra pole, the IDG propagator has a positive 
residue whereas the 4th derivative propagator has a negative residue. This negative
residue will produce an excitation with negative kinetic energy, or a ghost.
\newpage
\subsection{Quadratic variation of the action}
It is helpful to see what happens when we decompose the perturbation
to the metric $h_{\mu\nu}$ into scalar, vector and tensor parts as 
\cite{DHoker:1999bve}
\ba \label{eq:decompofh}
        h_{\mu\nu}=h^\perp_{\mu\nu}+\bar{\D}_\mu A_\nu+\bar{\D}_\nu A_\mu+\left(\bar{\D}_\mu \bar{\D}_\nu-\frac{1}{4}
\bar{g}_{\mu\nu}\bar{\Box}\right)B+\frac{1}{4} \bar{g}_{\mu\nu}h,
\ea
where $h^\perp_{\mu\nu}$ is the transverse ($\D^\mu h^\perp_{\mu\nu}=0$) and traceless spin-2 excitation, 
$A_\mu$ is a transverse vector field, and $B,~h$ are two scalar 
degrees of freedom which mix. It should be noted that the decomposition
\eqref{eq:decompofh} is only valid around backgrounds of constant curvature.

Upon decomposing the second variation of the action 
around maximally symmetric backgrounds such as (Anti) de Sitter, 
the vector mode and the double derivative scalar mode do not contribute to
the second variation of the action and we end up only with $h^\perp_{\mu\nu}$ and $\phi=\Box B-h$~\cite{Alvarez-Gaume:2015rwa,Biswas:2016etb},
i.e. 
\ba
        h_{\mu\nu} =  h^\perp_{\mu\nu} -\frac{1}{4} g_{\mu\nu} \phi. 
\ea        
For maximally symmetric backgrounds such as (A)dS and Minkowski, we can split the second variation of the action up into tensor and scalar
parts as $\delta^2S=\delta^2S(h^\perp_{\mu\nu})+\delta^2S(\phi)$, where~\cite{Biswas:2016etb}
\ba \label{eq:secondvariationoftheactionformaxsymm}
         \hspace{-1mm}\delta^2 S(h^\perp_{\mu\nu})\hspace{-1mm}
        &=& \hspace{-1mm}\frac{1}{2} \int d^4 x \sqrt{-g}\hspace{1.5mm} \widetilde{h}^\perp_{\mu\nu}\left( \Box  - \frac{\bar{R}}{6} \right)
        \Bigg[ 1 + \frac{2\lambda}{M^2_p} f_{1_0}  \bar{R} \nonumber\\
        &&+ \frac{\lambda}{M^2_p}\Bigg\{\left(\Box - \frac{\bar{R}}{6} \right) F_2 (\Box)  
        +  2\left( \Box - \frac{\bar{R}}{3}\right) F_3 \left(\Box +  \frac{\bar{R}}{3}\right) 
         \Bigg\}\Bigg]\widetilde{h}^{\perp\mu\nu},\non\\
        \delta^2 S(\phi)  &=&- \frac{1}{2} \int d^4 x \sqrt{-g}\hspace{0.4mm} \widetilde{\phi}\left(\Box+\frac{\bar{R}}{3}\right) \Bigg\{ 1
+ \frac{2\lambda}{M^2_p} f_{1_0} \bar{R} \nonumber\\
        &&  -\frac{\lambda}{M^2_p} \Bigg[ 2F_1 (\Box) \left(3\Box +\bar{R}\right) + \frac{1}{2} F_2 \left(
\Box + \frac{2}{3} R \right)\Box    \Bigg]\Bigg\}\widetilde{\phi}.
\ea      
where we have rescaled $h^\perp_{\mu\nu}$ and $\phi$ to $\tilde{h}^\perp_{\mu\nu}$ and $\tilde{\phi}$ by  multiplying by $\frac{M_P}{2}$ and 
$M_P\sqrt{\frac{3}{32}}$ respectively. 
A more detailed derivation is given in Appendix \ref{sec:2ndvaria}. Note that there are no mixed terms in \eqref{eq:secondvariationoftheactionformaxsymm}
i.e. the action has been decomposed into terms which are either quadratic in 
$\tilde{h}_{\mu\nu}$ or quadratic in $\tilde{\phi}$, with no terms which 
are linear in both $\tilde{\phi}$ and $\tilde{h}_{\mu\nu}$.

\subsubsection{Minkowski limit}
Around Minkowski space, the second variation of the action 
(\ref{eq:secondvariationoftheactionformaxsymm}) reduces to  
\ba \label{eq:Minkowskisecondvariationofaction}
        \delta^2 S(h^\perp_{\mu\nu})&=& \frac12 \int d^4x\sqrt{-\bar g}\hspace{1mm} h^\perp_{\mu\nu}\bar\Box\hspace{0.5mm} a(\bar\Box)\hspace{0.8mm} h^{\perp\mu\nu},
        \nonumber\\
        \delta^2 S(\phi)&=& - \frac12 \int d^4x\sqrt{-\bar g}\hspace{1mm} \phi\hspace{0.5mm}\bar\Box \hspace{0.7mm}c(\bar\Box)\hspace{0.8mm}\phi,        
\ea
where $a(\bar\Box)$ and $c(\bar\Box)$ are defined in \eqref{eq:acfdefinitionsmink}. 
Note that \eqref{eq:Minkowskisecondvariationofaction} corresponds to the 
propagator around a flat background, as in this case, the transverse traceless 
excitation corresponds to the $P^{(2)}$ projection operator and the scalar excitation
corresponds to the $P^{(0)}_s$ projection operator, 
although the correspondence between the propagator and the second variation does not hold true for the (A)dS background.

\chapter{Newtonian potential}
\label{chapter:potential}
Our next task is to investigate the effect of IDG on the Newtonian potential,
which is the simplest application
of General Relativity to a real-world system. Here we add a static point source to a background metric
and solve the perturbed field equations. The Newtonian potential
 is simple to calculate in GR, but results 
in a singular potential. Although it is a much more difficult calculation in IDG,
the result does not diverge.

\section{Perturbing the flat metric}
\label{sec:flatmetricpotential}
We want to find the weak-field limit, or Newtonian potential, which is the potential 
of a small \footnote{The mass must be small enough that
the potential produced is much less than 1, so that the weak field regime
is valid.} mass $\mu$ added to a flat background. We must therefore perturb flat Minkowski space as
\ba \label{eq:perturbedmiknowskiscalar}
        ds^2=-(1+2\Phi)dt^2 + (1-2\Psi)\left(dr^2 +r^2 d\Omega^2\right),
\ea
where $\Psi(r)$ and $\Phi(r)$ are the scalar potentials generated by the perturbation. 
Solving the linearised GR field equations for the metric 
\eqref{eq:perturbedmiknowskiscalar}, with the boundary conditions that
$\Phi=\Psi=0$ at infinity produces the potential \cite{Carroll:2004st}
\ba \label{eq:grpotential}
        \Phi(r)=\Psi(r)=-\frac{G\mu}{r},
\ea        
which is clearly divergent as $r\to0$. We will show that in contrast, IDG produces a 
non-singular potential.

\subsection{Method}
Note that we are perturbing the flat space metric $\eta_{\mu\nu}$ as
\ba
        g_{\mu\nu} &=& \eta_{\mu\nu} + h_{\mu\nu}, 
        \quad \quad g^{\mu\nu} = \eta^{\mu\nu} -h^{\mu\nu},
\ea
and therefore 
\ba
        h_{00}=h^{00} = - 2 \Phi, \quad \quad
        h_{ij} =-2\Psi\eta_{ij},\quad \quad
        h^{ij}=-2\Psi\eta^{ij},
\ea
which means the linearised components of the Ricci tensor and Ricci scalar \eqref{eq:linaerisedcurvmink} are
\ba \label{eq:linearisedcurvatures}
        r_{00}&=&\Delta \Phi(r),\non\\
        r_{ij}&=&\partial_i \partial_j \left(\Psi-\Phi\right)  +\eta_{ij} \Delta
\Psi, \non\\
        r &=& 4\Delta\Psi(r)-2\Delta \Phi(r),
\ea
where $\Delta\equiv \eta^{ij}\nabla_i \nabla_j$ is the Laplace operator. 
When calculating the Newtonian potential, we consider a non-relativistic system, 
so there is zero pressure and the only contribution to the stress-energy tensor 
comes from the energy density $\rho$. Therefore the 00-component
of the energy-momentum tensor is $T_{00} = \rho$, 
the trace is $T= - \rho$ and the other components vanish. 
Combining \eqref{eq:linearisedcurvatures} with
the linearised equations of motion around Minkowski \eqref{eq:minkowskieqns}, 
we find the metric potentials in terms of each other, and $\Phi(r)$
in terms of the density $\rho$
\ba \label{eq:psiintermsofphi}
        \Delta \Psi(r)  &=& -\frac{c(\Box)}{a(\Box)  -2  c(\Box)}  \Delta
\Phi(r),   \nonumber\\
         \Delta \Phi(r)    &=& \frac{a(\Box)
         -2  c(\Box)}{a(\Box)\left(a(\Box)
         -3  c(\Box)\right)} \kappa \rho,
\ea
where $\kappa=\frac{1}{M^2_P}$.
\subsubsection{Difference between $\Phi$ and $\Psi$}
Using \eqref{eq:linearisedcurvatures}, the $ij$ component (with $i\neq j$) of the equations of motion 
\eqref{eq:minkowskieqns} for the perturbed metric is
\ba 
        T_{ij}=\partial_i \partial_j \left[c(\Box)(2\Psi-\Phi)-a(\Box)\Psi\right],
\ea
 which means $T_{ij}$ accounts for the anisotropic stress.
It can be seen that $\Psi=\Phi$ if the following conditions are fulfilled
\begin{enumerate}
\item there is no anisotropic stress $T_{ij}$,

\item $a(\Box)=c(\Box)$, 

\item there is a boundary condition demanding that $\Psi$, $\Phi$ and their 
derivatives vanish at infinity. 
\end{enumerate}

We can also see this from \eqref{eq:psiintermsofphi} (which assumes $T_{ij}=0$). 
Clearly if $a\neq c$, as we later choose, then $\Phi$ and $\Psi$ are not
necessarily the same.


\subsection{Assuming a point source}
As our source is a point source of mass $\mu$, it is represented by a Dirac
delta function:
$\rho = \mu\hspace{0.7mm}\delta^3(\boldsymbol{r})$.
Thee Fourier transform of the Dirac delta function is
\ba
        \delta^3(\boldsymbol{r})=\int \frac{d^3k}{(2\pi)^3} 
        e^{i \boldsymbol{k}\cdot \boldsymbol{r}}.
\ea
The Newtonian potential is given by Fourier transforming \eqref{eq:psiintermsofphi}
\ba \label{eq:potentialitofr}
        \Phi(r)=- \frac{\kappa \mu}{(2\pi)^3} \int^\infty_{-\infty} d^3k \frac{a-2c}{a(a-3c)} 
        \frac{e^{i \boldsymbol{k}\cdot \boldsymbol{r}}}{k^2}  =-\frac{\kappa \mu}{4\pi^2 M^2_p} \frac{f(r)}{r},
\ea
where $f(r)$ is defined as (see Section \ref{sec:fouriertrans} for a derivation of this step)
\ba \label{eq:generaldefinitionoffofr}
       f(r)=\int_{0}^{\infty}dk\frac{a(-k^2)-2c(-k^2)}{a(-k^2)(a(-k^2)-3c(-k^2))}\frac{\sin(kr)}{k}. 
\ea  

\subsection{Choosing a form for $a(\Box)$ and $c(\Box)$}
The simplest choice is $a=c$ which produces the propagator 
\eqref{piforidgwithaequalc} with a clear limit back to GR when we set $a=1$. We choose $a(-k^2)=e^{\gamma(k^2/M^2)}$ 
where $\gamma(k^2/M^2)$ is an entire function and therefore
\eqref{eq:generaldefinitionoffofr} becomes
\ba \label{eq:integralforflatpoential}
       f(r)=\int_{0}^{\infty}dk~e^{-\gamma(k^2)}\frac{\sin(kr)}{2k}. 
\ea       
Any entire function can be written as a power series, i.e.
\ba \label{eq:entirefctpoly}
        \gamma(k^2/M^2)=  \sum_{n=0}^\infty c_n \left(\frac{k^2}{M^2}\right)^n,
\ea
where the $c_n$ are dimensionless coefficients.
It may seem like there is an infinite amount of choice in 
the function \eqref{eq:entirefctpoly}. However, there are two factors to bear in mind. 
Firstly, as long as $\gamma(k^2)>0$ for large $k$, then the integrand in
\eqref{eq:integralforflatpoential} will be exponentially damped for large $k$ and we will generate a 
non-singular potential \cite{Biswas:2011ar}.

\subsubsection{Rectangle function approximation}
Secondly, we can write 
\ba \label{eq:expansionofexpofentire}
        e^{-\tau(k^2/M^2)} = e^{-c_1\frac{k^2}{M^2}-c_2\frac{k^4}{M^4}-c_3\frac{k^6}{M^6}-\cdots}.
\ea
It is possible to approximate the contribution to the integrand of all the large $n$ terms in 
\eqref{eq:expansionofexpofentire} as a single rectangle function.
This is because 
$\exp\left[-c_n \left(\frac{k^2}{M^2}\right)^n\right]$ 
forms a rectangle function as long as $n$ is large and $c_n$ is positive: 
\ba \label{eq:rect}
        e^{-c_n \left(\frac{k^2}{M^2}\right)^n}\approx \text{Rect} \left(\frac{\sqrt[n]{c_n}k^2}{ M^2}\right)\equiv\left\{\begin{array}{lr}
        1, & \text{for } \frac{\sqrt[n]{c_n}k^2}{ M^2}< 1\\
        0, & \text{for } \frac{\sqrt[n]{c_n}k^2}{M^2}>1
        \end{array}\right\} .
\ea   

\begin{figure}[!htbp]
\centering
\includegraphics[width=121mm]{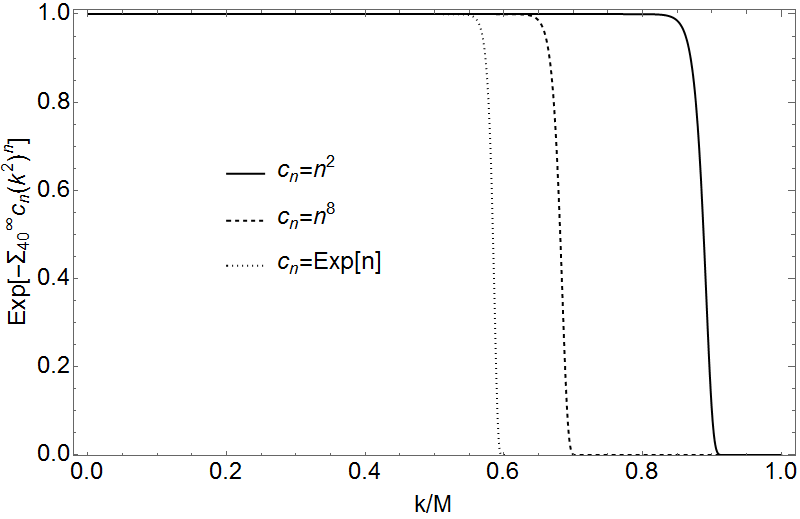}
\caption{$\exp\left[-\sum_{n=40}^{\infty}c_n\left(\frac{k^2}{M^2}\right)^n\right]$
with different choices of the coefficients $c_n$ showing that the rectangle function 
with a single unknown parameter will be a good approximation for the higher order 
part of \eqref{eq:expansionofexpofentire} for any form of $c_n$ \cite{Edholm:2018wjh}.}
\label{rectfctplot}
\end{figure}

It transpires that when we combine the terms with a large value of $n$, 
we can approximate them well as the rectangle function Rect($C k^2/M^2$),
where $C$ is another
 constant to be found.
This is because an infinite product of rectangle functions 
\ba
        \text{Rect}(a_1 x) \times \text{Rect}(a_2 x) \times \cdots,
\ea        
where $a_i>0$, is equal to the rectangle function given by the lowest value of $a_i$. The validity of this approximation is shown in Fig. \ref{rectfctplot},
where we can see that even if $c_n$ increases exponentially with $n$,
the approximation is still valid.

\subsection{Simplest choice of $\gamma(-k^2)$}
For the choice $\gamma=-k^2/M^2$, the integral in the function $f(r)$ in \eqref{eq:integralforflatpoential} gives an error function,
using
\ba \label{eqrorfunctionintegral}
        \int_{-\infty}^{\infty}dk~e^{-k^2/M^2}\frac{\sin(kr)}{k}
        = \pi \text{Erf} \left(\frac{Mr}{2}\right),
\ea    
which means the potential \eqref{eq:potentialitofr} is given by 
\ba \label{eq:erfpotential}
        \Phi(r)= -\frac{2G\mu}{r} \text{Erf}\left(\frac{Mr}{2}\right).
\ea   
The usual GR potential \eqref{eq:grpotential} receives a modification
in the form of the error function Erf$(r)$, which is defined as
\ba
       \text{Erf}(x)= \frac{2}{\sqrt{\pi}} \int^x_0 e^{-t^2}dt.
\ea        
Erf$(r)$ is proportional     
to $r$ for small $r$, so \eqref{eq:erfpotential} does not diverge as
$r\to 0$, while Erf$(r)\to 1$ for large $r$, meaning that $\Phi(r)$
returns to the GR prediction \eqref{eq:grpotential} at large distances.\footnote{IDG is
not the only (metric-tensor based) modified gravity theory to predict a non-singular 
potential, as higher derivative gravity models can also give
finite potentials at the origin (although
these models do generically generate ghosts \cite{Modesto:2014eta,Giacchini:2016xns}).}

By differentiating the potential \eqref{eq:erfpotential}, we find the force on a test mass of mass $\mu$, given by
$F(r)=-d\Phi(r)/dr$
\cite{Buoninfante:2018xiw}
\ba
            F(r)=-\frac{2G\mu }{r^2}\left[\text{Erf}\left(\frac{Mr}{2}\right)
            -\frac{e^{-\frac{M^2r^2}{4}}Mr}{\sqrt{\pi}}\right].
\ea     
It is possible to generalise the potential \eqref{eq:erfpotential} to rotating metrics
\cite{Cornell:2017irh} and charged systems \cite{Buoninfante:2018stt}.

The avoidance of the singularity at the origin can be related to the non-locality of the action. Non-locality means that interactions no longer take place
at a single point but instead are ``smeared out'' over a wider area, 
meaning the potential does not diverge at the point where the mass is \cite{Biswas:2011ar,Conroy:2017uds}.
This is also how string theory avoids singularities, by replacing point particles 
with non-local strings \cite{Calcagni:2017sdq,Blumenhagen:2013fgp}.

\subsection{More general choice of $\gamma(-k^2)$}
\subsubsection{Monomial choice of $\gamma(-k^2)$}

Next we want to generalise to different choices of $\gamma(-k^2)$. The simplest generalisation is to take 
\ba \label{eq:monomialchoicefortau}
        \gamma(-k^2) = -\frac{k^{2n}}{M^{2n}}.
\ea
The choice \eqref{eq:monomialchoicefortau} gives the general 
formula \cite{Edholm:2016hbt}\footnote{\eqref{eq:monomialfofrformula} is a 
generalisation of \cite{Frolov:2015usa}, which examined 
only even choices of $n$ in \eqref{eq:monomialchoicefortau}.} 
\ba \label{eq:monomialfofrformula}
       f(r)=\frac {Mr}{n}\sum_{p=0}^\infty(-1)^p\frac{\Gamma(\frac pn+\frac1{2n})}
       {(2p+1)!}(Mr)^{2p},  
\ea     
where the Gamma function is $\Gamma(x)\equiv (x-1)!$. 
\begin{figure}[ht]
\centering
\includegraphics[width=117mm]{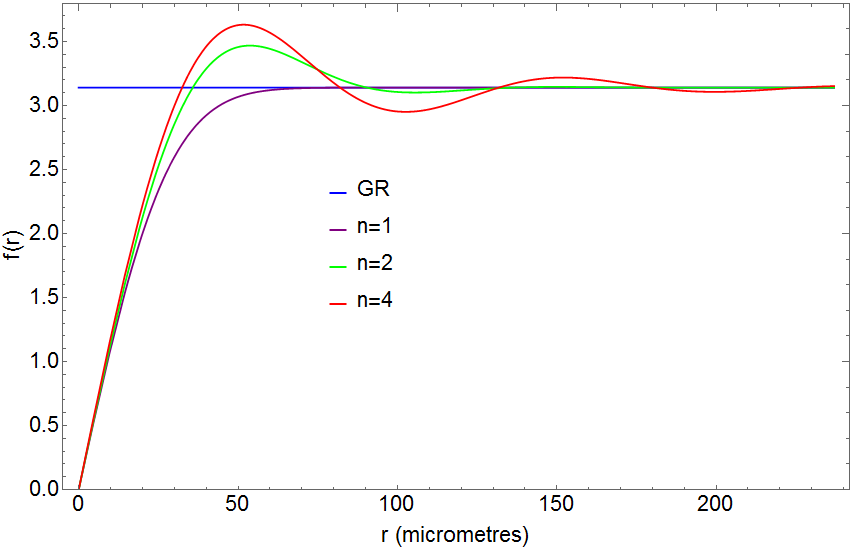}
\caption{A plot of \eqref{eq:monomialfofrformula} for various choices of
$n$ as well
as a comparison to GR  \cite{Edholm:2016hbt}. We see that for $n=1$, we have
the error function, while for higher $n$
we generate oscillations. These oscillations decay in amplitude as we move
further away from the source.
We have chosen $M$ to be the lower bound on our mass scale $M=4\times10^{-3}$
eV
from table-top experiments.}
\label{newtonianplot}
\end{figure}
The plot of $f(r)$ in Fig. \ref{newtonianplot} using \eqref{eq:monomialfofrformula}
resembles the error function, but with sinusoidal oscillations 
that decay in magnitude as we get further away from the source. 

\subsubsection{Polynomial choice of $\gamma(-k^2)$}
We can generalise even further by using a simple mathematical trick \cite{Edholm:2016hbt}. 
Any function $\gamma(-k^2)$ can be written 
as $\gamma(-k^2)=-k^2/M^2 - \rho(-k^2)$, where $\rho(-k^2)$ does
not contain a term which is quadratic in $k$. We can expand $e^{\rho(-k^2)}$ 
as the infinite sum $e^{\rho(-k^2)}\equiv\sum^\infty_{m=0}\rho_m (-1)^m k^{2m}/M^{2m}$,
where $\rho_m$ are dimensionless coefficients, 
without any loss of generality. This trick will allow us to utilise the identity
(\ref{eqrorfunctionintegral}) for more general integrals. We can then write
\eqref{eq:integralforflatpoential} as 
\ba \label{eq:fofrwithalpha}
        f(r) =\sum^\infty_{m=0} \rho_m \int_{-\infty}^{\infty}\frac{dk}{k} 
        \left(\frac{-k^2}{M^2}\right)^m e^{-k^2/M^2}\sin(kr).
\ea
By inserting an auxiliary field $\alpha$, which we will later set to 1,
 we can rewrite \eqref{eq:fofrwithalpha} in an advantageous form
\ba
        f(r) = \sum^\infty_{m=0} \rho_m \left[ \frac{\partial^m}{\partial \alpha^m} 
        \int_{-\infty}^{\infty}\frac{dk}{k}  e^{-\alpha k^2/M^2}\sin(kr)\right]\bigg|_{\alpha=1}. 
\ea
Now the integral is just that of (\ref{eqrorfunctionintegral}) and thus we simply have
\ba \label{eq:fofrwithalphasolved}
        f(r) = \pi\sum^\infty_{m=0} \rho_m \left[\frac{\partial^m}{\partial \alpha^m}  
        \text{Erf} \left(\frac{Mr}{2\sqrt{\alpha}}\right)\right]\bigg|_{\alpha=1}. 
\ea
We can write \eqref{eq:fofrwithalphasolved} either explicitly as \cite{Edholm:2016hbt}
\ba \label{eq:explicitformoffofr}
        f(r) =  \sum^\infty_{m,p=0}\rho_m (-1)^p  
        \frac{\Gamma(m+p+\frac{1}{2})}{(2p+1)!}  \left(Mr\right)^{2p+1},
\ea
or using Hermite polynomials $H_m(x)\equiv (-1)^m e^{x^2} \frac{d^m}{dx^m} e^{-x^2}$ as
\ba \label{newtonsinerhoshermitian}
        f(r)= \pi \mathrm{erf} \left(\frac{M r}{2}  \right) 
        -2 \sqrt{ \pi}  e^{-\frac{M^2r^2}{4}} \sum^{\infty}_{m=1} \rho_m (-1)^m 
         \frac{1}{4^m}  H_{2m-1}\left( \frac{M r}{2} \right)\,.       
\ea

Clearly, when $\gamma(-k^2)=-k^2/M^2$, therefore $e^{-\rho(-k^2)}=e^0$ and hence 
$\rho_0=1$ and $\rho_m=0$ for $m\neq 0$. Therefore we return to the error function 
given in (\ref{eqrorfunctionintegral}).
\begin{figure}[ht]
\centering
\includegraphics[width=130mm]{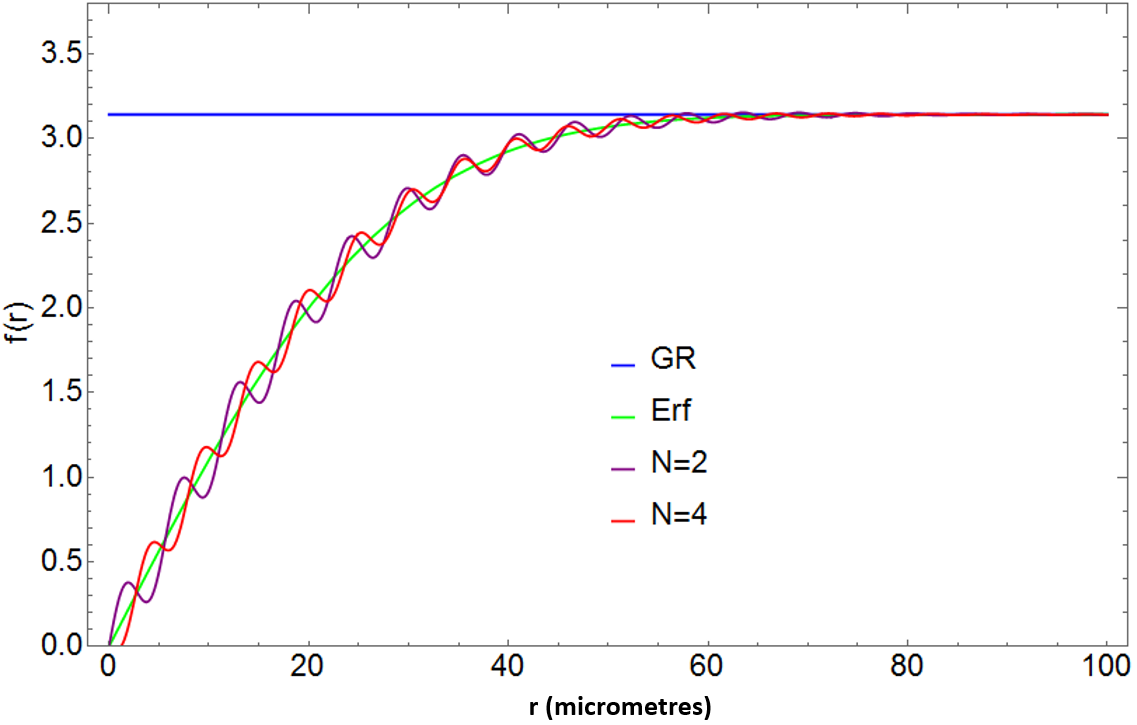}
\caption{A plot of \eqref{newtonsinerhoshermitian} with the binomial choice
\eqref{eq:binomialchoice} for various choices of $N$ as well
as the error function potential \eqref{eq:erfpotential} and a comparison to GR  \cite{Edholm:2016hbt}. 
We have chosen $a_2 =4.65\times 10^{-3}$, $a_4=1.24 \times 10^{-7}$ and $M$
to be the lower bound $M=4\times10^{-3}$ eV on our mass scale from table-top
experiments. We again see oscillations in the potential which disappear for
large $r$.} 
\label{newtonianplotbinomial}
\end{figure}
\subsubsection{Large distance limit}
We can check that as $r\to \infty$, \eqref{newtonsinerhoshermitian} 
converges for any sensible choice of $\tau(-k^2)$. 
For large values of $r$, we can simplify the second term by noting that
$H_n(x)=2^n x^n +O(x^{n-1})$. Therefore
the largest term in $H_{2m-1}\left( \frac{M r}{2} \right)$ will be
$\frac{1}{2}\left( M r \right)^{2m-1}$ and the second term in \eqref{newtonsinerhoshermitian} 
is proportional to \ba \label{eq:largevaluesofrlimitnewtonian}
  \lim_{m\to\infty} m!\left(\frac{4}{M^2r^2}\right)^m\rho_m (-1)^m  \frac{1}{4^m} \left( M r \right)^{2m-1}=
\lim_{m\to\infty}   \rho_m (-1)^m m!\frac{1}{M r},
\ea
which converges as long as $\rho_m$ decreases at least as fast as $\frac{(-1)^m}{m!}$. 
$\frac{(-1)^m}{m!}$ is in fact the $\rho_m$ given by the choice $\rho(-k^2)=-k^2$, 
and so long as $a(-k^2)$ fulfils the condition that in the UV limit that the propagator must be suppressed 
exponentially, i.e. $a(-k^2) \to \infty$ as $k \to \infty$, 
then we will always return to the GR prediction in the IR limit.

As an example, in Fig. \ref{newtonianplotbinomial} 
we plot \eqref{newtonsinerhoshermitian} for the binomial
choice\ba \label{eq:binomialchoice}
        \gamma(k^2) = -\frac{k^2}{M^2} -a_N \frac{k^{2N}}{M^{2N}},
\ea
where $a_N$ is some dimensionless constant. 
The choice \eqref{eq:binomialchoice} requires our coefficients $\rho_m$ to be of the form
\ba
        \rho_m = \frac{(-a_N)^{m/N}}{(m/N)!} \text{ for } \frac{m}{N} \in 
        \mathbb{N} \text{ and zero otherwise.}       
\ea

\subsection{Comparison to experiment}
\begin{figure}[H]
\centering
\includegraphics[width=60mm]{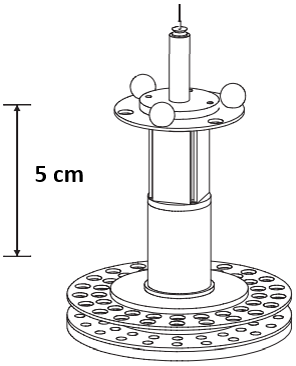}
\caption{A diagram of the apparatus used in tests of gravity
at short distances \cite{Kapner:2006si}. This equipment probed distances below 100 $\mu$m to 
look for modifications to gravity in the form of a Yukawa potential.
 A ``missing-mass'' torsion-balance was used, which is made up of two attractor rings and a detector ring. 
 The force of gravity between the upper
attractor ring and the detector ring generated a torque which would 
be cancelled out by the force caused by the lower attractor
ring. This cancellation would only occur if GR holds exactly, i.e. 
any residual torque would be due to modifications to GR.}
\label{experiment}
\end{figure}

\subsubsection{Constraining the error function potential}
Using the equipment shown in Fig. \ref{experiment}, Adelberger \cite{Kapner:2006si} found that the $1/r$ fall of the potential continued down to 
$5.6 \times 10^{-5}$\hspace{0.4mm}m. 
The experiment assumed that any modification to GR would be a Yukawa potential of the form 
\ba \label{eq:Yukawa}
        V(r)= \frac{V_0}{r} \left[ 1 + \exp(-r/\lambda)\right],
\ea
and ruled out the potential \eqref{eq:Yukawa} with
$\lambda > 5.6 \times 10^{-5}$\hspace{0.4mm}m. 
By seeing the  value of $M$ for which our potential gives a bigger divergence from GR 
than the Yukawa potential,
we see where our potential would be ruled out.
Taking $\gamma=-k^{2n}$,
and using  equation \eqref{eq:monomialfofrformula}, 
  we found a constraint of 
$M>$ 0.004, 0.02, 0.03, 0.05 eV for $n=$1, 2, 4, 8 respectively. 

\subsubsection{Constraining the oscillating potential}
We know that higher powers of $n$ give an oscillating function for $r>M^{-1}$, 
but work by Perivolaropoulos \cite{Perivolaropoulos:2016ucs} 
showed that for $n>10$ one can accurately parameterise \eqref{eq:monomialfofrformula}
as 
\ba \label{eq:parameterisation}
        f(r) &=& \alpha_1 M r, \quad\quad ~~~~~~~~~~~~0<Mr<1\non\\
        f(r) &=& 1+ \alpha_2 \frac{\cos(Mr +\theta)}{Mr}, \quad\quad Mr>1
\ea        
where $\alpha_1 =0.544$, $\alpha_2= 0.572$, $\theta = 0.855\hspace{0.8mm}\pi$. This 
parameterisation is a good approximation because for high $n$,
 $\exp(-ax^n)$ forms a rectangle function as shown in \eqref{eq:rect}. 

\begin{figure}[H]
\centering
\includegraphics[width=130mm]{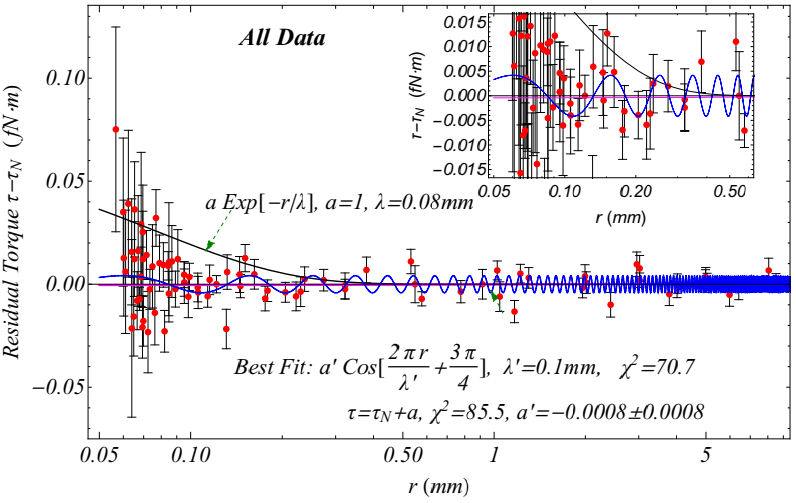}
\caption{A plot of the residual torque compared to the value predicted by
IDG using \eqref{eq:parameterisation} and a Yukawa
 potential \cite{Perivolaropoulos:2016ucs}.
The blue curve is the best fit of the oscillating 
potential \eqref{eq:parameterisation} to the data.
The \textit{residual} torque refers to the torque found
compared to that predicted by GR, 
i.e. if GR was correct then we would see $\tau=\tau_N$.}
\label{residualtorque}
\end{figure}

The best fit of \eqref{eq:parameterisation} 
actually provides a better fit to the data from the 
Adelberger experiment than the standard $1/r$ fall of the potential that GR
predicts by around 2$\sigma$ \cite{Perivolaropoulos:2016ucs}. Perivolaropoulos'
results are plotted in Fig. \ref{residualtorque}.
While the better fit than GR could be a case of fine-tuning, it certainly
gives experimentalists an impetus to look for other 
modified gravity theories when they 
undertake tests of gravity at short distances. Currently they largely test 
only for either a Yukawa potential or GR.

\newpage
\section{$a\neq c$}

While investigating the Newtonian potential, we have so far taken the choice $a=c$. However, this case doesn't allow defocusing of null rays~\cite{Conroy:2016sac}, 
as we will later see in Section \ref{sec:defocusingflat}, and therefore leads to 
geometric singularities through the Hawking-Penrose singularity theorems \cite{hawking_ellis_1973}.
We therefore should investigate what happens to the potential when we take $a\neq c$ 
by using \eqref{eq:choiceofawithonepole}         
from earlier. This fulfils the defocusing conditions which we later find in Section \ref{sec:defocusingflat}, and as we will later see, 
provides a similar prediction for the potential at high $n$.

Taking the choice of $c(\Box)$ given in \eqref{eq:choiceofawithonepole}, 
then (\ref{eq:generaldefinitionoffofr}) becomes
\ba \label{eq:generalfofrforaneqc}
        f(r)=\int^\infty_{-\infty} dk \left[ 4-\frac{m^2}{\tilde{a}(-k^2)(m^2+k^2)}\right]
\frac{\sin(kr)}{k~a(-k^2)}. 
\ea 

We recall that we require both $a(-k^2)$ and $\tilde{a}(-k^2)$
to be the exponential of entire functions so that no ghosts are generated, 
i.e. $a(-k^2)=e^{\gamma(-k^2)}$ and
$\tilde{a}(-k^2)=e^{\tau(-k^2)}$. Therefore \eqref{eq:generalfofrforaneqc} becomes
\cite{Conroy:2017nkc}\ba \label{eq:generalfofrforaneqcwithexponentials}
        f(r)=\int^\infty_{-\infty} dk \left[ 4-\frac{m^2}{(m^2+k^2)e^{\tau(-k^2)}}\right]
\frac{e^{-\gamma(-k^2)}\sin(kr)}{k}. 
\ea 
\subsubsection{Conditions on 
$\gamma(-k^2)$\ and $\tau(-k^2)$}
By considering the behaviour of the propagator \eqref{piforidgonepole}, we can find conditions on 
$\gamma(-k^2)$\ and $\tau(-k^2)$. 
In the ultraviolet regime, we want the propagator to be exponentially 
suppressed. The coefficient of the $P^2$ component is $e^{-\gamma(-k^2)}$
and therefore we require that $\gamma(k^2)>0$ (and that $\gamma$ is not 
a constant). 
Similarly the coefficient of the $P^0_s$ component is $e^{-\left(\gamma(-k^2)+\tau(-k^2)\right)}$
and therefore we require $\gamma(-k^2)+\tau(-k^2)>0$.
The defocusing condition for a static perturbation is found in Section \ref{sec:defocusingflatstatic}
and is given by \cite{Conroy:2016sac}
\ba 
        \left(1-\Box/m^2\right)\tilde{a}(\Box)r> r,
\ea 
where $r$ is the perturbed Ricci scalar. Given our earlier choice $\tilde{a}(-k^2)=e^{\tau(-k^2)}$, 
we see that moving to momentum space gives

\ba \label{eq:defocusingconditioninmom}
       \left(1+k^2/m^2\right)e^{\tau(-k^2)}\hspace{0.5mm} \tilde{r}> \tilde{r}.
\ea       
where $\tilde{r}$ is the Fourier transform of $r$.
If \eqref{eq:defocusingconditioninmom} holds true for all $k$, 
then we have the condition that $\tau(-k^2)\geq0$. 
The result of these three constraints is that the integral 
\eqref{eq:generalfofrforaneqcwithexponentials} converges.

We now have two undetermined entire functions $\tau(-k^2)$ and $\gamma(-k^2)$ and
two free mass scales. We will choose both our mass scales to be the Planck mass $M_P$ and
take $\tau(-k^2)=0$, so that all the model freedom is in the function $\gamma(-k^2)$.
The simplest choice for $\gamma$ is the 
monomial $\gamma(-k^2)=(Ck^2/M^2_P)^D$. We now have only two free parameters, $C$ and $D$,
and our integral becomes \cite{Conroy:2017nkc}
\ba \label{eq:generalfofrforaneqcwithexponentialsspecific}
        f(r)=\int^\infty_{-\infty} dk \left[ 4-\frac{M^2_P}{(M^2_P+k^2)}\right]
\frac{e^{-(Ck^2/M^2_P)^D}\sin(kr)}{k}. 
\ea 
While this choice is fairly restrictive, we obtain qualitatively similar results
for other choices of the mass scales and the function $\tau(-k^2)$.

\begin{figure}[H]
\centering
\includegraphics[width=13cm]{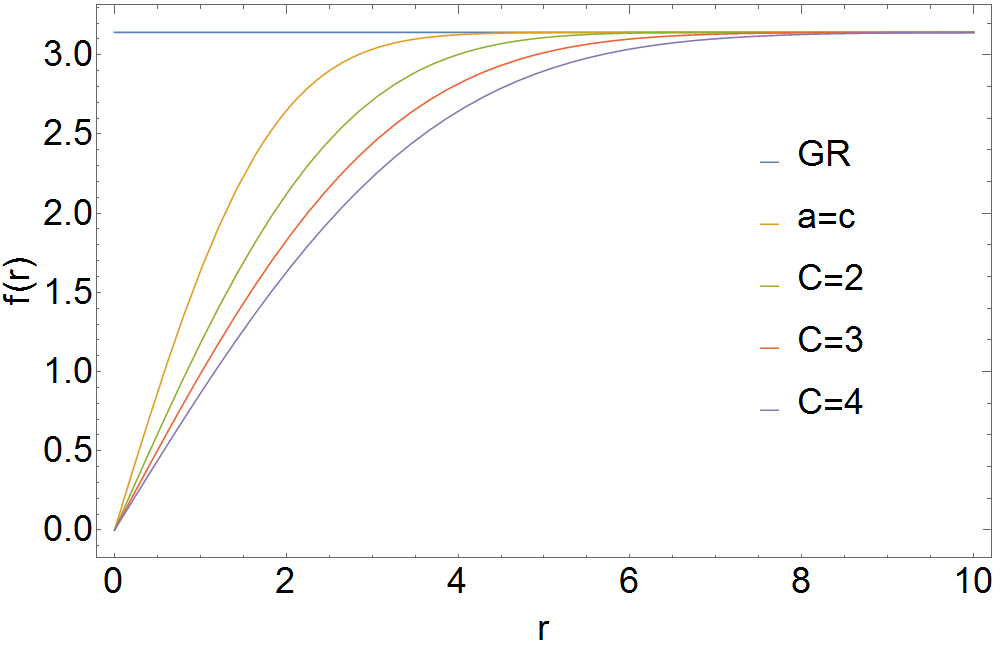} 
\caption{We set $D=1$ and vary $C$ in our plot of
$f(r)$ against $r$ from
\eqref{eq:generalfofrforaneqcwithexponentialsspecific}. 
We choose $M_p=1$m$^{-1}$ to illustrate our case. We also plot the
$a(\Box)=c(\Box)$ case from (\ref{eqrorfunctionintegral}).
As $C$ increases, the modification to GR is visible
further away from the source.}
\label{DconstantalterCfofr}
\end{figure}

\begin{figure}[H]
\centering
\includegraphics[width=11.5cm]{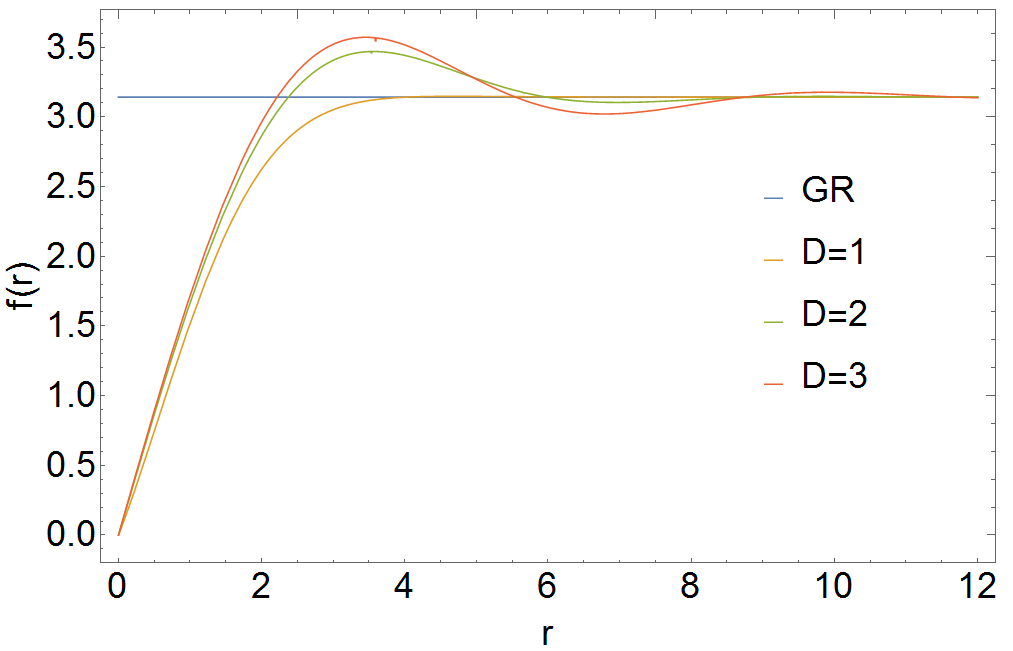}
\caption{Here we take $C=1$ and vary $D$ to see how $f(r)$ vs $r$ changes
using \eqref{eq:generalfofrforaneqcwithexponentialsspecific}.
For $D\geq 2$, $f(r)$ oscillates, with an increasing magnitude 
as $D$ increases, although it still returns to GR\ at large distances. 
We have again chosen $M_p=1$m$^{-1}$.
}
\label{CconstantalterDfofr}
\end{figure}

\newpage
We first take the case $D=1$ and plot 
\eqref{eq:generalfofrforaneqcwithexponentialsspecific} 
in Fig. \ref{DconstantalterCfofr} for different values of $C$ \cite{Conroy:2017nkc}. 
We can see that increasing $C$ makes the IDG
effect stronger at distances further away from the source. This effect
is akin to decreasing the 
mass scale $M$ for the $a=c$ case. 
\newpage
In Fig. \ref{CconstantalterDfofr}, we take $C=1$ and plot $f(r)$ 
for different $D$ \cite{Conroy:2017nkc}. As for the $a=c$ case, 
we find oscillations for $D>1$ which have approximately constant 
frequency and decaying amplitude 
as $r$ increases. To being with, these oscillations become larger in amplitude as we increase $D$,
but this effect stops for around $D>10$ because at this point 
\eqref{eq:rect} becomes a good approximation
and so increasing $D$ further has no effect on the amplitude of the oscillations. 

\begin{figure}[H]
\centering
\includegraphics[width=11.5cm]{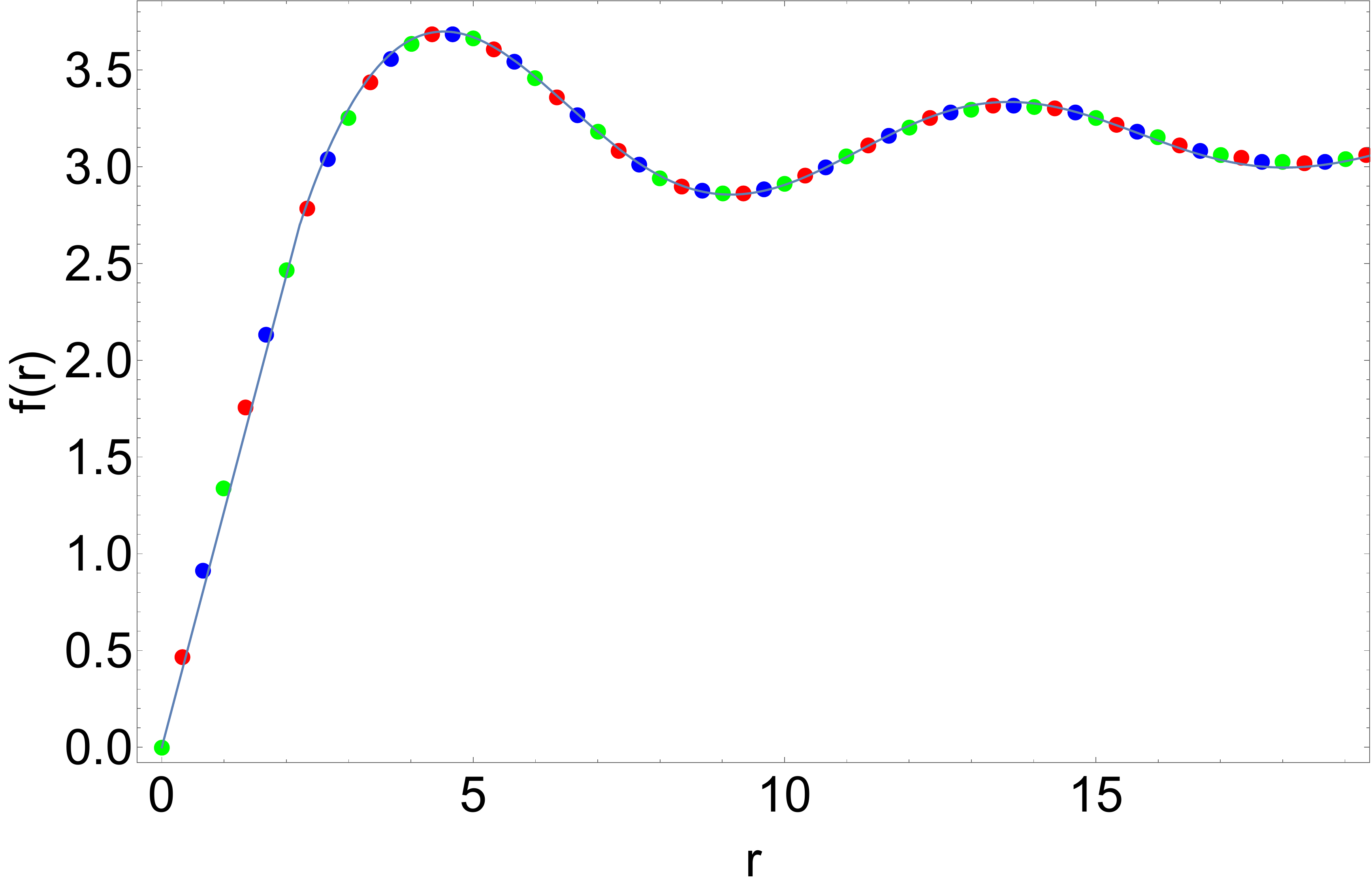}
\caption{We show that the black line, representing \eqref{eq:parameterisation}
is a valid approximation to 
the green, blue and red coloured dots, representing
$D$=11, 12, 13 in \eqref{eq:generalfofrforaneqcwithexponentialsspecific} respectively \cite{Conroy:2017nkc}.
We have chosen $\alpha_1=0.388$, $\alpha_2=0.6$ and $\theta=2.75$.
}
\label{approxparam}
\end{figure}

\newpage
We also show in the plot Fig. \ref{approxparam} that we can accurately approximate
the potential\footnote{In this plot, we obtain a slightly better match to
\eqref{eq:generalfofrforaneqcwithexponentialsspecific} by changing from the
first to the second case in \eqref{eq:parameterisation} at $Mr=2.2$ rather
than 1.} with the parameterisation \eqref{eq:parameterisation} 
which matches the experimental results
\cite{Conroy:2017nkc}. 

\newpage
\section{Raytracing}
\begin{figure}[H]
\centering
\includegraphics[width=112mm]{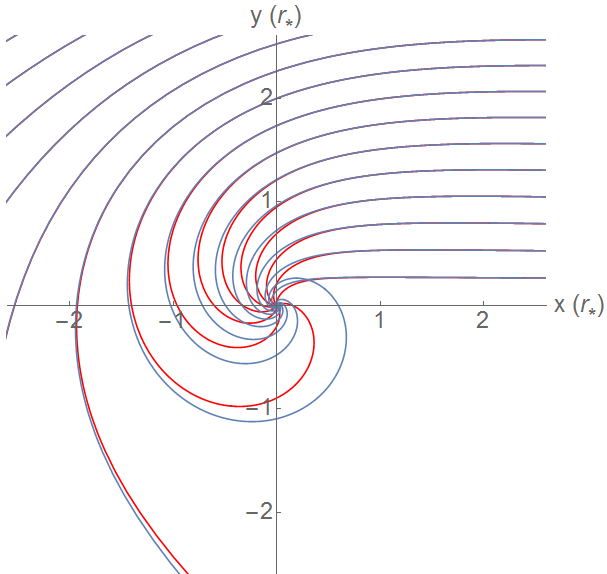}
\caption{The path of photons passing a point mass  of mass $m$ in both GR (red) and IDG
(blue). 
The photons start from the right hand side, travelling in the negative $x$
direction 
and are pulled towards the mass by 
its gravitational field. The axes have units of the GR Schwarzschild radius
$r_*=2Gm$.}
\label{raytracing}
\end{figure}    
In Appendix \ref{sec:raytracing} we give the method for calculating the trajectory
of photons passing a point mass of mass $m$ in GR against IDG. By making a coordinate transform, 
one can write the metric \eqref{eq:perturbedmiknowskiscalar} with the potential \eqref{eq:erfpotential}
as
\ba \label{eq:erfradialmetricmain}
        d\tau^2 =\left(1-\frac{r_*}{r} \text{Erf}\left(\frac{Mr}{2}\right)\right)dt^2-
\left( 1-\frac{r_*}{r} \text{Erf}\left(\frac{Mr}{2} \right)
\right)^{-1} dr^2 -r^2 d\phi^2.
\ea 
One finds that the
path of the photons
can be described by 
\ba       \label{eq:differentialray}
        u''(\phi) = - u(\phi)+\frac{3}{2}r_*u^2(\phi) \text{Erf}\left(\frac{M}{2u(\phi)}\right) 
        + \frac{r_*u^3(\phi)}{2} \text{Erf}'\left(\frac{M}{2u(\phi)}\right),
\ea 
where we have defined $u\equiv 1/r$, $r_*=2Gm$ and the primes represent differentiation
with respect to $\phi$. 
In Fig. \ref{raytracing} we plot various photon trajectories 
to see the effect IDG has on the photon trajectory in GR.

One can see that the photons are more strongly attracted to 
the mass in GR than in IDG, especially when they are closer to the mass.

\section{Potential around a de Sitter background}
We could use the de Sitter background as written in \eqref{eq:desittermetricwithexp}. 
However, if we perturb \eqref{eq:desittermetricwithexp}
 with a point source which generates a potential $\Phi(r)$, 
we will have an arbitrarily large number of
 d'Alembertian operators 
  acting on functions of both $r$ and $t$, 
which is non-trivial to solve. 

However, we can write the de Sitter metric so that its coefficients have no time dependence, 
whilst retaining the same equations of motion and the maximal 
symmetry \cite{Spradlin:2001pw}. 
The perturbed de Sitter background can then be written as
\ba \label{eq:desittermetricperturbedphi}
        ds^2=-\left(1+2\Phi(r)-H^2 r^2\right)dt^2 +
        \left(1-2\Psi(r)\right)\left[\left(1+H^2r^2\right)dr^2 +r^2d\Omega^2\right],~~
\ea 
where we are working in the regime $H^2 r^2 \ll 1$ to simplify the 
ensuing analysis \cite{Jin:2006if}. This is justified because we are interested in 
probing IDG at very short distances from the object producing the potential compared to
the size of the cosmological horizon.

In this approximation, the background de Sitter d'Alembertian operator $\bar{\Box}$ acting on a function of $r$ takes on a pleasing form
\ba
        \bar{\Box} X(r)&=&\bigg[ \frac{1}{1+H^2r^2} \partial_r
        - \frac{H^2r}{1+H^2r^2}\left(\frac{1}{1-H^2r^2}
        +\frac{1}{1+H^2r^2}\right)
        +\frac{2}{r}\frac{1}{1+H^2r^2}
        \bigg]\partial_r X(r)\non\\
        &\approx&  \left(\partial_r^2 
        +\frac{2}{r}\partial_r\right) X(r)\non\\
        &=& \Delta X(r),
\ea        
where we have defined $\Delta\equiv \eta^{ij}\nabla_i \D_j$ as the Laplace operator for a flat background.       

We will need the following perturbed Ricci 
curvatures
for the metric \eqref{eq:desittermetricperturbedphi} up to linear order in $\Phi$ and $\Psi$ (and their derivatives).
\ba \label{eq:generalperturbedmetricriccitt} 
        R_{tt}&\approx& \Delta\Phi(r) - 3H^2 \left(1+2 \Psi(r)-\Phi(r)\right),\non\\
        R_{rr}&\approx& \frac{2 \left(r \Psi  ''(r)+\Psi  '(r)\right)}{r}-\Phi''(r)
        +H^2 \bigg[3-r\Phi'(r) -2\Phi(r)+4\Psi (r)\bigg],\non\\
        R&\approx&4\Delta\Psi(r)-2\Delta\Phi (r)     
        +12 H^2  \left(1+3\Psi(r)-\Phi  (r)\right).~~~
\ea
\begin{figure}[H]
\centering
\hspace{-4mm}\includegraphics[width=130mm]{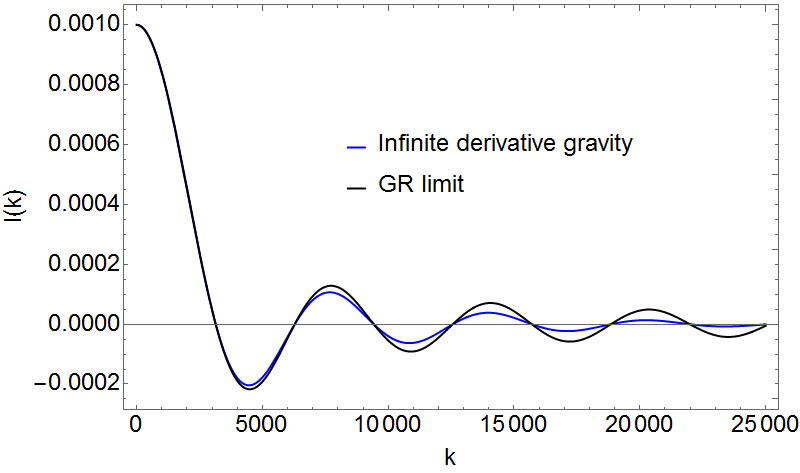}
\caption{A plot of the integrand $I(k)$ given in \eqref{eq:finalintegration}
for $H=7.25\times$ $10^{-27}$~m$^{-1}$, $M=1.79\times10^{4}$~m$^{-1}$ 
and $r=1$m \cite{Edholm:2018wjh}. It can be seen that the difference from GR is
negligible for $k\lessapprox 10^3$.
 $M$ is chosen to be the lower bound found by experiment~\cite{Edholm:2016hbt,Kapner:2006si}
and we choose the coefficients $f_n$ such that around a flat background
$a=c=e^{-\Box/M^2}$ as in~\cite{Edholm:2016hbt}. 
If we take a higher $M$, IDG matches GR up to even higher energy scales. 
The value of $k$
at which IDG ceases to be a good 
approximation to GR is independent of the choice of $r$.}
\label{difffromGR}
\end{figure}

By substituting the perturbed Ricci curvatures into the equations of motion
\eqref{eq:dsfieldeqnssimplified}, we find that
\ba \label{eq:dsphiintermsofpsi}
        \hspace{-2.3mm}\frac{\Psi(r)}{\Phi(r)}\hspace{-0.7mm}=\hspace{-0.7mm}\frac{a(\Delta)(
 \Delta     
         + 3H^2)-\left(3c(\Delta)+\Delta f(\Delta)\right)\left(\Delta   
 +6 H^2  \right)}{a(\Delta)(4  \Delta   + 30H^2)-2\left(3c(\Delta)+\Delta
f(\Delta)\right)\left(\Delta+9 H^2  \right)}, ~~~~~~
\ea             
and we can find $\Phi$ in terms of the density $\rho$ of our source
 \ba \label{Phiintermsofdensity}
        \left(
 \Delta ^2    
         + 15H^2\Delta +63H^4\right)\Phi(r)=\frac{\rho}{M^2_P}\frac{a(\Delta)( 2\Delta
  + 15H^2)-\left(3c(\Delta)+\Delta
f(\Delta)\right)\left(\Delta+9 H^2  \right) }
{a(\Delta)\left[2a(\Delta)-4c(\Delta)-\Delta f(\Delta)\right]}.\hspace{0.4mm}~~~~~
\ea

The energy density of a point source of mass $\mu$ is $\rho=\mu\hspace{0.8mm}\delta^3(\textbf{x})$.
We again go into momentum space in order to write
$\Phi(r)$ as an integral 
\ba \label{eq:finalintegration}
        &&\Phi(r)=\frac{\mu}{2\pi^2M^2_Pr}I(k)\non\\
        && \hspace{9.55mm}=\frac{\mu}{2\pi^2M^2_Pr}\non\\
        &&\cdot\int^\infty_{-\infty}dk~ k\sin(kr)\frac{a(-k^2)(15H^2-2k^2
  )+\left(3c(-k^2)-k^2
f(-k^2)\right)\left(k^2+9 H^2  \right)}{a(-k^2)\left(2a(-k^2)-4c(-k^2)+k^2
f(-k^2)\right)\left(  k^4    
         -15H^2k^2 +63H^4\right)}.~~~~~~~~~
\ea
\begin{figure}[H]
\centering
\hspace{-4mm}\includegraphics[width=130mm]{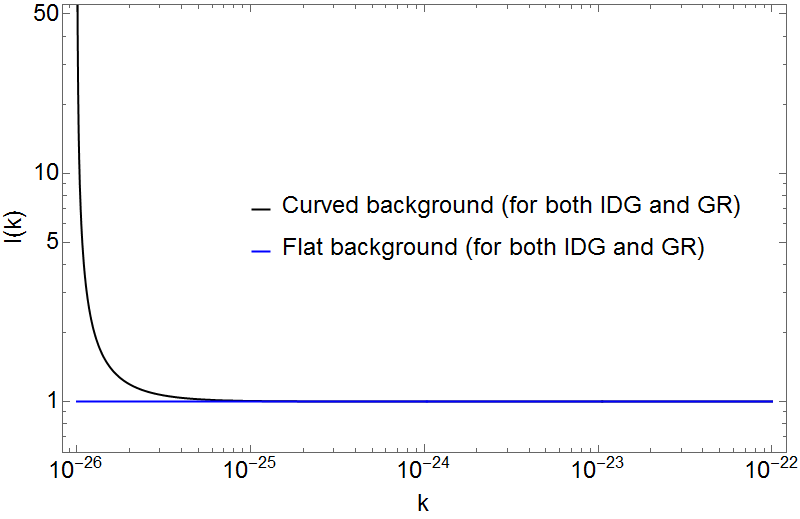}
\caption{A plot of the  the integrand $I(k)$ given in \eqref{eq:finalintegration}
for $H=7.25\times$ $10^{-27}$~m$^{-1}$, $M=1.79\times10^{4}$~m$^{-1}$ and $r=1$m, as compared to a flat background \cite{Edholm:2018wjh}. Note the plot is the same no matter if we take GR or IDG. The difference between the flat background and a curved background is negligible for $k\gg H$.}
\label{difffromflat}
\end{figure}
\newpage
We note that the current value of the Hubble parameter, 
$H=7.25\times$ $10^{-27}$~m$^{-1}$ is many orders of magnitude smaller than even the lower bound on 
the mass scale of IDG \cite{Edholm:2016hbt}, $M=1.79\times10^{4}$~m$^{-1}$. At current times, i.e. when we conduct experiments
on the potential generated by IDG, we can 
therefore ignore the background curvature, 
as detailed further in Fig. \ref{difffromGR} and Fig. \ref{difffromflat},
and use our results from Section \ref{sec:flatmetricpotential}.

\section{Summary}
In this chapter we investigated the Newtonian potential, which is the potential generated
by a point mass in a background.
We first found the potential around a Minkowski background for various choices 
of the form of the propagator assuming that it had no poles. Regardless of the choice made, we always generate
a non-singular potential that returns to GR at large distances. For certain choices,
we obtained an oscillating potential.
Using our results, it was found that the oscillating potential
could provide a better fit to experimental data than GR \cite{Perivolaropoulos:2016ucs}. 

We further showed that if the propagator has a single pole,
it still generates a non-singular potential which returns to GR at large distances
and can produce an oscillating potential. 
We generalised our method to a de Sitter background and found 
that the low value of the Hubble parameter today means that it 
makes a negligible difference to the potential
and we can safely use the calculation from a flat background.

\chapter{Defocusing and geodesic completeness}
\label{sec:defocusing}
\section{Notion of a singularity}
A singularity is traditionally defined as \textit{a place where the
curvature blows up}, i.e. becomes infinite \cite{Wald:1984rg}. 
In other theories, a singular potential
is a relatively easy concept to understand. In electromagnetism,
the Coulomb potential is infinite in various scenarios, for example the potential
given by a point charge diverges at the point itself. The issue with extending 
the idea of a singular potential to
GR is that we need to solve for the metric of spacetime itself. If the metric 
potential diverges, then we do not have a complete description of spacetime and
it is difficult to understand what a ``place'' means in our definition of a singularity.
In other words, if we cannot define the manifold, we cannot say where the ``place''
is that the curvature diverges.

\subsubsection{Geodesic completeness}
The solution to our dilemma lies in the idea of geodesic past-completeness.
In a non-singular spacetime, one can take a geodesic and trace its path back
to null past infinity. The property of being able to trace the path back
is the definition of \textit{geodesic completeness},
which we will later use to find the conditions for IDG to avoid the singularities
produced in GR.

One can imagine a spacetime with `holes' in its fabric, defined by geodesics
being unable to traverse them, and which would thus be geodesically incomplete \cite{Conroy:2017uds}. A particle which is falling freely in such a spacetime would cease to exist at a finite proper time. It must be mentioned that
the definition of geodesic completeness is not as precise as we would perhaps wish. It is possible for our imaginary spacetime to be geodesically incomplete for spacelike or timelike vector fields
and still be geodesically complete for the other two \footnote{In full, a spacetime can be geodesically incomplete for spacelike vector fields while still complete for timelike and null vector fields, or could be incomplete
for timelike vector fields while incomplete for spacelike and null vector fields. Further details of the possible permutations are given in \cite{Geroch:1968}.} \cite{Conroy:2017uds}. However, it is clear that if a spacetime is geodesically incomplete for null or timelike fields, then we will not be able to fully describe the spacetime geometry. 
The definition of geodesic incompleteness  to be used in this chapter is based on the Hawking-Penrose singularity theorems, which tell us whether a spacetime will be geodesically complete.

\section{Raychaudhuri equation}
The Raychaudhuri equation is a geometric equation which is model-independent in that it depends on  
gravity only through its effect on  the curvature of spacetime~\cite{Wald:1984rg,poisson_2004}.

If there exists a congruence of  null rays, then the tangent vectors to
 the congruence $k_\mu$  satisfy $k^\mu k_\mu=0$. 
These are defined through $k^\mu\equiv \frac{dx^\mu}{d\lambda}$ where $\lambda$ is an affine parameter~\cite{Wald:1984rg,poisson_2004}.

We can define several parameters of the congruence using these tangent vectors. 
The expansion $\theta$, shear $\sigma_{\mu\nu}$ and rotation/twist $\omega_{\mu\nu}$ are 
\ba
        \theta\equiv \D^\mu k_\mu,\quad \sigma_{\mu\nu}=\D_{(\mu} k_{\nu)}, 
        \quad \omega_{\mu\nu} =\D_\mu k_\nu - \D_\nu k_\mu.
\ea 
The Raychaudhuri equation for null geodesic congruences is
\ba
        \frac{d\theta}{d\lambda} +\frac{1}{2} \theta^2 = - \sigma_{\mu\nu} \sigma^{\mu\nu} + \omega_{\mu\nu} \omega^{\mu\nu} 
        - R_{\mu\nu} k^\mu k^\nu,
\ea        
using $k^\lambda \D_\lambda \theta = \frac{d\theta}{d\lambda}$. 
The Hawking-Penrose singularity theorem states that unless 
$\theta$ is both positive and increasing,
a singularity will be generated~\cite{hawking_ellis_1973}.
By choosing the congruence wisely\footnote{Strictly 
speaking the congruence of null rays must be orthogonal to a hypersurface.},
we can discard the twist tensor term and we can note that the 
shear tensor is purely spatial and so 
$\sigma_{\mu\nu} \sigma^{\mu\nu}$ is non-negative. 

This leads us to the \textit{null convergence condition} (null CC) 
\ba \label{eq:nullc}
        \frac{d\theta}{d\lambda} +\frac{1}{2} \theta^2 \leq - R_{\mu\nu} k^\mu k^\nu.
\ea
If this is fulfilled, singularities are generated. The absolute minimum 
necessary condition to avoid these singularities, known as 
the minimum defocusing condition is
\ba \label{eq:generalmindefo}
        R_{\mu\nu} k^\mu k^\nu <0.
\ea        

The Null Energy Condition (NEC)  means that in GR, \eqref{eq:generalmindefo} cannot be satisfied and so
singularities are generated. The NEC states that 
the stress-energy tensor contracted with the tangent vectors is positive, 
i.e. $T_{\mu\nu} k^\mu k^\nu \geq 0$.
By inserting the Einstein equation \eqref{eq:einsteinequation}
into \eqref{eq:generalmindefo}, 
it can be seen that the defocusing condition
will not be satisfied in GR as long as the NEC holds, 
i.e. as long as we do not introduce exotic matter which violates the NEC.
However in IDG, it is possible to satisfy the defocusing 
condition due to our modification to the Einstein equation.

\section{Defocusing around a flat background}
\label{sec:defocusingflat}
We now examine the necessary condition for defocusing for perturbations around 
a flat background, in the context of singularity avoidance, especially
in the early universe.
We contract the linearised equations of motion around a 
flat background \eqref{eq:minkowskieqns} with $k^\mu k^\nu$ to give 
the 
defocusing condition \cite{Conroy:2016sac}
\ba  \label{eq:flatdefoccond}
        k^\mu k^\nu r_{\mu\nu} = \frac{1}{a(\Box)}\left( M^{-1}_P T_{\mu\nu} k^\mu k^\nu 
        + \frac{1}{2} k^\mu k^\nu f(\Box)\D_\mu\D_\nu r\right)<0.
\ea
where $r_{\mu\nu}$ and $r$ are the perturbed Ricci tensor and perturbed Ricci
scalar respectively.  As in Section
\ref{sec:propflat}, we require $a(-k^2)>0$ to avoid  a Weyl ghost
~\cite{Biswas:2016etb,Biswas:2016egy,Conroy:2017uds}.
Due to the Null Energy Condition (NEC) $T_{\mu\nu} k^\mu k^\nu\geq0$,
the minimum condition to avoid a singularity is therefore
\ba \label{eq:generalminkdefoc}
        k^\mu k^\nu f(\Box)\D_\mu\D_\nu r<0.
\ea
\subsection{Homogenous perturbation}
We examine the case where the perturbed curvature depends 
only on the cosmic time $t$. 
This is useful for considering cosmological singularities by 
using a perturbation around a Minkowski background
as an approximation to the FRW metric \eqref{eq:FRWmetric}. Using
$\Box r(t)= - \partial_t^2  r(t)$ for a flat background, \eqref{eq:generalminkdefoc} 
simplifies to
\ba  \label{eq:homogenousdecomp}
        \left(k^0\right)^2 f(\Box)\Box r(t)>0,
\ea
and using $f(\Box)\Box=a(\Box)-c(\Box)$ from \eqref{eq:acfdefinitionsmink}, 
\eqref{eq:homogenousdecomp} becomes
\ba \label{eq:minimumdefocusingconditionhomogenous}
         \left(a(\Box)-c(\Box)\right) r(t)>0
\ea      
The simple choice $a(\Box)=c(\Box)$ does not allow defocusing \cite{Conroy:2016sac}. We can make a choice of $a(\Box)\neq c(\Box)$, 
but this choice introduces an extra pole in the propagator,
which in theory could produce a ghost.\footnote{We could choose
$a=A c=e^{\gamma(k^2)}$ where $A$ is a constant, but if we want to return to GR\ in the 
low energy limit then we are forced to choose $A=1$.} However, we showed in Section 
\ref{sec:propflat} that a single extra pole was allowed
and we introduced the choice \eqref{eq:choiceofawithonepole}
which generates one extra pole in the propagator.

With the choice \eqref{eq:choiceofawithonepole}, the defocusing condition becomes 
\ba
        \left(1-\Box/m^2\right)\tilde{a}(\Box)r< r.
\ea        
\subsection{Static perturbation}
\label{sec:defocusingflatstatic}
We next examine the case where the perturbation is a function only 
of a single spatial coordinate $x$. 
In this case,  the defocusing condition \eqref{eq:generalminkdefoc} is
\ba
        k^i k^j f(\Box)\partial_i \partial_j \hspace{0.4mm} r(x)<0.
\ea        

 Using \eqref{eq:acfdefinitionsmink} and $\Box r(x)=\partial_i^2r(x)$,
we find the minimum defocusing condition 
\ba
         \left(a(\Box)-c(\Box)\right) r(x)<0.
\ea
Note the change of sign from the homogenous case \eqref{eq:minimumdefocusingconditionhomogenous}.
Once again, $a(\Box)=c(\Box)$ does not allow defocusing so 
we take the choice \eqref{eq:choiceofawithonepole} giving \cite{Conroy:2017nkc}
\ba \label{eq:staticdefocusingcond}
        \left(1-\Box/m^2\right)\tilde{a}(\Box)r> r.
\ea 
This is the condition we used to derive \eqref{eq:defocusingconditioninmom}.
\subsection{Inhomogenous perturbations}
Lastly we examine perturbations which are functions of time and a single Cartesian
coordinate, which we take to be $x$ here. The defocusing condition \eqref{eq:generalminkdefoc}
becomes 
\ba
       \left((k^0)^2 f(\Box) \partial_t^2 + (k^x)^2 f(\Box) \partial_x^2 
       \right) r(t,x)<0 
\ea       
From the condition $g_{\mu\nu} k^\mu k^\nu=0$ (i.e. the requirement that 
the vector $k_\mu$ is null), we deduce that $(k^0)^2=(k^x)^2$,
so noting that $(k^0)^2$ is strictly positive and assuming that the stress energy tensor is a perfect fluid\footnote{If $T_{\mu\nu}$
is a perfect fluid, it has no off-diagonal terms and therefore 
there are no cross terms in \eqref{eq:flatdefoccond}.}, 
\ba
        f(\Box)\left( \partial_t^2 + \partial_i^2 
       \right) r(t,x_i)<0. 
\ea

\section{Defocusing around backgrounds expanding at a constant rate}
\subsection{(Anti) de Sitter}
Around a curved (Anti) de Sitter background, the linear 
equations of motion become more complicated but are nonetheless
tractable. Contracting the equations of motion 
\eqref{eq:dsfieldeqnssimplified} with the tangent vectors $k_\mu k^\nu$, and the defocusing condition becomes
\ba \label{eq:defocusimpleds}
         r^\mu_\nu k_\mu k^\nu = \frac{1}{a(\Box)} \left[ M^{-2}_P T^\mu_\nu k_\mu k^\nu +
         \frac{1}{2} k_\mu k^\nu \nabla^\mu \partial_\nu f(\Box)r\right]<0,
\ea
where $r$ is the perturbed Ricci tensor. We already know that $T^\mu_\nu k_\mu k^\nu$ must be positive by the NEC (as long as we have non-exotic matter) and 
that (in momentum space) $a(-k^2)$ must 
be positive, otherwise the background de Sitter spacetime will have a negative entropy~\cite{Conroy:2015wfa,Conroy:2017uds}. \eqref{eq:defocusimpleds} can then
be simplified to give the minimum defocusing condition \cite{Edholm:2017fmw} 
\ba \label{eq:dsdefocusingfull}
         k_\mu k^\nu \nabla^\mu \partial_\nu f(\Box)r<0.
\ea        
If the perturbation is homogenous, i.e. $r=r(t)$, then we can expand the covariant derivatives
and note that the d'Alembertian acting on a scalar function of time is
 $\Box = - \partial_t^2 - 3H \partial_t$, so \eqref{eq:dsdefocusingfull}  becomes
\ba
        \Box f(\Box)r(t)>-4H\partial_t f(\Box) r(t).
\ea        

\subsection{Anisotropic backgrounds}
\subsubsection{AdS-Bianchi I metric} 
We now look at anisotropic backgrounds. The observable universe is almost
exactly isotropic (looks the same in different spatial directions)~\cite{Wald:1984rg}, 
but it is interesting
to consider anisotropy. This is  because while the anisotropy in the observable universe is
very small, it is unlikely to be exactly zero so we should consider its effects. Furthermore,  
just because our observable universe is isotropic does not mean the rest of the universe is as well.
We study the anisotropic metric
\ba \label{eq:anisotropiclineelement}
        ds^2 = -dt^2 + e^{2At} dx^2 + e^{2Bt} dy^2 + e^{2Ct} dz^2, 
\ea        
where  $A$, $B$, and $C$ are positive constants, which we call an (A)dS-Bianchi I metric. 
The observable universe is highly isotropic~\cite{Wald:1984rg}, so we assume that 
$A$, $B$, and $C$ are almost equal.

The (A)dS-Bianchi I metric \eqref{eq:anisotropiclineelement} produces a constant background Ricci tensor $\bar{R}_{\mu\nu}$ and a constant positive Ricci scalar
$\bar{R}=2 \left(A^2+B^2+C^2+AB+AC+B C\right)$. 

\subsubsection{Equations of motion}
Studying the reduced action \footnote{We have set $F_2(\Box)=F_3(\Box)=0$ to simplify our calculations.} 
\begin{eqnarray}\label{reducedaction}
S=\frac{1}{16\pi G}\int d^{4}x \, \sqrt{-g}\left[R+R\mathcal{F}(\Box)R\right],
\end{eqnarray}
around the background \eqref{eq:anisotropiclineelement} gives the equations of motion 
\ba \label{eq:dsbianchieoms}
        T^\alpha_\beta =\left(M^2_p+2 f_0 \bar{R}\right) \left(r^\alpha_\beta
         - \frac{1}{2} \delta^\alpha_\beta
        r \right)   +  \left(2\bar{S}^\alpha_\beta + \frac{1}{2} \delta^\alpha_\beta \bar{R}
        -2 \nabla^\alpha
\nabla_\beta +2 \delta^\alpha_\beta \Box\right)F(\Box)r.~~~~
\ea        
These are the same as the (A)dS field equations for the action \eqref{reducedaction} 
apart from the additional term 
with the traceless curvature tensor $\bar{S}^\alpha_\beta\equiv \bar{R}^\alpha_\beta-\frac{1}{4}
\delta^\alpha_\beta \bar{R}$.

To find the necessary conditions to avoid singularities, we use the same method as in Section \ref{sec:defocusingflat}, contracting \eqref{eq:dsbianchieoms} with $k^\beta k_\alpha$ 
to give the defocusing condition for a generic null tangent vector $k^\alpha$  \cite{Edholm:2017fmw} 
\ba
         k^\beta k_\alpha r^\alpha_\beta = \frac{1}{M^2_P + 2 f_0 \bar{R}}\bigg(k^\beta k_\alpha T^\alpha_\beta- 2 \lambda k^\beta k_\alpha\bar{R}^\alpha_\beta
F(\Box)r+2 \lambda k^\beta k_\alpha \nabla^\alpha
\nabla_\beta F(\Box)r\bigg)<0.~~~~
\ea  

\subsubsection{Choice of $k_\alpha$}
Thus far we have not had to choose $k_\alpha$, because the condition $k^\alpha k_\alpha=0$ 
and the isometry of the Minkowski and (Anti) de Sitter
background has ensured that we can easily find $(k^0)^2$ in terms of $(k^i)^2$. 
However, that is no longer true when we introduce anisotropy. We therefore have to choose 
$k^\alpha$ carefully. We must both satisfy the geodesic equations and 
ensure that the rotation term vanishes.

If $k_\mu$ is of the form,  $k_\mu=(k_0,k_x,0,0)$,
the non-trivial geodesic equations for the null ray are 
\ba \label{eq:nontrivgeode}
        \frac{d^2t}{d\lambda^2} + A e^{2At}\hspace{0.4mm} \frac{dx}{d\lambda} \frac{dx}{d\lambda}=0,\quad
        \quad \frac{d^2x}{d\lambda^2} + 2A \hspace{0.2mm}\frac{dx}{d\lambda} \frac{dt}{d\lambda}=0,
\ea         
where $\lambda$ is the affine parameter. By dividing $g_{\mu\nu} dx^\mu dx^\nu =0$
(the condition for the vector $k_\mu\equiv \frac{dx_\mu}{d\lambda}$ to be null)
by $d\lambda^2$, we see that 
\ba \label{eq:identityfrommetric}
        \frac{dt}{d\lambda}=e^{At} \frac{dx}{d\lambda}.
\ea        
Combining \eqref{eq:nontrivgeode} and \eqref{eq:identityfrommetric} leads to
\ba
        \frac{dt}{d\lambda}=k^0=C e^{-At}, \quad \quad
        \frac{dx}{d\lambda} = k^x = e^{-2At}.
\ea
For simplicity we choose $C=1$ to give 
\ba \label{eq:choiceofkmmu}
        k_\mu=(-e^{-At},1,0,0).
\ea     
For the rotation term to vanish, we require~\cite{poisson_2004}\footnote{The Christoffel
symbols generated by the covariant derivatives in \eqref{eq:rotationtermignorecond} 
cancel out due to the symmetry in the lower indices of the Christoffel symbol, 
so that only partial derivatives remain.} 
\ba \label{eq:rotationtermignorecond}
        \omega_{\alpha\beta} \equiv \nabla_\alpha k_\beta - \nabla_\beta k_\alpha,
\ea         
to be zero, 
which is satisfied by \eqref{eq:choiceofkmmu}.
If we assume the perturbation depends only on time, then our defocusing condition becomes 
(using the NEC) \cite{Edholm:2017fmw}\footnote{We
assume here that $f_0$ is positive, as we later use IDG as an extension to Starobinsky $R+R^2$ inflation. We
also note that $\bar{R}$ is non-negative.}
\ba \label{eq:finaldefocusingconditionfordsbianchi}
        \bigg[ \partial_t^2
        - A \partial_t +(B^2 + C^2)-A (B+C)
        \bigg]F(\Box)r(t)<0,
\ea   
How does \eqref{eq:finaldefocusingconditionfordsbianchi}
compare with the (A)dS case \eqref{eq:dsdefocusingfull}? If we take $H=A$, then we can 
see that the addition of anisotropy has generated
an additional term 
\begin{equation}\label{eq:extra}
        \left[B^2+C^2- A(B+C)\right]F(\Box)r(t).
\end{equation}
Thus anisotropy can have a significant effect on whether a spacetime 
can avoid the null CC condition \eqref{eq:nullc}, especially for slowly
evolving perturbations, and therefore whether 
the spacetime will generate singularities.
\section{Defocusing in more general spacetimes}
\subsection{A constraint on the curvature adapted to IDG}
Even for the reduced action\footnote{More
generally there are more terms, but 
for an FRW metric with spatial curvature $\kappa=0$, the Weyl tensor is zero
and we can discard the
Ricci tensor term using the extended Gauss-Bonnet identity \cite{Li:2015bqa}.}, 
\begin{eqnarray}\label{reducedactionfrw}
S&=&\frac{1}{16\pi G}\int d^{4}x \, \sqrt{-g}\left[R+
R\mathcal{F}(\Box)R-2\Lambda\right],\non\\
&&\text{where } \mathcal{F}(\Box)=\sum_{n=0}^\infty f_n \left(\frac{\Box}{M^2}\right)^n,
\end{eqnarray}

it is almost impossible to solve the equations of motion for IDG for the more 
general Friedmann-Robertson-Walker (FRW) metric
\ba \label{eq:FRWmetric}
        ds^2= -dt^2 + \frac{a^2(t)}{1-\kappa r^2}\left(dx^2 +dy^2 + dz^2\right)
\ea
where $a(t)$ is the scale factor of the universe and $\kappa$ is the spatial curvature.

The FRW metric has the Christoffel symbols
\ba \label{eq:FRWchristoffel}
        \Gamma^t_{ij} = a^2 H \delta_{ij},\quad
        \quad \Gamma^i_{tj}=H \delta^i_j,
\ea        
where $\delta_{ij}$ is the Kronecker delta function, which is 1 if $i=j$ and zero otherwise. 
By using an ansatz we can simplify the equations of
motion \cite{Biswas:2005qr}, to make it possible to generate a
solution for the scale factor. We choose the ansatz 
\ba \label{eq:anstaz}
        \Box R = r_1 R + r_2,
\ea
where $r_1$ and $r_2$ are constants, which means that for $n>0$, $\Box^n R = r_1^n\left(R +\frac{r_2}{r_1}\right)$.
Using this ansatz, it is possible to show that $a(t)=\cosh(\sigma t)$ and 
$a(t)=e^{\lambda t^2}$ are solutions
to the IDG equations of motion for the action, where $\sigma$ and $\lambda$ are constants
\cite{Biswas:2005qr,Biswas:2010zk,Koshelev:2012qn,Koshelev:2013lfm,Koshelev:2017tvv}.
It is notable that both these scale factors generate bouncing universes, meaning
that at some scale, gravity becomes repulsive such that the
universe is forced to stop contracting and start expanding.

In this section we will test whether such bouncing solutions 
can avoid the Hawking-Penrose singularity within the IDG framework and 
are thus geodesically complete. While one might expect that any 
FRW spacetime with a non-singular metric would be geodesically complete, 
this is not necessarily the case \cite{Borde:1993xh,Borde:1997pp,Borde:2001nh}.

Here we will focus on FRW spacetimes, but one should note 
(as found by Sravan Kumar\footnote{This work is so far unpublished.}) that
anisotropic metrics of the form 
\ba
        ds^2=-dt^2 + e^{2\sigma t^2} dx^2 + e^{2\gamma t^2} \left(dy^2 +  dz^2 \right),
\ea        
where $\sigma$ and $\gamma$ are positive constants, would also 
fulfil the ansatz \eqref{eq:anstaz} which could be a topic of further study.
\subsection{Defocusing with the ansatz}
By contracting the equations of motion for the action \eqref{reducedactionfrw} 
with $k^\mu k^\nu$, 
the defocusing condition \eqref{eq:generalmindefo} for any general metric is
\ba \label{eq:defocusingcondition}
         k^\alpha k^\beta R_{\alpha\beta}&=&\frac{1}{M^2_p 
+ 4 F(\Box)R}\bigg[k^\beta k^\alpha T_{\alpha\beta} 
                 + 4 k^\beta k^\alpha\nabla_\alpha
\nabla_\beta  F (\Box)R \nonumber\\
&&+ 2 k^\beta k^\alpha \sum^\infty_{n=1} \frac{f_{n}}{M^{2n}} \sum^{n-1}_{l=0}\left(\partial_\alpha
\Box^l R\right) \partial_\beta \Box^{n-l-1} R\bigg]<0.
\ea
Using the ansatz \eqref{eq:anstaz}, \eqref{eq:defocusingcondition} becomes
\ba \label{eq:mindefocusingconditionwithansatz}
         &&\frac{1}{M^2_p 
+4 F(r_1)\left(R+\frac{r_2}{r_1}\right)-4 f_{0}r_2/r_1 
}\bigg[ k^\beta k^\alpha T_{\alpha\beta} 
                 +4 k^\beta k^\alpha \sum_{n=0}^\infty \frac{f_n}{M^{2n}}\nabla_\alpha
\nabla_\beta \left(r^n_1 R\right) \nonumber\\
&&+2k^\beta k^\alpha \sum^\infty_{n=1}\frac{f_{n}}{M^{2n}}\sum^{n-1}_{l=0}
\partial_\alpha  \left(r^l_1 R\right)\partial_\beta \left(r_1^{n-l-1} R\right)\bigg]<0.
\ea
We note that $r_1$ is a constant so we can pull these terms out of the derivatives, 
which means the last term depends on 
\ba
                \sum^{n-1}_{l=0}\left(\partial_\alpha   R\right)\left(\partial_\beta  R\right)=
                n\left(\partial_\alpha   R\right) \left(\partial_\beta R\right),
\ea
which means \eqref{eq:mindefocusingconditionwithansatz} simplifies to 
\ba \label{eq:simplefocusfrwwoo}
         &&\frac{1}{M^2_p 
+4 F(r_1)\left(R+\frac{r_2}{r_1}\right)-4 f_{0}r_2/r_1 
}\bigg[ k^\beta k^\alpha T_{\alpha\beta} 
                 +4 \sum_{n=0}^\infty \frac{f_n}{M^{2n}}r^n_1 k^\beta k^\alpha\nabla_\alpha
\nabla_\beta  R\non\\
&& +2\sum^\infty_{n=1}\frac{nf_{n}}{M^{2n}}r^{n-1}_1 \left(k^\alpha \partial_\alpha R\right)^2 \bigg]<0.
\ea
We can write \eqref{eq:simplefocusfrwwoo} more succinctly as 
\ba \label{eq:defocusingconditionforgenericf}
        \frac{1}{M^2_P +4 F( r_1) (R+\frac{r_2}{r_1}) -4f_0\frac{r_2}{r_1}} 
        \bigg[ k^\beta k^\alpha T_{\alpha\beta} 
                 +4 F\left(r_1\right) k^\beta k^\alpha \nabla_\alpha
\nabla_\beta R +2 F'(r_1)\left(k^\alpha \partial_\alpha R\right)^2 
        \bigg]<0,~~~~
\ea
where $F'(\Box)$ is defined as
\ba
        F'(\Box)\equiv  \sum^\infty_{n=1}\frac{nf_n}{M^{2n}}\Box^{n-1}.        
\ea

\subsubsection{Simplifying our condition}
Until now, we have assumed only that the metric satisfies the ansatz \eqref{eq:anstaz}.
If we assume an FRW metric \eqref{eq:FRWmetric}, then we can make some simplifications
because the Ricci scalar depends only on $t$ even if the spatial curvature $\kappa$ is 
non-zero.
Furthermore, the tangent vector $k^\mu$ satisfies $k^\mu k_\mu=0$, which means
\ba
        k^\beta k^\alpha \nabla_\alpha
\nabla_\beta R(t) 
&=& (k^0)^2\left( \partial_0^2 - H \partial_0
\right)R(t)\non\\
&=& -(k^0)^2\left(\Box +4H \partial_0\right)R(t),
\ea
where we have used that the d'Alembertian of the FRW\ metric
acting on a homogenous scalar  is 
$\Box S(t)= -\partial_0^2 S(t) -3H\partial_0
S(t)$ to get to the second line.
We can further simplify \eqref{eq:defocusingconditionforgenericf} by using
the vacuum equations of motion for a metric fulfilling the ansatz. The trace vacuum equations of motion can be 
written as~\cite{Biswas:2010zk,Koshelev:2012qn,Koshelev:2013lfm}
\ba \label{eq:traceequationforansatz}
        A_1 R + A_2 \left(2 r_1 R^2 +( \partial_\mu R)  \partial^\mu R \right)
+ A_3 = 0, 
\ea
where the coefficients $A_i$ are defined as
\ba
        A_1 &=& -M^2_P + 4 F'(r_1) r_2 - 2 \frac{r_2}{r_1} \left(F(r_1) -f_0\right)
+ 6 F(r_1)r_1, \quad A_2 =  F'(r_1),\non\\ 
        A_3 &=& 4 \Lambda + \frac{r_2}{r_1}\left( M^2_P+ A_1\right) - 2 \frac{r_2^2}{r_1}F'(r_1).
\ea
\subsubsection{Simple choice of solution}
The obvious simple choice of solution to \eqref{eq:traceequationforansatz} is to set $A_i=0$
which gives the identities
\ba \label{eq:simplesolutiontoeoms}
         F'(r_1)=0, \quad \quad r_2 = - \frac{r_1\left(M^2_P - 6 F(r_1)r_1\right)}
         {2\left(F(r_1)-f_0\right)},\quad
\quad \Lambda =-\frac{r_2M^2_P}{4r_1}= M^2_P \frac{\left(M^2_p-6 F(r_1)r_1\right)}
{8\left(F(r_1)-f_0\right)}.~~~~
\ea       
These identities allow us to eliminate $f_0$, the lowest order of
$\mathcal{F}(\Box)$ defined in \eqref{reducedactionfrw}, 
from \eqref{eq:defocusingconditionforgenericf}. 
This simplifies the requirement to avoid the Hawking-Penrose singularity to \cite{Edholm:2018dsf}
\ba \label{eq:defocusingconditionfinalforFRWsimplesoln}       
        \frac{ 2F\left(r_1\right) \left( (r_1+4H\partial_t) R+r_2\right)-(\rho+p)}{ F( r_1)
(4R+6r_1)-M^2_P } >0,
\ea

We will now apply \eqref{eq:defocusingconditionfinalforFRWsimplesoln} to various 
choices of the scale factor which satisfy the ansatz, produce a bouncing universe 
and are non-singular, i.e. $a(t)$ is strictly positive.

We will eliminate $F(r_1)$ using 
\ba
        F(r_1)= \frac{2f_0 r_2 -r_1M^2_P}{2r_2-6r_1^2},
\ea        
derived from the middle equality in \eqref{eq:simplesolutiontoeoms}.
\subsubsection{Toy model}
The simplest model we will examine is a toy model with zero spatial curvature 
$\kappa$ \cite{Conroy:2014dja}
\ba \label{eq:toymodelmetric}
        a(t)=1+a_2t^2,
\ea        
which fulfils the ansatz \eqref{eq:anstaz} near the bounce as $R=12a_2 -12a_2^2t^2$ and $H\approx a_2t$. 
We find $\Box R= -6a_2 R + 48a_2^2$. 
The condition for the metric \eqref{eq:toymodelmetric} to fulfil the defocusing condition is (neglecting matter 
and assuming the solution \eqref{eq:simplesolutiontoeoms})
\ba \label{eq:defocusingconditionfinalforFRWsimplesolntoy}       
        \frac{ M^2_P+16a_2f_0-\frac{5}{12}(\rho+p)}{M^2_P +96
f_0a_2} >0,
\ea 
which is satisfied for $-\frac{M^2_P}{16}<a_2f_0<-\frac{M^2_P}{96}$.
\subsubsection{Cosh scale factor}
The bouncing hyperbolic cosine model
\ba \label{eq:coshscalefactor}
        a(t)= \cosh(\sigma t)
\ea
was examined in \cite{Biswas:2005qr,Biswas:2010zk}. 
It was found that an FRW metric \eqref{eq:FRWmetric} with the scale factor \eqref{eq:coshscalefactor} fulfils the ansatz \eqref{eq:anstaz}, even if the spatial
curvature $\kappa$ is non-zero, with 
$r_1=2\sigma^2$\ and $r_2=-24\sigma^4$.        
The condition for this metric to fulfil the defocusing condition is
\ba \label{eq:defocusingfrwcoshfdashvanishedwithsecondconditionsmalltimes}
        \frac{24\sigma ^2  \left(\kappa-\sigma ^2 \right)F\left(r_1\right)
        -(\rho+p)}{M^2_P +24 F(r_1)
        \left(\kappa-\sigma ^2 \right) +48\sigma^2 f_0}   <0.
\ea 
Taking the simple solution \eqref{eq:simplesolutiontoeoms} to the 
equations of motion and neglecting matter, 
\eqref{eq:defocusingfrwcoshfdashvanishedwithsecondconditionsmalltimes}
is fulfilled for zero spatial curvature $\kappa$ if 
$-\frac{1}{24}\frac{M^2_P}{\sigma^2}<f_0<-\frac{1}{96}\frac{M^2_P}{\sigma^2}$.
If matter is included, and gives a very large contribution such
that $\rho+p>24\sigma ^2  \left(\kappa-\sigma ^2 \right)F\left(r_1\right)$, 
then defocusing occurs for $f_0<-\frac{1}{96}\frac{M^2_P}{\sigma^2}$. 

\subsubsection{Exponential bouncing solution}
Finally we will examine the scale factor $a(t)=\exp(\lambda t^2)$ where $\lambda$ is a constant, which
gives $R=3\lambda\left(1+\lambda t^2\right)$. An FRW metric with this scale factor
satisfies the ansatz with $r_1=-6\lambda$ and $r_2=24\lambda^2$ as long as the 
spatial curvature $\kappa=0$.
The condition for this metric to fulfil the defocusing condition is 
\ba \label{eq:defocusingconditionfinalforFRWsimplesolnexpbouncingnearbounce}       
        \frac{ 5\lambda^2F\left(r_1\right)+(\rho+p)}{M^2_P +20\lambda F( r_1)  -8
f_0\lambda} <0.
\ea
and with the simple solution to the equations of motion \eqref{eq:simplesolutiontoeoms}, 
we find that there is defocusing for $-\frac{1}{4}M^2_P<\lambda f_0 <\frac{1}{28}M^2_P$ 
if we neglect matter, or $\lambda f_0 <\frac{1}{28}M^2_P$ if the contribution 
of matter is large.

We already knew that IDG produced bouncing FRW\ solutions which do not generate metric singularities, but this 
does not necessarily mean that singularities are avoided due to the Hawking-Penrose singularity theorems.
  In this section, we have shown that all of these bouncing scale factors can also avoid 
the Hawking-Penrose singularity under certain conditions.

\chapter{Inflation}
\label{chapter:inflation}
Primordial inflation is a period of accelerated expansion
in the early universe~\cite{Guth:1980zm,Linde:1981mu,Albrecht:1982wi}. 
It was developed as a way to solve the horizon and flatness problems, as well as to generate 
temperature anisotropy in the Cosmic Microwave Background (CMB)\footnote{The anisotropy
is generated through quantum fluctuations of the inflaton \cite{Langlois:2004de}.}.

\subsubsection{Horizon problem}
The horizon problem is that the universe appears to be 
homogeneous across regions of space which would not be causally connected 
unless there was a period of accelerated expansion 
in the early universe \cite{Kolb:1990vq}. 
In other words, if inflation had not occurred, light would not have had enough time to travel
between these regions, yet they appear to have exchanged information
and settled to an equilibrium state.
One can observe that at angular scales of around one degree, the CMB is isotropic to one part in 10,000 
and appears to have the spectrum of a thermal blackbody. 
Assuming that the universe is approximately homogeneous 
means any fundamental observer in the universe would see this 
high degree of isotropy in the CMB. 

On the other hand, within the inflationary paradigm, regions can 
exchange information and then be rapidly expanded away from each other to
give the homogeneity across large distances that we see today.

\subsubsection{Flatness problem}
The flatness problem is that the universe appears to be fine-tuned to have almost exactly the correct
density of matter and energy to generate a flat universe \cite{Kolb:1990vq}.
To be exact, the universe can be described with an FRW metric \eqref{eq:FRWmetric} with spatial curvature $\kappa\approx 0$. The energy density needed to achieve a completely flat FRW metric is given by $\rho=\rho_{\text{crit}}$. We cannot observe 
$\kappa$ directly, but we can constrain the density parameter $\Omega=\rho/\rho_{\text{crit}}$ to be almost exactly 1 through observing the anisotropies in the CMB \cite{Aghanim:2018eyx}. 

The density parameter is given by 
\ba
        \Omega_\kappa =\frac{-\kappa}{a^2H^2}.
\ea
During the radiation (matter) domination phase of the universe's expansion,
$a\propto t^{1/2} (t^{2/3})$ and $H\propto 1/t$ in both. Therefore $\Omega_\kappa$
is increasing in both regimes. We require accelerated expansion so that the
density parameter shrinks as time progresses \cite{Senatore:2016aui}.
In particular, if we are in a quasi-de Sitter regime, then $a\approx e^{Ht}$ 
and $H$ is roughly constant so that the density parameter is exponentially suppressed.

\subsubsection{Mechanisms for inflation} 
There are several mechanisms for inflation, but Starobinsky inflation~\cite{Starobinsky:1980te}
is possibly the most
successful when compared to observables \cite{Akrami:2018odb}.
Starobinsky inflation is generated by $R+R^2$ gravity, 
a subset of $f(R)$ gravity \eqref{eq:fofraction}. 
The extra $R^2$ term compared to GR can be recast as an extra scalar 
degree of freedom which drives an accelerated expansion of the universe 
when the curvature is large, i.e. in the early universe \cite{Pallis:2016mvm}.  
Note that $R+R^2$ gravity does not allow both defocusing
and a positive curvature \cite{Conroy:2016sac}, so Starobinsky
inflation, also called Starobinsky gravity, cannot
simultaneously generate inflation and resolve the Big Bang\ Singularity.

IDG contains $R+R^2$ gravity within it, and can thus be written as an 
extension to Starobinsky gravity \cite{Craps:2014wga,Koshelev:2016xqb,Koshelev:2017tvv},
 so we should check what 
effect the extra terms have on the inflationary observables and how any modification to Starobinsky inflation affects the compatibility of the observables 
with data.

One can find the quantum scalar and tensor perturbations
for IDG around an inflationary background
 \cite{Craps:2014wga,Koshelev:2016xqb}. 
The scalar perturbations receive no modification 
when we add the extra terms in IDG to the Starobinsky action. 
However, the tensor perturbations are changed by a factor dependent 
upon the IDG mass scale.

In this chapter, we extend the previous work \cite{Craps:2014wga,Koshelev:2016xqb}
 by explicitly computing the
scalar spectral tilt, which is how the scalar power spectrum changes with
the wavenumber, and use observational data to constrain our 
mass scale $M$.  

\section{Method}
We recast the action \eqref{fullaction2} so that the Riemann tensor squared term
 $R_{\mu\nu\rho\sigma} F_3(\Box) R^{\mu\nu\rho\sigma} $
is replaced by the Weyl tensor squared term $C_{\mu\nu\rho\sigma} \mathcal{F}_3(\Box)
C^{\mu\nu\rho\sigma}$ giving
\ba \label{eq:occurenceunreducedaction}
           S = \frac{1}{2} \int d^4 x \sqrt{-g} \Big[ M^2_p R + R {\cal F}_1 (\Box) R 
           + C_{\mu\nu\rho\sigma} {\cal F}_3(\Box) C^{\mu\nu\rho\sigma}  \Big],
\ea     
where we have used the extended Gauss-Bonnet identity\footnote{$GB_n \equiv R \Box^n R\ -4 R_{\mu\nu} \Box^n R^{\mu\nu}
+ R_{\mu\nu\alpha\beta} \Box^n R^{\mu\nu\alpha\beta}$ does not contribute to the 
equations of motion.} \cite{Li:2015bqa} to discard $F_2(\Box)$.
Using a reduced action, without the Weyl term 
in \eqref{eq:occurenceunreducedaction}, one can examine the 
power spectrum of a bouncing FRW solution in the context of super-inflation
(where $\dot{H}>0$)
to give an estimate of $M\sim$ $10^{15}$~GeV~\cite{Biswas:2013dry}.

We again use the ansatz $\Box R = c_1 R$ used in \eqref{eq:anstaz} and the simple solution to the equations of motion
 \eqref{eq:simplesolutiontoeoms}. We set $c_2=0$ in
\eqref{eq:anstaz}, 
which by \eqref{eq:simplesolutiontoeoms} is equivalent
to setting $\Lambda=0$ in order to 
embed \eqref{eq:occurenceunreducedaction} in the Starobinsky framework \cite{Koshelev:2016xqb}.
This ansatz gives ${\cal F}_1(\Box)R = F_1 R$, where $F_1$ is a constant.
We use the solutions to the equations of motion which generate
the identities \eqref{eq:simplesolutiontoeoms}.
\subsection{Scalar fluctuations around an inflationary background}
Using the second scalar variation of the action 
\eqref{eq:secondvariationoftheactionformaxsymm},
we can see that the Weyl term does not affect the scalar perturbations 
and so the action \eqref{eq:occurenceunreducedaction} predicts the same 
scalar perturbations as Starobinsky gravity \cite{Koshelev:2016xqb}.

It can therefore be shown that the scalar power spectrum of 
\eqref{eq:occurenceunreducedaction}
is \cite{Craps:2014wga,Koshelev:2016xqb}
\ba 
        |\delta_\Phi (\textbf{k},\tau)|^2 = \frac{k^2}{16 \pi^2 a^2} 
        \frac{1}{3 F_1 \bar{R}}.
\ea    
which in the long wavelength limit $k\ll H$ becomes
\ba \label{eq:scalarpowerspectrumintermsofphi}
        |\delta_\Phi (\textbf{k},\tau)|^2 = \left.\frac{H^2}{16 \pi^2 } 
        \frac{1}{3 F_1 \bar{R}}\right|_{k=aH}.
\ea   
By multiplying by $H^4/\dot{H}^2$, we obtain the measured power spectrum of the
gauge-invariant co-moving curvature perturbation
$\mathcal{R}\equiv \Psi  +\frac{H}{\dot{\bar{R}}} \delta R_{GI}$ where $\delta R_{GI}$ is the linear variation of the Ricci scalar.
During inflation $\dot{H}\ll H^2$, so 
$\mathcal{R} \approx - H^2 \Phi/\dot{H}$ \cite{Craps:2014wga,Koshelev:2016xqb}.
We will rewrite \eqref{eq:scalarpowerspectrumintermsofphi} in terms of the number of e-folds $N=\ln(aH)$ from the time the
perturbation crosses the horizon until the end of inflation, 
i.e. the number of times the scale factor increases by a factor of $e$ \cite{Mazumdar:2010sa}.
In the Starobinsky framework, $N$ is given by \cite{Koshelev:2016xqb}
\ba \label{eq:defnofnefolds}
        N = - \frac{1}{2} \frac{H^2}{\dot{H}},
\ea 
and using the background Ricci scalar $\bar{R}\approx 12H^2$ during inflation, the power spectrum is
given by \cite{Edholm:2016seu} 
\ba \label{eq:powerspectrumintermsofN}
        P_s=|\delta_\mathcal{R}|^2 \approx \frac{N^2}{24 \pi^2 } 
        \frac{1}{6 F_1 }.
\ea 
Note that upon taking $ F_1=1/(6M_s^2)$, i.e. the limit back to Starobinsky gravity $R+\frac{1}{6M^2_s}R^2$, 
we return to the Starobinsky result $P_s= \frac{M_s^2N^2}{24 \pi^2 } 
        $~\cite{Huang:2013hsb}. 
There is no contribution to the scalar perturbations 
in \eqref{eq:secondvariationoftheactionformaxsymm}
from the Weyl tensor term, so the result \eqref{eq:powerspectrumintermsofN}
being equivalent to the Starobinsky prediction is expected.

Finally, we calculate the scalar spectral tilt $n_s$, or how the scalar power spectrum 
changes with the wavenumber $k$, which is generally written in terms of 
$k$ using $n_s=1+\frac{d(\ln P_s)}{d(\ln k)}$, but we write $n_s$ in terms 
of the number of e-folds $N$ using 
\ba
        \frac{d(\ln P_s)}{d(\ln k)} = \frac{1}{P_s} \frac{dP_s}{dN} \frac{dN}{d(\ln k)},
\ea
and $\frac{dN}{d(\ln k)} = \frac{d(-\ln k)}{d(\ln k)}=-1$. Therefore
\cite{Edholm:2016seu,SravanKumar:2018dlo}
\ba \label{eq:scalarspectraltilt}
        n_s= 1- \frac{1}{P_s} \frac{dP_s}{dN} = 1-\frac{2}{N}.
\ea

 The spectral tilt is one of the 
observables used to constrain theories of inflation and we will use it later
when comparing our predictions to data from the Planck satellite.          

\subsection{Tensor perturbations}

The tensor part of the second variation of the action is given by 
\eqref{eq:secondvariationoftheactionformaxsymm}, which we can manipulate 
using the identities \eqref{eq:simplesolutiontoeoms} into
\ba \label{eq:tensorperturbationaction}
        \delta^2 S_\perp =\frac{1}{4} \int d^4 x \sqrt{-\bar{g}}
 h^\perp_{\mu\nu} \left(\Box - \frac{\bar{R}}{6} \right) F_1 \bar{R} 
        \left[1 + \frac{1}{ F_1 \bar{R}}\left( \Box - \frac{\bar{R}}{3}
\right) {\cal F}_3 \left( \Box + \frac{\bar{R}}{3} \right) \right]
        h^{\perp\mu\nu}.~~
\ea
\eqref{eq:tensorperturbationaction} is the result for the Starobinsky action, 
multiplied by the factor in square brackets. 
Note that it retains the standard GR\ root at 
$\Box=\bar{R}/6$ \footnote{It might seem that 
returning to GR by setting $F_1=0$ would make 
$\delta^2 S_\perp$ vanish, but this is not the case 
due to the identities \eqref{eq:simplesolutiontoeoms}.}. 
We do not want to
add any extra poles into the propagator, so we do not allow any zeroes from the term in
the square brackets. As before, we therefore choose this term to be the exponential of 
an entire function, i.e. \cite{Koshelev:2016xqb}
\ba \label{eq:defnpbox}
       P(\Box)= 1 + \frac{1}{ F_1 \bar{R}}\left( \Box - \frac{\bar{R}}{3}
\right) {\cal F}_3 \left( \Box + \frac{\bar{R}}{3} \right),
\ea
where $P(\Box)$ is the exponential of an entire function. The simplest choice is 
\ba \label{eq:choiceofPbox}
        P(\Box)=e^{\omega(\Box)},
\ea         
meaning that $\mathcal{F}_3(\Box)$ takes the form
\ba \label{eq:f3boxforpoorchoice}
        \mathcal{F}_3(\Box)= F_1 \bar{R} \frac{e^{\omega(\Box-\bar{R}/3)}-1}
        {\Box-\frac{2}{3}\bar{R}}.
\ea       
The factor \eqref{eq:choiceofPbox} is always positive, which will be important later 
when we look at the tensor-scalar ratio.

The Starobinsky tensor power spectrum is modified by the inverse of
the factor \eqref{eq:choiceofPbox}
evaluated at $\Box=\bar{R}/6$, i.e. the root of \eqref{eq:tensorperturbationaction},
and becomes \cite{Koshelev:2016xqb}
\ba  \label{eq:tensorpowerspec}
        |\delta_h|^2 = \frac{H^2}{2\pi^2 F_1 \bar{R}} e^{-\omega(\bar{R}/6)},
\ea
Using \eqref{eq:powerspectrumintermsofN} and \eqref{eq:tensorpowerspec}, the ratio between the tensor and scalar power spectra
$r = 2 \frac{|\delta_h|^2}{|\delta_R|^2}$
 is given by\footnote{The factor of 2 accounts for the two polarisations of the tensor modes.}
\ba \label{eq:tensortoscalarratioN}
        r  = 48 \frac{\dot{H}^2}{H^4}  e^{-\omega(\bar{R}/6)}
        = \frac{12}{N^2}e^{-\omega(\bar{R}/6)}.
\ea        
where the right hand side is written terms of the number of e-foldings $N$ defined in \eqref{eq:defnofnefolds}.
The tensor-scalar 
ratio for pure Starobinsky inflation is given by \cite{Ade:2015xua}
\ba
        r_{\text{Staro}} = \frac{12}{N^2},
\ea
i.e. the extra Weyl tensor term 
in \eqref{eq:occurenceunreducedaction} 
adds the modification $e^{-\omega(\bar{R}/6)}$.
This result, derived in  \cite{Koshelev:2016xqb}, is extremely important. It will allow us to constrain the IDG mass scale in this chapter, or even one day be able to  confirm the presence of IDG when more precise observations can be made. 
\section{Constraints from CMB data}
\label{eq:constrtaintscmbs}
From observations, the bound on the tensor-scalar ratio is $r<0.07$ \cite{Ade:2015xua}, so from 
\eqref{eq:tensortoscalarratioN}, 
\ba
        \frac{12}{N^2}e^{-\omega(\bar{R}/6)}<0.07,
\ea
which leads to 
\ba
       -\omega(\bar{R}/6) < 2 \log(N) -5.14.
\ea
During Starobinsky inflation, the universe was in a quasi-de Sitter 
spacetime \cite{Ozkan:2015iva} with
\ba
        \bar{R}\equiv 12H^2 = \frac{9.02\times 10^{32}}{N^2} \text{GeV}^2,
\ea
giving the constraint
\ba
       -\omega\left(\frac{9.02\times 10^{32}}{6M^2 N^2} \text{GeV}^2\right) < 2 \log(N) -5.14.
\ea                        
A priori, $\omega(\Box)$ in \eqref{eq:choiceofPbox} could be any polynomial, giving a huge amount
of functional freedom.\ We have the sole condition that $\omega(-k^2)\to \infty$
in the $UV$\ limit $k^2 \to \infty$, meaning that the propagator is exponentially suppressed.
This condition means that $\omega$ must be built up from terms of the form $(-\Box)^n$, where $n>0$.
When we evaluate these terms at $\Box=\bar{R}/6$, they become $(-\bar{R}/6)^n
$ which for odd $n$ means that the argument of the exponential in \eqref{eq:choiceofPbox} is negative, meaning that $r$ will be 
larger than the Starobinsky prediction, and vice-versa for even $n$.
 
\subsection{Example functions}
We will choose a few simple functions to demonstrate the effects of IDG on the tensor-scalar 
ratio, looking only at odd powers as we do not have a lower bound on $r$
and therefore cannot test predictions of a decrease in $r$. We first take 
\begin{itemize}
\item
$\omega(\bar{R}/6)= -(\bar{R}/6M^2)^n$, i.e. a monomial function

Using this function and choosing odd $n$, we generate a bound on $M$ of 
\ba
        M>\frac{\sqrt{3/2}}{N} \left[(2\log(N) -5.14)\right]^{-1/2n}
        \times 10^{16}\text{ GeV}.
\ea        
Taking the simplest case $n=1$ and perturbations which left the horizon $N=60$ e-foldings before the end of inflation\footnote{55-60 e-foldings corresponds to the CMB scale \cite{Barnaby:2011qe}. We have chosen 60 e-foldings here, but similar results are obtained with any $N$ in the region  $55<N<60$.}  gives
\ba
        M>1.17 \times 10^{14} \text{ GeV}.
\ea        
with $n=3$ and again taking $N=60$, we obtain
\ba
        M>1.69 \times 10^{14} \text{ GeV}.
\ea
Choosing $N=60$ but keeping $n$ general, we find
\ba \label{eq:constraintfrominlationngeneral}
        M> 2.04 \times (0.573)^{1/n} \times 10^{14} \text{ GeV}.
\ea
\eqref{eq:constraintfrominlationngeneral} is only weakly dependent on $n$, which one might have expected -
the exponential enhancement of \eqref{eq:tensortoscalarratioN} means that once 
$M$ decreases past $M\approx \bar{R}/6$,
the extra factor very quickly blows up and takes the tensor-scalar ratio
out of the allowed region.

\item
$\omega(\bar{R}/6)= \bar{R}/6M^2+  (\bar{R}/6M^2)^a$, using a binomial form of
$\omega$. We assume that the coefficients of the two terms are the same, 
so can be absorbed within $M$.
We obtain a lower bound of $M>1.85\times 10^{14}$ GeV for $a=3$, increasing to
$M>2.01 \times 10^{14}$ GeV for $a=16$ and 
$2.03 \times 10^{14}$ GeV for $a=81$. We again see the very weak dependence of the
constraint on $M$ on our choice of $a$. 

\item
$\omega(\bar{R}/6)= \sum_{a=1}^\infty(\bar{R}/6M^2)^a$,
i.e. $\omega$ is a sum over all orders of odd $a$, assuming that the coefficients 
of the terms are the same. Using this choice we generate a constraint of 
$M>2.21 \times 10^{14}$ GeV.

\end{itemize}
As one might expect, the constraint on $M$ is very mildly dependent on our choice
of the function $\omega$. Once $M$ becomes of the order of $10^{14}$ GeV, then 
the exponential enhancement means that $r$ rapidly becomes larger than the upper bound.

\begin{figure}[!htbp]
\centering
\hspace{-4mm}\includegraphics[width=130mm]{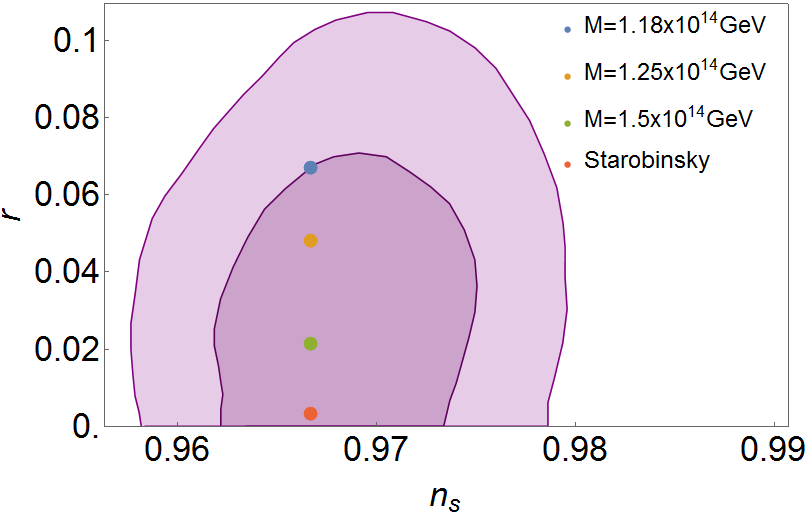}
\caption{The tensor-scalar ratio $r$ against the scalar spectral index $n_s$ for
different choices of our mass scale $M$ \cite{Edholm:2016seu}. We have taken the number of 
e-foldings since the start of inflation to be 60 and plotted against 
the Planck data \cite{Ade:2015xua}.}
\label{inflationplot}
\end{figure}

We can compare the predictions of IDG compared to Starobinsky gravity 
against the data graphically. 
Combining \eqref{eq:scalarspectraltilt} 
and \eqref{eq:tensortoscalarratioN} gives the tensor to scalar ratio in terms
of the scalar spectral tilt
\ba \label{eq:rintermsofns}
        r&=&\frac{12}{N^2} e^{-\omega(\bar{R}/6)}\non\\
        &=& 3(1-n_s)^2 e^{-\omega(\bar{R}/6)},
\ea        
which is compared to the Planck data \cite{Ade:2015xua}
in Fig. \ref{inflationplot}, 
assuming the form 
$\omega(\bar{R}/6)= -\bar{R}/6M^2$.
By visual inspection of Fig. \ref{inflationplot}, we obtain a very slightly stronger constraint 
of $M>1.18\times 10^{14}$ GeV.

\newpage      
\section{Conclusion}
In this chapter we have seen that IDG can provide a mechanism for inflation. 
We have described the differences from Starobinsky inflation and constrained 
the IDG mass scale while providing a possible explanation if further observations 
show an increased tensor-scalar ratio compared to the Starobinsky prediction.
Further work in this area could look at other inflationary observables. 
Recently \cite{SravanKumar:2018dlo} looked into the effect of IDG on the tensor spectral tilt, 
and found that the Starobinsky prediction for the spectral tilt would be modified by 
a factor depending on \eqref{eq:choiceofPbox}.

\chapter{Gravitational radiation}
\label{chapter:radiation}
The detection of gravitational waves was one of the most successful 
events in the history of gravitational physics. The merger of two black 
holes   caused a ripple in spacetime that travelled for 
over a billion years to reach the LIGO detectors \cite{Abbott:2016blz}.
Two years later, the detection of gravitational waves caused by a neutron
star merger and the associated electromagnetic waves 
confirmed that gravitational waves 
travel at the speed of light \cite{TheLIGOScientific:2017qsa}, 
placing extremely tight constraints on many modified gravity theories
which try to explain the accelerated expansion of the 
universe \cite{Nojiri:2017hai,Amendola:2015ksp,Creminelli:2017sry,Lombriser:2015sxa,Ezquiaga:2017ekz,Sakstein:2017xjx}.

However, a less well-known test of gravitational waves
has been ongoing for the past forty years. The Hulse-Taylor binary is made up 
of a pulsar in orbit with another neutron star \cite{Hulse:1974eb}. 
The pulsar sends out radio pulses
every 59 milliseconds, which are detected slightly earlier or later depending on the 
stage of the orbit. Hulse and Taylor  \cite{Hulse:1974eb} showed that the
orbit had a period of 7.75 hours. 

As with any objects in orbit around another,
gravitational wave (GW) radiation is produced, causing the system to lose energy and
the orbit to gradually shrink. The fact that we can very accurately measure 
the orbital period of the binary system gives us another test of GR \cite{Weisberg:2016jye}.
The decrease in orbit over the past forty years is almost exactly what 
GR would predict. We will investigate whether IDG affects this prediction.

\section{Modified quadrupole formula}
The quadrupole formula describes the change in the metric caused by a particular 
system with a given stress-energy tensor. Using certain approximations
and working in the linearised regime, the quadrupole formula
 is relatively easy to calculate 
in GR, but the addition of infinite derivatives means the calculation is more
difficult. 

We start by writing the IDG equations of motion in a neat form.
In the de Donder gauge, $\partial_\mu h^{\mu\nu} =\frac{1}{2}\partial^\nu h,$ the linearised equations of motion \eqref{eq:lineareomsintermsofh} become
\ba \label{eq:lineareomsinddgauge}
        -2\kappa T_{\mu\nu}=
          a(\Box) \left(\Box h_{\mu\nu} - \frac{1}{2}\partial_\mu \partial_\nu h \right)
        + c(\Box)\left(\frac{1}{2} \partial_\mu \partial_\nu h  - \frac{1}{2}g_{\mu\nu} \Box h \right).
\ea         
If we take $a(\Box)=c(\Box)$ so that there are no extra poles in the propagator, then \eqref{eq:lineareomsinddgauge} becomes 
\ba  \label{eq:lineareomsinddgaugeaequalsc}
        -2\kappa T_{\mu\nu} = a(\Box)\Box \bar{h}_{\mu\nu},
\ea
where $ \bar{h}_{\mu\nu} \equiv h_{\mu\nu}-\frac{1}{2}g_{\mu\nu}h$. 
\eqref{eq:lineareomsinddgaugeaequalsc} returns to the GR result 
if we set $a= 1$. The method for deriving the 
quadrupole formula for GR is well established in the literature \cite{Carroll:2004st} and here we
go through the same method but with the additional $a(\Box)$ factor. 
         
We will use the Fourier transform with respect to time $t$, 
\ba
        \tilde{\phi}(k,\textbf{x}) &=&\frac{1}{\sqrt{2\pi}}
        \int dt \hspace{2mm} e^{ik t}~ \phi(t,\textbf{x}),\non\\
        \phi(t,\textbf{x})&=&\frac{1}{\sqrt{2\pi}}
        \int dk\hspace{1mm} e^{-ik t} \tilde{\phi}(k,\textbf{x}).
\ea        
We take the transform of the metric perturbation \eqref{eq:lineareomsinddgaugeaequalsc} and invert $a(-k^2)$ to obtain
\ba
        \tilde{\bar{h}}_{\mu\nu} = 4 G  
        \int d^3 y \hspace{1mm}e^{i k |\textbf{x}-\textbf{y}|} 
        \frac{\tilde{T}_{\mu\nu}(k,\textbf{y})}{|\textbf{x}-\textbf{y}|a(-k^2)}.
\ea
We assume that the source is isolated, far away from 
the observer and made up of non-relativistic matter (and hence
slowly moving).
 The source can therefore be considered to be centred
 at a distance $r$, with the different parts of the source
 at distances $r+\delta r$ where $\delta r\ll r$.
Since the source is moving slowly, the radiation emitted will mostly
be at frequencies $k$ such that $\delta r \ll k^{-1}$.
Using these approximations, as for the GR method  \cite{Carroll:2004st}, we can replace $\frac{e^{i k |\textbf{x}-\textbf{y}|}}{ |\textbf{x}-\textbf{y}|}$
with $\frac{e^{i k r}}{r}$, which can be taken outside the integral to give 
\ba \label{eq:firstdefinitionquadhfour}
        \tilde{\bar{h}}_{\mu\nu} = 4 G \frac{e^{i k 
        r}}{r} 
        \int d^3 y \frac{\tilde{T}_{\mu\nu}(k,y)}{a(-k^2)}.
\ea
\subsubsection{Quadrupole moment}
The mass quadrupole moment gives the lowest-order contribution 
to gravitational radiation \cite{Thorne:1980ru}. 
In linearised gravity, the mass monopole and the mass dipole moment 
are both conserved quantities, corresponding to the 
mass-energy in the system and the center of mass of the system respectively.  
The quadrupole moment is defined by 
\ba \label{eq:quadrupolemoment}
        I_{ij} = \int d^3y~T^{00} (y) y^i y^j . 
\ea
Upon using integration by parts on $\int d^3 y \hspace{0.4mm}\tilde{T}^{ij}$,
one produces two terms. The first is a surface integral term, which vanishes because the source 
is isolated. The second term can be simplified using the Fourier space identity
\ba
        -\partial_m \tilde{T}^{m\mu}=i k\hspace{0.2mm} \tilde{T}^{0\mu},
\ea
where $m$ is a spatial index. Using this identity twice produces 
\ba
        \int d^3 y\hspace{0.4mm} \tilde{T}^{ij} = -\frac{k^2}{2} \int y_i y_j \tilde{T}^{00} d^3y. 
\ea
Inserting \eqref{eq:quadrupolemoment} into the $ij$ component of 
\eqref{eq:firstdefinitionquadhfour} gives
\ba \label{eq:firstdefinitionquadhfourI}
        \tilde{\bar{h}}_{ij} = -2 G k^2\frac{e^{i k r}}{r} 
        \frac{\tilde{I}_{ij}(k)}{a(-k^2)}.
\ea
Transforming back to $t$, we obtain 
\ba \label{eq:finaleqnwithgenerala}
        \bar{h}_{ij}= \frac{ - G }{\pi} \frac{1}{r}\frac{d^2}{dt^2}\int dk~ dt'_r e^{-ik(t'-r')} e^{ik(t-r)}  
        \frac{1}{a(-k^2)} I_{ij} (t'-r').
\ea
If we define the retarded time $t_r=t-r$, then \eqref{eq:finaleqnwithgenerala} simplifies to
\ba \label{eq:finaleqnwithgeneralatr}
        \bar{h}_{ij} =\frac{ -2 G }{2\pi} \frac{1}{r}\frac{d^2}{dt^2}\int dk ~dt'_r  
        \frac{e^{ik(t_r-t'_r)}}{a(-k^2)} I_{ij} (t'_r).
\ea
\eqref{eq:finaleqnwithgeneralatr} is the general formula for any choice of $a(-k^2)$. 
However, to calculate the integral
we have to make a choice for $a(-k^2)$. 
\subsubsection{Simplest choice of $a(-k^2)$} 
The simplest ghost-free choice is $a(-k^2)=e^{k^2/M^2}$. 
The inverse Fourier transform of a Gaussian distribution $e^{-k^2/a}$ is 
\ba \label{eq:identityinversefofgauss}
        \int^\infty_{-\infty} e^{-k^2/a}e^{-ikx} dk = \sqrt{\pi a} \hspace{1mm} e^{-ax^2/4}.
\ea
Inserting \eqref{eq:identityinversefofgauss} into \eqref{eq:finaleqnwithgeneralatr} produces        
\ba \label{eq:finaleqnwithspecificatr}
        \bar{h}_{ij} = - \frac{  G }{\sqrt{\pi}} \frac{M}{r}\frac{d^2}{dt^2}\int dt'_r  
        e^{-M^2\left(t_r(t)-t_r^{\prime 2}\right)} I_{ij} (t'_r),
\ea
where we have noted that $t_r$ is a function of $t$. We have calculated the modified quadrupole formula for the simplest case of ghost-free IDG. 
For any quadrupole moment $I_{ij}$, we can calculate the corresponding perturbation to
the metric. 

We take the example of a binary system of two stars in a circular orbit and examine the modification 
to the 1-1 component of the perturbed metric. The 1-1 component of the 
quadrupole moment is given by 
\ba
        I_{11}=M_sR^2 \left(1+\cos(2\omega t)\right),
\ea
where $M_s$ is the mass of each star in the binary system, $R$\ is the distance between the stars and $\omega$
is their angular velocity. The perturbation to the 1-1 component of the metric is given 
by
\ba
        \bar{h}_{11}  = \frac{4GM_s^2R^2}{r} \left(1+e^{-4\omega^2/M^2} \cos(\omega t_r)
        \right)      
\ea
We can see there is an additional factor $e^{-4\omega^2/M^2}$, which tends to 1 in 
the GR limit $M\to \infty$. If $\omega$ is of the same order as $M$
or bigger, this additional term will suppress the oscillating part of the perturbation.
\newpage
\section{Backreaction equation} 
\label{sec:backreaction}
When gravitational waves (GWs) propagate through space, they create a second order effect 
known as the backreaction.  This is caused because gravity is nonlinear and therefore couples to itself, which means that GWs can be said to gravitate. 
The backreaction was previously studied in the framework of
 Effective Quantum Gravity \cite{Kuntz:2017pjd}, 
 which attempts to find the lowest order corrections to GR\ produced 
 by any quantum gravity theory. This generates the action \eqref{fullaction} 
 but where the $F_i(\Box)$ are given by $a_i + b_i \log(\Box/\mu^2)$
 \cite{Calmet:2016sba,Jenkins:2018ysa}.

Here we will reproduce the calculation of \cite{Kuntz:2017pjd} but with more general 
$F_i(\Box)$, and further generalise to an (Anti) de Sitter background. We can 
focus on \eqref{fullaction} without the Riemann term using the Gauss Bonnet identity and 
a similar expression for the higher order terms \cite{Li:2015bqa}.

Upon averaging the equations of motion over spacetime, we find the following conditions
for short-wavelength quantities \cite{Stein:2010pn}:

\begin{itemize}
\item
The average of an odd product vanishes 

\item
The derivative of a tensor or scalar averages to zero

\item
One can therefore integrate by parts and exchange derivatives, e.g. \\
$\left< S_\beta \D_\mu R^\mu_\alpha\right> = - \langle R^\mu_\alpha \D_\mu S_\beta \rangle$

\end{itemize}

We will perturb around a de Sitter background $\bar{g}_{\mu\nu}$ with background curvature
$\bar{R}=12H^2$ where $H$ is the Hubble constant. If we write 
$\bar{g}_{\mu\nu} \to \bar{g}_{\mu\nu} + h_{\mu\nu}$, the linearised curvatures  
are given by \eqref{eq:linearisedcurvds}. In the transverse-traceless (TT) gauge 
$\partial^\nu h_{\mu\nu}=0$ and 
$h=0$, these become
\ba
        r^\mu_\nu &=& - \frac{1}{2}
        \Box h^\mu_\nu  - 3H^2 h^\mu_\nu,\non\\
        r &=&    -(\Box+3H^2)h +\D_\tau \D^\sigma h^\tau_\sigma.
\ea
To quadratic order in $h_{\mu\nu}$, we obtain
\ba
        r^{(2)}_{\mu\nu} &=& \frac{1}{4} \left(h^{\alpha\beta}\D_\mu \D_\nu 
        h_{\alpha\beta} - 2 h_{\alpha(\nu} (\Box-4H^2) h^\alpha_{\mu)}\right), \non\\
        r^{(2)} &=& -\frac{1}{4} h_{\mu\nu} \left(\Box-8H^2\right) h^{\mu\nu}. 
\ea

In the TT gauge, the linear equations of motion in a vacuum \eqref{eq:dsfieldeqns} become
\ba \label{eq:lineareomsinttgauge}
        (\Box-2H^2)^2 F_2(\Box) h^\mu_\nu = - \left(1+24M_P^{-2}H^2f_{1_0} \right) (\Box-2H^2)h^\mu_\nu,
\ea
and by using the above conditions for spacetime averages, we find that 
the second order equations of motion (for the extra IDG terms) are
\ba \label{eq:quadraticeomsinttgauge}
        \kappa t^\mu_\nu{}^{(2)\text{IDG}}= \frac{1}{2} \left< h^\mu_\sigma F_2(\Box) (\Box-2H^2)^2
        h^\sigma_\nu \right> - \frac{1}{8} \delta^\mu_\nu \left< h^\sigma_\tau F_2(\Box)
        (\Box-2H^2)^2h^\sigma_\tau\right>.~~
\ea
If we substitute \eqref{eq:lineareomsinttgauge} into \eqref{eq:quadraticeomsinttgauge},       
we find that 
\ba \label{eq:quadraticeomsinttgaugesubst}
        t^\mu_\nu{}^{(2)\text{IDG}}= - \left(M^2_P +24H^2 f_{1_0}\right)
        \left[\frac{1}{2} \left< h^\mu_\sigma (\Box-2H^2)
        h^\sigma_\nu \right> - \frac{1}{8} \delta^\mu_\nu \left< h^\sigma_\tau
        (\Box-2H^2)h^\sigma_\tau\right>\right].~~~~
\ea
\eqref{eq:quadraticeomsinttgaugesubst} is the additional backreaction produced by gravitational 
waves propagating through a de Sitter background due to IDG. 
The energy density is given by
\ba \label{eq:energydensity}
        \rho_{dS} = t_{00}^{(2)}= \left(M^2_P +24H^2 f_{1_0}\right) \left[
        \frac{1}{2} \left< h_{0\sigma} (\Box-2H^2)
        h^\sigma_0 \right> + \frac{1}{8}  \left< h^\sigma_\tau 
        (\Box-2H^2)h^\sigma_\tau\right>\right].~~~~~
\ea        
\subsection{Plane wave example}
We now look at the example of a plane wave. A plane wave propagating through
de Sitter space (to linear order in $H^2$) has the 
solution \cite{Nowakowski:2008de,Arraut:2012xr}
\ba \label{eq:planewave}
        h_{\mu\nu}=\epsilon_{\mu\nu} \cos(\omega t- kz) + \xi_{\mu\nu},
\ea
where $\epsilon_{\mu\nu}$ is the polarisation tensor and $\xi_{\mu\nu}$
is the de Sitter correction, given by 
\ba \label{eq:desittercorrection}
        \xi_{00} = -\Lambda t^2, \quad \xi_{0i} = \frac{2}{3} \Lambda t x_i,         \quad \xi_{ij} = \Lambda t^2 \delta_{ij} +\frac{1}{3}\Lambda \epsilon_{ij},
\ea
where  $\epsilon_{ij} = x_i x_j$ for $i\neq j$ and 0 otherwise. 
For the solution \eqref{eq:planewave},
the energy density \eqref{eq:energydensity} becomes\footnote{There is no correction from the terms in \eqref{eq:desittercorrection} because inserting them into \eqref{eq:planewave} generates terms which are linear in 
$\cos$ or $\sin$, which have a vanishing spacetime average.}
\ba
        \rho_{dS} =\frac{1}{4}\left(M^2_P +24H^2 f_{1_0}\right)
        \bigg\{\omega^2\epsilon^2
        +2\left( 4 \epsilon^\sigma_0
        \epsilon_{0\sigma}+\epsilon^2  \right)\left(8H^2+ \omega^2-k^2\right)\bigg\}.
\ea  

\subsection{Gravitational memory}
When gravitational waves pass through a medium, they cause a permanent displacement
of the medium which endures after the wave has passed through \cite{Favata:2010zu}.
The implications of IDG on the gravitational 
memory effect were studied in \cite{Kilicarslan:2018yxd,Kilicarslan:2018unm}, where it was found that 
the effect was suppressed for small distances, as one would perhaps expect, 
as we have seen in previous sections that IDG suppresses the strength of 
gravity at short distances.

\section{Power emitted on a Minkowski background}

In this section we can finally apply our calculations to produce a testable prediction.
In analogy with electromagnetic radiation, the power emitted by a system is given by the integral of the energy flow in the radial direction $t_{0\mu}$, i.e.
\cite{Carroll:2004st}
\ba \label{eq:defnpoweremitted}
                 P=\int_{S^2_\infty} t_{0\mu} n^\mu r^2 d\Omega,
\ea                 
which can be calculated using the backreaction we studied earlier. We calculate the integral \eqref{eq:defnpoweremitted} over the two-sphere at spatial infinity
$S^2_\infty$ and take $n^\mu$ to be the spacelike normal vector to the two-sphere,
which in polar coordinates is $n^\mu=(0,1,0,0)$. The form of $n^\mu$ means
that the only non-zero component of $t_{0\mu} n^\mu$ will be $t_{0r}$.

Returning to a Minkowski spacetime background and including the GR expression, 
the backreaction \eqref{eq:quadraticeomsinttgaugesubst} becomes 
\ba \label{eq:quadraticeomsinttgaugesubstmink}
        t_{\mu\nu}{}^{(2)\text{IDG}}= \frac{1}{64\pi G}
        \left[2\left< \partial_\mu h^{TT}_{\alpha\beta} \p_\nu h^{\alpha\beta}_{TT}\right>
        +4 \left< h^{TT}_{\sigma(\mu} \Box
        h^\sigma_{\nu)}{}^{TT} \right> - \eta_{\mu\nu} \left< h^{TT}_{\sigma\tau}
        \Box h^{\sigma\tau}_{TT}\right>\right].
\ea
We can discard the second and third terms when calculating the power because
$h^{TT}_{0\nu}=\eta_{0r}=0$. 
$h^{TT}_{ij}$\ is traceless, so
\ba \label{eq:modifiedquadtt}
        h^{TT}_{ij} = \bar{h}^{TT}_{ij} = - \frac{G}{r} \frac{M}{\sqrt{\pi}}
        \frac{d^2}{dt^2} \int^\infty_{-\infty}dt'_r e^{-M^2(t_r-t'_r)^2} I_{ij}^{TT}(t'_r),
\ea
where we recall that we defined the retarded time $t_r=t-r$. To write \eqref{eq:modifiedquadtt} more succinctly, we define $\hat{I}_{ij}$ as the integral in \eqref{eq:modifiedquadtt}, i.e. 
\ba \label{eq:definitionofhatI}
        \hat{I}_{ij}(t_r)=\int^\infty_{-\infty}dt'_r e^{-M^2(t_r-t'_r)^2} I_{ij}^{TT}(t'_r).
\ea       
In the limit $r\to \infty$, the first derivatives of $\hat{I}_{ij}(t_r)$ are
\ba
        \partial_0 h_{ij}^{TT}=-\partial_r h_{ij}^{TT}
        = -\frac{ G}{r}\frac{M}{\sqrt{\pi}}\frac{d^3}{dt^3}\hat{I}_{ij}(t_r),
\ea
where the limit means that we can ignore the derivative of the $\frac{1}{r}$ factor.

Therefore the necessary expression for calculating the power given a certain quadrupole moment is
\ba \label{eq:necessaryexpressionpower}
        t_{0\mu}n^\mu = -\frac{ GM^2}{32\pi^2 r^2}\left<\frac{d^3}{dt^3}
        \left(\hat{I}_{ij}(t_r)\right)
        \frac{d^3}{dt^3}\left(\hat{I}^{ij}(t_r)\right)\right>.
\ea     
\eqref{eq:necessaryexpressionpower} is precisely the GR result with $I_{ij}$ 
replaced with $\hat{I}_{ij}$, which was defined in \eqref{eq:definitionofhatI}.

\subsubsection{Reduced quadrupole moment}
For more complicated systems, 
it is easier to use the reduced quadrupole moment 
$J_{ij} = I_{ij} - \delta_{ij} \delta^{kl}I_{kl}$ \cite{Carroll:2004st}. 
In terms of the reduced quadrupole moment, \eqref{eq:necessaryexpressionpower}
is
\ba
        t_{0\mu}n^\mu =\frac{- GM^2}{32\pi^2 r^2}\bigg<\frac{d^3\hat{J}_{ij}}{dt^3}
        \frac{d^3\hat{J}^{ij}}{dt^3}- 2\frac{d^3\hat{J}_i{}^j}{dt^3}\frac{d^3\hat{J}^{ik}}{dt^3}
        n_j n_k + \frac{1}{2}\frac{d^3\hat{J}^{ij}}{dt^3}
        \frac{d^3\hat{J}^{kl}}{dt^3}n_i n_j n_k n_l\bigg>,~~~~~
\ea      
where
\ba \label{eq:definitionofJhat}
         \hat{J}_{ij}= \int^\infty_{-\infty} dt'_r  
         e^{-M^2(t_r-t'_r)^2} J_{ij} (t'_r).
\ea
We can then use the identities \cite{Carroll:2004st}
\ba
        \int \hspace{-1mm} d\Omega = 4\pi, \quad 
        \int \hspace{-1mm} n_i n_j d\Omega =\frac{4\pi}{3} \delta_{ij}, \quad
        \int \hspace{-1mm} n_i n_j n_k n_l d\Omega =\frac{4\pi}{15} \left( \delta_{ij} \delta_{kl} 
        + \delta_{ik}\delta_{jl} + \delta_{il} \delta_{jk}\right),~~~~
\ea
together with \eqref{eq:defnpoweremitted} to give the formula for the power emitted by a system with
reduced quadrupole moment $J_{ij}$
\ba \label{eq:generalradiationpowerformula}
        P = - \frac{G}{5} \left<\frac{d^3\hat{J}_{ij}}{dt^3}
        \frac{d^3\hat{J}^{ij}}{dt^3}\right>.
\ea
Using the formula \eqref{eq:generalradiationpowerformula}, 
we can predict the power emitted through gravitational radiation by a system once we know its quadrupole moment. 
We will apply this formula to binary systems with both circular and elliptical 
orbits.

\subsubsection{Modification to orbits}
Before looking at the modification to the power produced by a system, we must first see whether IDG will have a significant effect on the orbital motion of the system, and thus the quadrupole moment. The  orbital frequency $\omega$ in an inspiraling binary system
is (see \cite{Calmet:2018qwg})
\ba
        \omega^2 = - \frac{\Phi}{r^2} \propto \frac{\text{Erf}\left(\frac{Mr}{2}\right)}{r^3}.
\ea         
We have already shown $Mr \gg 1$ for $r>10^{-4}$\hspace{0.5mm}m, i.e. distances large enough for 
the inspiral to be a good approximation, and thus we can safely use the Newtonian prediction for the
orbital motion of the system.

\subsection{Circular orbits}
The reduced  quadrupole 
moment $J_{ij}$ for a binary system with a circular orbit
in polar coordinates is given by  \cite{Carroll:2004st}
\ba \label{eq:quadforbinarysystem}
        J_{ij}=\frac{M_s R^2}{3}
        \begin{pmatrix}
   \left(1+3\cos(2\omega t)\right) & 3\sin(2\omega t) &0\\
   3\sin (2\omega t) &\left(1-3\cos(2\omega t)\right) & 0\\
   0 & 0 & -2
  \end{pmatrix},~~~~~~~
\ea
where $R$\ is the distance between the stars of mass $M_s$ and 
$\omega$ is the angular velocity.
Inserting \eqref{eq:quadforbinarysystem} into our formula \eqref{eq:generalradiationpowerformula}, the power is 
\ba
        P &=& - \frac{G}{5} \left<(g^{rr})^2\left(\frac{d^3\hat{J}_{rr}}{dt^3}\right)^2
        + (g^{\theta\theta})^2\left(\frac{d^3\hat{J}_{\theta\theta}}{dt^3}\right)
        + (g^{\phi\phi})^2\left(\frac{d^3\hat{J}_{\phi\phi}}{dt^3}\right)\right>\non\\
        &\approx&  - \frac{GR^2M_s^4}{45} \left<\left(\frac{d^3}{dt^3}\int  dt'_r  
         \frac{\left(1+3\cos(2\omega t'_r)\right)}{e^{M^2(t_r-t'_r)^2}}\right)^2 \right>,~~~~~~
\ea        
where we again use the limit $r\to \infty$ to discard the second and third terms. 

Evaluating the integral 
using $\left<\sin ^2(x)\right>\equiv\frac{1}{2}$ gives
\ba \label{eq:powercircrad}
       P&=& - \frac{128}{5}  GR^2M_s^4 \omega^6 e^{-2 \omega ^2/M^2}. 
\ea  
\eqref{eq:powercircrad} is simply the GR result with 
an extra factor $e^{-2 \omega ^2/M^2}$.
As $M$ decreases, i.e. as we move further away from GR, the amount of radiation emitted from a binary system of stars
in a circular orbit decreases exponentially.

\subsection{Generalisation to elliptical orbits}
One might naively think that the power emitted by a binary system
in an elliptical orbit could be well approximated by 
a system in a circular orbit. However, the power strongly 
depends on the eccentricity $e$ of the orbit.
This dependency can be expressed as $P_{\text{GR}}= P^{\text{circ}}_{\text{GR}} f^{\text{GR}}(e)$
where $f^{\text{GR}}(e)$ is the \textit{enhancement factor}
\ba
        f^{\text{GR}}(e)= \frac{1+\frac{72 e^2}{24}+\frac{37 e^4}{96}}
        {\left(1-e^2\right)^{7/2}},
\ea         which reaches $10^3$ at $e=0.9$~\cite{Peters:1963ux}. 

We will therefore calculate the effect of elliptical orbits 
on the gravitational radiation produced in IDG.
The necessary components of the reduced quadrupole moment for a binary 
system in an elliptical orbit are \cite{Peters:1963ux}
\ba
        J_{xx}&=& \mu d^2 \left(\cos^2(\psi) - \frac{1}{3}\right),    \non\\
        J_{yy}&=& \mu d^2 \left(\sin^2(\psi) - \frac{1}{3}\right),
\ea        
where $\mu$ is the reduced mass $m_1 m_2/(m_1 +m_2)$ and
the distance between the two stars is given by
\ba
        d= \frac{a(1-e^2)}{1+e\cos(\omega t)},
\ea 
where $a$ is the semimajor axis and $e$ is the eccentricity of the orbit~\cite{Peters:1963ux}.

The change in angular position over time is       
\ba
        \dot{\psi}&=&\frac{\left[G(m_1+m_2)a(1-e^2)\right]^{1/2}}{d^2}.
\ea 
To find the $xx$ component of \eqref{eq:definitionofJhat}, we require
\ba \label{eq:xxcomponentfirstint}
        \hat{J}_{xx}=\mu a^2(1-e^2)^2 \int^\infty_{-\infty} dt'_r e^{-M^2 (t_r-t'_r)^2} 
        \frac{\cos^2(\psi( t'_r))-\frac{1}{3}}
        {\left(1+e\cos(\psi (t'_r))\right)^2}.~~~~~~
\ea  
\subsubsection{Evaluating our integral}
At first glance, \eqref{eq:xxcomponentfirstint} might seem an impossible integration. 
However, we can perform the coordinate change $z=M(t_r-t'_r)$, 
which allows us to generate a Taylor expansion in $\frac{1}{M}$, 
i.e. the GR value plus a small correction (recall that $M\to \infty$ is the GR limit). 

We define $y=\frac{1}{M}$, and the function we want to expand as $f(y)$
\ba
        f\left(y\right)=\frac{\cos^2(\psi( t_r-zy))-\frac{1}{3}}{\left(1+e\cos(\psi
(t_r-zy))\right)^2},
\ea
such that 
\ba
        \hat{J}_{xx} = - \mu a^2(1-e^2)^2 \int^\infty_{-\infty} dz e^{-z^2} \left[f(0)+f'(0) y + f''(0)y^2 + \cdots \right].
\ea
Each derivative of $f(y)$ will give us an extra power of $z$,  and we can see from the
identities 
\ba
        &&\int^{-\infty}_\infty e^{-z^2}dz=-\sqrt{\pi},\non\\
        && \int^{-\infty}_\infty e^{-z^2}z~ dz=0, \non\\
        &&\int^{-\infty}_\infty e^{-z^2}z^2~ dz=-\frac{\sqrt{\pi}}{2},
\ea
that we must go to second order in $\frac{1}{M}$ to calculate the correction term. We find
\ba
        \frac{df(y)}{dy}= \frac{2 z \psi'(t-y z) \sin (\psi(t-y z)) (a+3 \cos (\psi(t-y z)))}{3 (a \cos (\psi(t-y z))+1)^3},
\ea
and
\ba
        \frac{d^2f(y)}{dy^2}&=& \frac{z^2}{6 (1+e \cos (\psi(t-y z)))^4}     \bigg\{ \psi'(t-y z)^2 \bigg(4 \left(e^2-3\right) \cos (2 \psi(t-y z))\non\\
        &&-8 e^2-19 e \cos (\psi(t-y z))+3 e \cos (3 \psi(t-y z))\bigg)\non\\
        && -2 \psi''(t-y z) \sin (\psi(t-y z)) \bigg[2 \left(e^2+3\right) \cos (\psi(t-y z))\non\\
        &&+e (3 \cos (2 \psi(t-y z))+5)\bigg]\bigg\}.~~~~~~
\ea

We can then write
\ba
       \hat{J}_{xx}&=& -\sqrt{\pi} \mu 
        \frac{\cos^2(\psi)-\frac{1}{3}}{\left(1+e\cos(\psi)\right)^2}\non\\
        &&+ \frac{\sqrt{\pi}\mu}{12 (1+e \cos (\psi))}  \frac{ G(m_1+m_2)}{\left(a(1-e^2)\right)^{3}}   \bigg\{  \left(4 e^3+19e\right) \cos (\psi)+4 \left(2 e^2+3\right) \cos (2 \psi)\non\\
        &&+\frac{3}{2} (e^2 (\cos (4 \psi)+7e)-6 \cos (3 \psi))\bigg\}.
\ea    
Therefore the third derivative of the extra term is given by
\ba
        \frac{d^3\hat{J}^{IDG}_{xx}}{dt^3}&=& \frac{6 \left(4 a^2+19\right) \psi '(t) \psi ''(t)}{1+e \cos (\psi (t))}-\frac{12 \left(4 e^2+19\right) \psi '(t) \left(\left(e^2-1\right) \psi ''(t)+e \psi '(t)^2 \sin (\psi (t))\right)}{(1+e \cos (\psi (t)))^3}\non\\
        &&+\frac{2 \left(4 e^2+19\right) \left(e \left(\psi '(t)^3-\psi ^{(3)}(t)\right) \sin (\psi (t))-9 \psi '(t) \psi ''(t)\right)}{(1+e \cos (\psi (t)))^2}\non\\
        &&-\frac{12 e \left(4 e^4+15 e^2-19\right) \psi '(t)^3 \sin (\psi (t))}{(1+e \cos (\psi (t)))^4}\non\\
        &&+9 e \left(\psi ^{(3)}(t)-9 \psi '(t)^3\right) \sin (3 \psi (t))+13 e \left(\psi ^{(3)}(t)-\psi '(t)^3\right) \sin (\psi (t))\non\\
        &&+39 e \psi '(t) \psi ''(t) \cos (\psi (t))+81 e \psi '(t) \psi ''(t) \cos (3 \psi (t))\non\\
        &&+24 \left(\psi ^{(3)}(t)-4 \psi '(t)^3\right) \sin (2 \psi (t))+144 \psi '(t) \psi ''(t) \cos (2 \psi (t)).
\ea
By calculating a similar 
expression for the $yy$ component and comparing with the expression for a circular orbit, we find that we can write 
\ba
        P\approx P_{\text{GR}} + P_{\text{IDG}}=P_{\text{GR}}^{\text{circ}} f^{\text{GR}}(e)+
        P_{\text{IDG}}^{\text{circ}} f^{\text{IDG}}(e),~~~
\ea
where $P_{\text{IDG}}^{\text{circ}}$ is given by\footnote{This is the lowest order correction
to GR of \eqref{eq:powercircrad}.} $\frac{256}{5}  
\frac{\omega ^8}{M^2}GR^2M_s^4 $ and the enhancement factor for the IDG term
is (up to order  $e^{20}$)
\ba \label{eq:fullidgenhancementfactor}
        &&f^{\text{IDG}}(e)
        =1-\frac{(9299+111168 \pi ) e}{12282+155520 \pi }
        +\frac{(753298+4783383 \pi ) e^2}{18423+233280 \pi }\non\\
        &&   +\frac{(1347719-15413436 \pi ) e^3}{147384+1866240 \pi }
        +\frac{(152362163+521885160 \pi ) e^4}{294768+3732480 \pi }\non\\
        &&   -\frac{(6051611+36789444 \pi ) e^5}{72 (2047+25920 \pi )}
        +\frac{(666697961+1567922058 \pi ) e^6}{294768+3732480 \pi }\non\\
        && 
        -\frac{15 (1908618+1108133 \pi ) e^7}{32752+414720 \pi }
         +\frac{(344524449+556982911 \pi ) e^8}{65504+829440 \pi }\non\\
        &&
        -\frac{(5826870871+2360357712 \pi ) e^9}{1152 (2047+25920 \pi )}
        +\frac{(37373085170+45561968109 \pi ) e^{10}}{4716288+59719680 \pi}\non\\
        && -\frac{(45892881151+15257013132 \pi ) e^{11}}{6144 (2047+25920 \pi )}
         +\frac{(685593299971+742716547416 \pi ) e^{12}}{36864 (2047+25920 \pi )} \non\\
        &&-\frac{(18923346001+5812048566 \pi ) e^{13}}{2304 (2047+25920 \pi )} +\frac{(1406663203279+1486964224080 \pi ) e^{14}}{73728 (2047+25920 \pi )}\non\\
        &&
        -\frac{(612225325649+186007875390 \pi ) e^{15}}{73728 (2047+25920 \pi )}+\frac{(1879563787501+1982636168004 \pi ) e^{16}}{98304 (2047+25920 \pi )}\non\\
        &&
        -\frac{5 (108886731499+33068066736 \pi ) e^{17}}{65536 (2047+25920 \pi )}+\frac{(7518767717389+7930544672016 \pi ) e^{18}}{393216 (2047+25920 \pi )}\non\\
        &&
        -\frac{(9799832804557+2976126006240 \pi ) e^{19}}{1179648 (2047+25920 \pi )}
        +\frac{(15037546015045+15861089344032 \pi ) e^{20}}{786432 (2047+25920 \pi )}\non\\
        &&+ O(e^{21}).
\ea

\begin{figure}[tbp]\hspace{-6mm}\includegraphics[width=130mm]{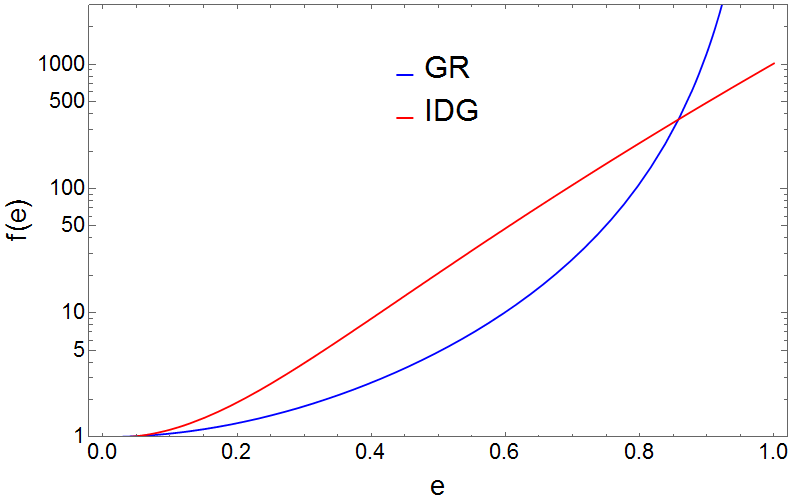}
\centering
\caption{The enhancement factor for IDG, $f^{\text{IDG}}(e)$ given by 
\eqref{eq:fullidgenhancementfactor}
as well as the enhancement factor for the GR term $f^{\text{GR}}(e)$ versus the eccentricity $e$, where the 
power emitted is $P_{\text{GR}}^{\text{circ}} f^{\text{GR}}(e)+
        P_{\text{IDG}}^{\text{circ}} f^{\text{IDG}}(e)$ \cite{Edholm:2018qkc}. This enhancement
         factor describes
how the power emitted changes with respect to the eccentricity.}
\label{enhancementfactorplot}
\end{figure}
\newpage

\begin{figure}[!htbp]
\centering
\includegraphics[width=130mm]{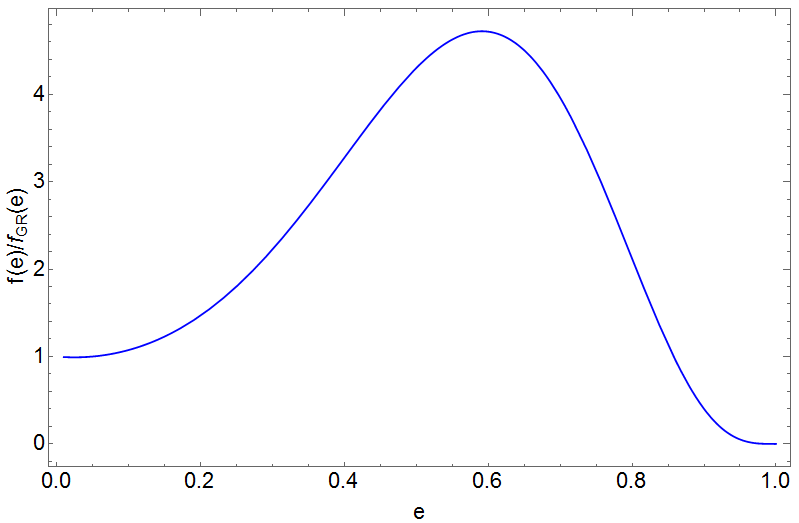}
\caption{The ratio between the IDG enhancement factor and the GR enhancement factor \cite{Edholm:2018qkc}. The extra IDG term has the strongest effect at around $e=0.7$, 
which coincidentally is roughly the value for the Hulse-Taylor binary. Note that this
enhancement factor is independent of $M$.}
\label{ratioenhancement}
\end{figure}

\subsection{Hulse-Taylor binary}
We now examine the case of the  Hulse-Taylor binary. This is a binary system of a pulsar and 
another neutron star, which allows us to observe the change in the orbital period 
due to gravitational radiation very accurately \cite{Hulse:1974eb}. The power emitted from the 
binary, which has an ellipticity of 0.617
and a period of 7.5 hours, is $0.998\pm 0.002$ of the GR prediction \cite{Weisberg:2016jye}. 
We find the constraint on our mass scale $M$ to be
\vspace{-1mm}
\ba \label{eq:massconstraintradiation}
        M>2.1 \times 10^{-40} M_P= 3.1 \times 10^{-13} \text{eV.}  
\ea    
    
This constraint is much lower than that from lab experiments.
The previous lower bound\footnote{Assuming IDG generates
inflation, one obtains a lower bound of $\sim 10^{14}$
GeV using Cosmic Microwave Background 
data as in Section \eqref{eq:constrtaintscmbs}.}
is $\sim$0.01 eV from lab-based experiments.
To improve this constraint using observations of gravitational radiation, 
one would have to probe systems with orbital frequency\footnote{The frequency
of the radiation produced is twice the orbital frequency of 
the system \cite{Abbott:2016bqf}.} of $10^{-4}$ seconds.
Unfortunately, these frequencies are higher than
the regions that the LIGO detector and LISA (a future space-based GW detector) are most sensitive to
(15-150 Hz \cite{Abbott:2016bqf} and 
$10^{-4}$-$10^{-1}$\hspace{0.2mm}Hz \cite{Audley:2017drz} 
respectively). A system with such a high orbital frequency would also
be out of the weak field limit we took to simplify our 
calculations.
Therefore it seems that the tightest constraints on our mass scale will 
come from lab-based experiments and CMB data.

\chapter{Conclusion}
We do not yet have a singularity-free description of gravity.
Infinite Derivative Gravity provides a way out of this problem without resorting
to an unknown theory of quantum gravity. We have shown that we recover the predictions of GR
when looking at length scales above the IDG mass scales, while resolving 
the issues of GR at short distances.

\subsubsection{Outline of results}
We reviewed the existing concepts of IDG in Chapter \ref{chap:idg}, so that we had a solid base of knowledge to build on in the following chapters. 
We looked at both the general equations of motion and the linearised equations of motion around various backgrounds. Finally we examined the 
second variation of the action around constantly curved backgrounds
and looked at the propagator around a Minkowski background.

The impact of IDG on the Newtonian potential, the potential caused by a point source, was investigated in Chapter \ref{chapter:potential}, where we found that in contrast to GR, a non-singular potential is generated. The potential returns to GR at large distances, as IDG only has an effect at small distances. Using experimental data, we can constrain our mass scale to be at least the meV scale. We found that the potential caused by a point source in a de Sitter background could be approximated by a flat background.

GR generates singularities, where our description of spacetime breaks down, due to the Hawking-Penrose singularity theorems. We show how IDG can avoid this fate in Chapter \ref{sec:defocusing}, by examining the conditions necessary to be able to trace null rays back to past infinity. This was carried out for perturbations around a flat background and constantly expanding backgrounds. We also examined background FRW metrics which could explain the Big Bang singularity problem with non-singular bouncing metrics.

Primordial inflation is an accelerated expansion in the early universe,  used to solve the horizon problem and flatness problem. IDG contains within it Starobinsky inflation, which
is one of the most successful theories of inflation. We examine the effects of the extra terms compared to Starobinsky inflation in Chapter \ref{chapter:inflation}.
The modification does not affect scalar perturbations in the Cosmic Microwave Background, but does modify the tensor perturbations.
Therefore the tensor-scalar ratio, a key observable for inflationary theories, can be differentiated from Starobinsky inflation. When the mass scale falls below around $10^{14}$ GeV, the tensor-scalar ratio is exponentially enhanced and leaves the allowed region, allowing us to constrain the mass scale.   
 
Finally we saw the effect of IDG on gravitational radiation in Chapter \ref{chapter:radiation}.
In particular, IDG modifies the quadrupole formula, backreaction of gravitational waves and the power emitted through gravitational waves. We looked at binary systems with both circular and eccentric orbits to show the difference caused by in the power produced by gravitational radiation when comparing IDG to GR. Taking the example of the Hulse-Taylor binary, we found that IDG would match the GR predictions to a very high level of precision.    

We gained different bounds on the mass scale of IDG. We found a relatively weak bound from laboratory experiments, but assuming that IDG was
responsible for cosmological inflation produced a much stronger constraint.
Finally looking at the Hulse-Taylor binary gave us a very weak bound because
IDG\ almost exactly matched the GR\ predictions. However, future investigations
of the effect of IDG on black hole binaries, where the masses involved are much greater and the distances much shorter, could give stronger 
constraints as the system would be in the strong field regime.

\newpage

There are several possible directions for future work on IDG. 

\vspace{5mm}
\hspace{-5mm}
\textit{Black hole solution}

We know that IDG can resolve the singularity generated by a small test mass
in GR at the linearised level. The most important single step we could make would be to find a  modified Schwarzschild black hole solution.
At the linearised level, IDG\ weakens gravity at short distances, so one would expect that the same effect would occur with the non-linear equations of motion.

If this solution is found to give a non-singular metric potential, it can be checked whether the singularity generated by the Hawking-Penrose singularity theorems is still produced.

\vspace{5mm}
\hspace{-5mm}\textit{Gravitational waves}

In Chapter \ref{chapter:radiation}, we used the de Donder gauge to find the modified quadrupole
formula, i.e. the same gauge as for the GR calculation. 
It is also possible to modify the gauge itself, 
following the method of \cite{Berry:2011pb,Naf:2011za}. One can define  
\ba
        \gamma_{\mu\nu} = a(\Box) h_{\mu\nu} - \frac{1}{2}\eta_{\mu\nu}  
 c(\Box) h -\frac{1}{2}\eta_{\mu\nu} f(\Box) \partial_\alpha 
 \partial_\beta h^{\alpha\beta}, 
\ea
and choose the gauge $\partial^\mu \gamma_{\mu\nu}=0$.
This produces the equation of motion $-2\kappa T_{\mu\nu}=\Box \gamma_{\mu\nu}$
 \cite{Edholm:2018qkc}. It would be interesting to reproduce the calculations of 
\cite{Naf:2011za} for IDG.

\vspace{5mm}
\hspace{-5mm}\textit{Inflation}

For the choice $a\neq c$, IDG produces the result that the potentials $\Psi$
and $\Phi$ are not equal to each other at short distances. While 
the Cassini experiment showed that $\Phi$ is almost exactly equal to $\Psi$ \cite{Will:2005va},
this is only valid at large distances and significant differences could still exist 
at smaller length scales. 

One important implication of a difference between
$\Psi$ and $\Phi$ at short distances would be the effect on
inflationary perturbations.

\newpage

Whilst it is still in its infancy, we have shown that Infinite Derivative Gravity 
is a viable alternative to Einstein's theory of General Relativity.
The nature of IDG means that many calculations are all but impossible without 
taking restrictive choices of our functions, but by taking the simplest 
and most natural choices, we are able to produce predictions.

The most important feature of IDG is that it can ameliorate the 
behaviour of gravity at short distances without introducing ghosts.
If we want to avoid singularities, we know that at some stage, 
GR must be an approximation to some underlying theory of gravity.
Many would suggest that we should not
investigate classical solutions to the singularity problem
because we would still need a full theory of quantum 
gravity to describe gravity below the Planck length.
However, our experiments are still decades away from testing 
gravity anywhere near these length scales - 
there are still thirty orders of magnitude between our best laboratory 
tests of gravity and the Planck length.

If future experiments should produce divergences from GR in the vast gulf 
of opportunity before we reach quantum gravity scales, then
it is imperative that we have alternative theories of gravity that could 
provide an explanation for such phenomena. We would need to
discover whether those divergences were indeed a classically modified theory
of gravity, or some effective field theory for an unknown quantum completion
of gravity such as string theory or loop quantum gravity.

Indeed, it is very possible that IDG is in fact the effective field theory
description for quantum gravity itself - the Effective Quantum Gravity 
framework mentioned in Sec.~\ref{sec:backreaction} also predicts 
a similar action with d'Alembertian operators acting on quadratic curvature terms.

While there is still a huge amount of work to be done, 
it is clear that IDG can provide the basis for an 
overarching non-singular description of gravity, and a possible explanation
for any differences from GR discovered by future experiments.


\begin{appendices}

\chapter{Appendices}
\section{Notation and identities}
\subsection{Notation}
We take a metric with signature ($-$,+,+,+). We use Greek indices (e.g. $\mu$, $\nu$) to denote four-dimensional coordinate labels, 
where we take 0 to be the time index and 1, 2, 3 to be the spatial indices, and Latin indices (e.g. $i$, $j$, $k$) to denote three-dimensional
spatial coordinate labels. 

We work in ``God-given" units, so set both $\hbar$, 
the reduced Planck constant, and the speed of light $c$ equal to 1.
We use the Levi-Civita connection, so that we will only look at theories which are metric-compatible and torsion free.

We define symmetric and antisymmetric tensors with an additional factor of $\frac{1}{2}$, i.e.
\ba
        S_{(\mu\nu)} = \frac{1}{2} \left(S_{\mu\nu} + S_{\nu\mu}\right), \quad \quad 
        A_{[\mu\nu]} = \frac{1}{2} \left(A_{\mu\nu} - A_{\nu\mu}\right).
\ea
\subsection{Curvature tensors}
The Christoffel connection is given by
\ba
        \Gamma^\lambda_{\mu\nu}=\frac{1}{2}g^{\lambda\tau} \left(\p_\mu g_{\nu\tau}
        +\p_\nu g_{\mu\tau} - \p_\tau g_{\mu\nu}\right).
\ea        
The Riemann tensor is given by 
\ba
        R^\lambda{}_{\mu\sigma\nu} = \p_\sigma \Gamma^\lambda_{\mu\nu} - \p_\nu \Gamma^\lambda_{\mu\sigma}
        +\Gamma^\lambda_{\sigma\rho} \Gamma^\rho_{\nu\mu}-\Gamma^\lambda_{\nu\rho}
        \Gamma^\rho_{\lambda\mu} , 
\ea        
and is antisymmetric in the following indices
\ba \label{eq:riemannantisymm}
        R_{(\mu\nu)\rho\sigma} = R_{\mu\nu(\rho\sigma)}=0.
\ea        

The Ricci tensor (which is symmetric) is the contraction of the rst and third
indices of the Riemann tensor:
\ba
        R_{\mu\nu} = R^\lambda{}_{\mu\lambda\nu} =\p_\lambda \Gamma^\lambda_{\mu\nu}
        -\p_\nu \Gamma^\lambda_{\mu\lambda} +
        \Gamma^\lambda_{\lambda\rho} \Gamma^\rho_{\nu\mu}
        - \Gamma^\lambda_{\nu\rho} \Gamma^\rho_{\lambda\mu}.
\ea
The Ricci scalar is
\ba
         R =g^{\mu\nu} R_{\mu\nu} = g^{\mu\nu} \p_\lambda \Gamma^\lambda_{\mu\nu}
         -\p^\mu \Gamma^\lambda_{\mu\lambda} + g^{\mu\nu} \Gamma^\lambda_{\lambda\rho}
         \Gamma^\rho_{\nu\mu}- g^{\mu\nu} \Gamma^\lambda_{\nu\rho}
         \Gamma^\rho_{\lambda\mu}.         
\ea
The Weyl tensor is
\ba
        C^\lambda{}_{\alpha\nu\beta} &\equiv& R^\mu{}_{\alpha\nu\beta} - \frac{1}{2}
        \left(\delta^\mu_\nu R_{\alpha\beta} -\delta^\mu_\beta R_{\alpha\nu}
        +R^\mu_\nu g_{\alpha\beta} - R^\mu_\beta g_{\alpha\nu}\right)
        +\frac{R}{6} \left( \delta^\mu_\nu g_{\alpha\beta}
        -\delta^\mu_\beta \right), \non\\
        &&\text{ where } C^\lambda{}_{\alpha\lambda\beta} =0.
\ea        
The general formula for the commutator of covariant derivatives is
given by 
\ba
        [\D_\rho, \D_\sigma] X^{\mu_1....\mu_k}{}_{\nu_1 ... \nu_l}
        &=& R^{\mu_1}{}_{\lambda\rho\sigma} X^{\lambda \mu_2 ...\mu_k}
        {}_{\nu_1 ... \nu_l} + R^{\mu_2}{}_{\lambda\rho\sigma} 
        X^{\mu_1 \lambda \mu_3 ... \mu_k}{}_{\nu_1 ... \nu_l} + \cdots\non\\
        && - R^\lambda{}_{\nu_1 \rho\sigma} X^{\mu_1...\mu_k}{}_{\lambda...\nu_l}
        - R^\lambda{}_{\nu_2\rho\sigma} X^{\mu_1... \mu_k} 
        {}_{\nu_1\lambda\nu_3 ... \nu_l} - \cdots. ~~~~~~~~
\ea        
The first Bianchi identity is
\ba
        R_{\mu\nu\lambda\sigma} + R_{\mu\lambda\sigma\nu} + R_{\mu\sigma\nu\lambda}=0.        
\ea        
The second Bianchi identity is
\ba \label{eq:secondbianchi}
        \D_\kappa R_{\mu\nu\lambda\sigma} + \D_\sigma R_{\mu\nu\kappa\lambda}
        + \D_\lambda R_{\mu\nu\sigma\kappa} = 0.
\ea      
\subsection{Variation of the curvatures}
\label{sec:variationcurvature}
From the definitions of the Riemann and Ricci tensor,
\ba
        \delta R^\lambda{}_{\mu\sigma\nu} &=&
        \D_\sigma \left(\delta \Gamma^\lambda_{\mu\nu}\right)
        - \D_\nu\left(\delta \Gamma^\lambda_{\mu\sigma}\right),\non\\
        \delta R_{\mu\nu} &=& \D_\lambda \delta \Gamma^\lambda_{\mu\nu}
        - \D_\nu \delta \Gamma^\lambda_{\mu\lambda}, \non\\
        \delta \Gamma^\lambda_{\mu\nu} &=& \frac{1}{2} 
        \left(\D_\mu h^\lambda_\nu + \D_\nu h^\lambda_\mu
        - \D^\lambda h_{\mu\nu}\right).
\ea         
Inserting the varied Christoffel symbols gives
\ba
        \delta R^\lambda{}_{\mu\sigma\nu} &=& 
        \frac{1}{2}\left( \D_\sigma \D_\mu h^\lambda_\nu 
        - \D_\sigma \D^\lambda h_{\mu\nu} - \D_\nu \D_\mu h^\lambda_\sigma
        + \D_\nu \D^\lambda h_{\mu\sigma} \right), \non\\
        \delta R_{\mu\nu} &=& \frac{1}{2} \left( \D_\lambda \D_\mu
        h^\lambda_\nu + \D_\nu \D^\lambda h_{\mu\lambda}
        -\Box h_{\mu\nu} - \D_\mu \D_\nu h\right).
\ea         
The variation of the scalar curvature is 
\ba
        \delta R &=& \delta (g^{\mu\nu} R_{\mu\nu})\non\\
        &=& \delta g^{\mu\nu} R_{\mu\nu} +g^{\mu\nu} \delta R_{\mu\nu}\non\\
        &=& - h_{\alpha\beta} R^{\alpha\beta} + \D^\alpha \D^\beta
        h_{\alpha\beta} - g^{\alpha\beta} \Box h_{\alpha\beta},
\ea
using the definitions
\ba
        h_{\mu\nu} = - h^{\alpha\beta} g_{\alpha\mu} g_{\beta\nu}, 
        \quad \quad h = g^{\mu\nu} h_{\mu\nu},
        \quad \quad h_{\mu\nu} = \delta g_{\mu\nu}.
\ea         
\newpage
\subsection{Variation of the d'Alembertian acting on the curvatures}
\label{sec:variationdalembertianactingcurvature}
Using the definition of the d'Alembertian
\ba
        \Box = g^{\mu\nu} \D_\mu \D_\nu, 
\ea
the variation of the d'Alembertian acting on the Ricci scalar     
is
\ba
        \delta(\Box)R &=& \delta g^{\mu\nu} \D_\mu \D_\nu R 
        + g^{\mu\nu} \delta(\D_\mu) \D_\nu R 
        + g^{\mu\nu} \D_\mu \delta(\D_\nu) R \non\\
        &=& - h_{\alpha\beta}  \D^\alpha \D^\beta R 
        + \frac{1}{2} g^{\alpha\beta} (\D_\lambda R) \D^\lambda h_{\alpha\beta}
        - (\D^\alpha R) \D^\beta h_{\alpha\beta},
\ea
The variation of the d'Alembertian acting on the Ricci tensor
is        
\ba                
        \delta(\Box) R_{\mu\nu} &=& - h_{\alpha\beta} \D^\alpha \D^\beta R_{\mu\nu}
        - (\D^\beta h_{\alpha\beta} ) \D^\alpha R_{\mu\nu} +\frac{1}{2}
        g^{\alpha\beta} (\D^\sigma h_{\alpha\beta}) \D^\sigma
        \D_\sigma R_{\mu\nu} \non\\
        && -\frac{1}{2} \left[ \delta^\beta_{(\mu} R^\alpha_{\nu)}
        \Box h_{\alpha\beta} - \delta^\beta_{(\mu} R_{\tau\nu)}  
        \D^\tau \D^\alpha h_{\alpha\beta} + R^\alpha_{(\nu} \D^\beta \D_\mu
        h_{\alpha\beta}\right] \non\\
        && - (\D^\beta R^\alpha_{(\nu}) \D_{\mu)} h_{\alpha\beta}
        -\delta^\beta_{(\mu} (\D^\lambda R^\alpha_{\nu)} 
        \D_\lambda h_{\alpha\beta} + \delta^\beta_{(\mu} 
        (\D^\alpha R_{\tau\nu)}) \D^\tau h_{\alpha\beta}. ~~~~~~~~~
\ea        
The variation of the d'Alembertian acting on the Riemann tensor
is        
\ba
        &&\delta (\Box) R_{\mu\nu\lambda\sigma}
        \hspace{-0.5mm}= \hspace{-0.5mm} - h_{\alpha\beta} \D^\alpha \D^\beta R_{\mu\nu\lambda\sigma}
        - (\D^\beta h_{\alpha\beta}) \D^\alpha R_{\mu\nu\lambda\sigma}
        + \frac{1}{2} (\D^\tau h) \D_\tau R_{\mu\nu\lambda\sigma} \\\non
        && -\frac{1}{2}g^{\alpha\tau} \left[  (\D^\beta \D_\mu h_{\alpha\beta})
        R_{\tau\nu\lambda\sigma}  \hspace{-0.5mm}+ \hspace{-0.5mm} (\D^\beta \D_\nu h_{\alpha\beta})
        R_{\mu\tau\lambda\sigma}  \hspace{-0.5mm}+  \hspace{-0.5mm}(\D^\beta \D_\lambda  h_{\alpha\beta})
        R_{\mu\nu\tau\sigma}  \hspace{-0.5mm}+  \hspace{-0.5mm}(\D^\beta \D_\sigma  h_{\alpha\beta})
        R_{\mu\nu\lambda\tau}\right] \\\non
        && - g^{\alpha\tau} \left[ (\D_\mu h_{\alpha\beta})
        \D^\beta R_{\tau\nu\lambda\sigma} \hspace{-0.5mm}+\hspace{-0.5mm} (\D_\nu h_{\alpha\beta})
        \D^\beta R_{\mu\tau\lambda\sigma} \hspace{-0.5mm}+\hspace{-0.5mm} (\D_\lambda h_{\alpha\beta})
        \D^\beta R_{\mu\nu\tau\sigma} \hspace{-0.5mm}+\hspace{-0.5mm} (\D_\sigma h_{\alpha\beta})
        \D^\beta R_{\mu\nu\lambda\tau}\right].
\ea  
\newpage
\subsection{Second variation of the Ricci scalar for the Newtonian potential}
The variation of the Ricci tensor for the metric \eqref{eq:perturbedmiknowskiscalar}
 is (to second order in $\Psi$, $\Phi$ and their derivatives)\ba
        \delta^{(2)}R_{00} &=&2 \Psi \Delta \Phi 
        -\Phi ' \left(\Phi '+\Psi '\right)\non\\
        \delta^{(2)}R_{22} &=& 2 \Phi \Phi ''-\Phi ' \Psi '+\Phi '^2
        +4 \Psi '^2+4 \Psi \Delta\Psi \non\\
        \delta^{(2)}R_{33} &=& r^2\left[  \Psi ' \left(\Phi '+\Psi '\right)
        +\frac{2}{r}( \Phi  \Phi '+ \Psi  \Psi ')+2 \Psi \Delta  \Psi \right] \non\\
        \delta^{(2)}R_{44} &=&2 r^2\sin^2\theta\left[\Psi ' \left(\Phi '+\Psi '\right)
        +\frac{2}{r}( \Phi  \Phi '+ \Psi  \Psi ')+2 \Psi \Delta  \Psi\right]
\ea  
and second order variation of the Ricci scalar is
\ba
        \delta^{(2)}R= 2  \left(\Phi ' \Psi '
        +\Phi '^2+3 \Psi '^2\right)
        +16 \Psi  \Delta \Psi+4 (\Phi - \Psi)
        \Delta \Phi ~~~~~
\ea
If $\Phi(r)=\Psi(r)$, the Ricci tensor becomes
\ba
        \delta^{(2)}R_{00} &=&2 \Psi  \Delta \Psi-2 \Psi '^2\non\\
        \delta^{(2)}R_{22} &=& 4\Psi '^2+\Psi  \left(6\Delta \Psi 
        +8 \frac{\Psi '}{r}\right)\non\\
        \delta^{(2)}R_{33} &=&2 r^2\left[ \Psi '^2+\Psi  \left(\Delta \Psi 
        +2 \frac{\Psi '}{r}\right)\right] \non\\
        \delta^{(2)}R_{44} &=&2 r^2\sin^2\theta\left[ \Psi '^2+\Psi  \left(\Delta \Psi 
        +2 \frac{\Psi '}{r}\right)\right]
\ea        
and the Ricci scalar becomes
\ba
        \delta^{(2)}R&=&10 \Psi '(r)^2+16 \Psi (r) \Delta \Psi(r)
\ea        

\section{Second variation of the action}
\label{sec:2ndvaria}
In this section we give an overview of the derivation of the second variation of the action around maximally symmetric backgrounds. Around these backgrounds, one can  decompose the perturbation
to the metric $h_{\mu\nu}$ into scalar, vector and tensor parts as 
\cite{DHoker:1999bve}
\ba \label{eq:decompofhapp}
        h_{\mu\nu}=h^\perp_{\mu\nu}+\bar{\D}_\mu A_\nu+\bar{\D}_\nu A_\mu+(\bar{\D}_\mu
\bar{\D}_\nu-4
\bar{g}_{\mu\nu}\bar{\Box})B+4 \bar{g}_{\mu\nu}h,\label{ligodecomphabh}
\ea
where $h^\perp_{\mu\nu}$ is the transverse ($\D^\mu h^\perp_{\mu\nu}=0$)
and traceless spin-2 excitation, 
$A_\mu$ is a transverse vector field, and $B,~h$ are two scalar 
degrees of freedom which mix. 

Upon decomposing the second variation of the action 
around maximally symmetric backgrounds such as (A)dS, 
the vector mode and the double derivative scalar mode vanish identically,
and we end up only with $ h^\perp_{\mu\nu}$ and $\phi=\Box B - h$~\cite{Alvarez-Gaume:2015rwa,Biswas:2016etb},
i.e. 
\ba
        h_{\mu\nu} =  h^\perp_{\mu\nu} -\frac{1}{4} g_{\mu\nu} \phi. 
\ea  

One can see that inserting the decomposition \eqref{eq:decompofhapp} into the definition of the linearised Riemann tensor 
\ba
        r^{\sigma\mu}{}_{\nu\rho}&=& \frac{\bar{R}}{24}\left(\delta^\mu_\mu         h^\sigma_\rho - \delta^\mu_\rho h^\sigma_\nu - \delta^\sigma_\nu h^\mu_\rho +\delta^\sigma_\rho h^\mu_\nu \right)\non\\
        && + \frac{1}{2} \left(\bar{\D}_\nu \bar{\D}^\mu h^\sigma_\rho
        - \bar{\D}_\nu \bar{\D}^\sigma h^\mu_\rho - \bar{\D}_\rho \bar{\D}^\mu
        h^\sigma_\nu +\bar{\D}_\rho \bar{\D}^\sigma h^\mu_\nu \right),
\ea  
leads to the tensor part simply by replacing $h_{\mu\nu}$ with $h^\perp_{\mu\nu}$
\ba \label{eq:tensorpartofriemann}
        r^{\sigma\mu}{}_{\nu\rho}(h^\perp_{\mu\nu})&=& \frac{\bar{R}}{24}\left(\delta^\mu_\nu
        h^{\perp\sigma}_\rho - \delta^\mu_\rho h^{\perp\sigma}_\nu - \delta^\sigma_\nu
h^{\perp\mu}_\rho +\delta^\sigma_\rho h^{\perp\mu}_\nu \right)\non\\
        && + \frac{1}{2} \left(\bar{\D}_\nu \bar{\D}^\mu h^\sigma_\rho
        - \bar{\D}_\nu \bar{\D}^\sigma h^{\perp\mu}_\rho - \bar{\D}_\rho \bar{\D}^\mu
        h^{\perp\sigma}_\nu +\bar{\D}_\rho \bar{\D}^\sigma h^{\perp\mu}_\nu \right),~~~~~
\ea  
and
the scalar part is found by replacing $h_{\mu\nu}$ with $-\frac{1}{4}g_{\mu\nu}\phi$ 
\ba \label{eq:scalarpartofriemann}
        r^{\sigma\mu}{}_{\nu\rho}(\phi)\hspace{-0.4mm}= \hspace{-0.4mm}\frac{\bar{R}}{48}\left(\delta^\mu_\nu
        \delta^\sigma_\rho \hspace{-0.4mm}-\hspace{-0.4mm} \delta^\mu_\rho \delta^\sigma_\nu \right)\phi \hspace{-0.4mm}-\hspace{-0.4mm} \frac{1}{8} \left(\bar{\D}_\nu \bar{\D}^\mu \delta^\sigma_\rho
        \hspace{-0.4mm}-\hspace{-0.4mm} \bar{\D}_\nu \bar{\D}^\sigma \delta^\mu_\rho \hspace{-0.4mm}-\hspace{-0.4mm} \bar{\D}_\rho \bar{\D}^\mu
        \delta^\sigma_\nu \hspace{-0.4mm}+\hspace{-0.4mm}\bar{\D}_\rho \bar{\D}^\sigma \delta^\mu_\nu \right)\phi,~~~~~
\ea  
Contraction of \eqref{eq:tensorpartofriemann} with $\delta^\nu_\sigma$ gives
\ba
        r_{\mu\nu}(h^\perp_{\mu\nu})=-\frac{\bar{R}}{12}
h^{\perp\mu}_\rho - \frac{1}{2} \bar{\Box} h^{\perp\mu}_\rho,
\ea         
so $r(h^\perp_{\mu\nu})=0$, and contraction of of \eqref{eq:scalarpartofriemann} with $\delta^\nu_\sigma$ leads to 
\ba \label{eq:scalarpartofricciten}
        r^{\mu}{}_{\rho}(\phi)\hspace{-0.4mm}= \hspace{-0.4mm}\frac{1}{8}
\left(\hspace{-0.4mm} \bar{\Box} \delta^\mu_\rho
\hspace{-0.4mm}+2\hspace{-0.4mm} \bar{\D}_\rho \bar{\D}^\mu
         \hspace{-0.4mm} \right)\phi-\frac{\bar{R}}{16}\delta^\mu_\rho
        \phi,~~~~~
\ea  
so that 
\ba
        r(\phi)=\frac{3}{8}\bar{\Box} \phi - \frac{1}{4} \bar{R}\phi.
\ea              
For maximally symmetric backgrounds such as (A)dS and Minkowski,
the background traceless Einstein tensor and the background Weyl tensor are zero, so the only contribution from these terms comes from the linear part squared. We also know that the quadratic variation of the Einstein-Hilbert action with a cosmological constant $\Lambda = \frac{M^2_P}{4} \bar{R}$ is given by 
\ba
        \delta^2 S_{\text{EH}+\Lambda}\hspace{-1mm}&=&\hspace{-1mm}\int d^4x \sqrt{-\bar{g}} \frac{M^2_P}{2} \delta_0, \non\\
        \delta_0\hspace{-1mm} &=&\hspace{-1mm} \frac{1}{4} h_{\mu\nu} \bar{\Box}h^{\mu\nu} \hspace{-0.4mm}-\hspace{-0.4mm} \frac{1}{4} h \bar{\Box}         h \hspace{-0.4mm}+\hspace{-0.4mm}\frac{1}{2} h \bar{\D}_\mu \bar{\D}_\rho h^{\mu\rho} \hspace{-0.4mm}+\hspace{-0.4mm}\frac{1}{2}
        \bar{\D}_\mu h^{\mu\rho} \bar{\D}_\nu h^\nu_\rho  \hspace{-0.4mm}-\hspace{-0.4mm}\frac{\bar{R}}{48}
        (h^2 \hspace{-0.4mm}+\hspace{-0.4mm}2h^\mu_\nu h^\nu_\mu ).~~~~~~~~~
\ea
It further emerges that the $F_1(\Box)$ term can be reset in terms of $\delta_0$ and the linear variation of the Ricci scalar squared, i.e.
\ba
        \delta^2 S (R F_1(\Box) R) =  \frac{1}{2} \int d^4 x \sqrt{-\bar{g}} \left(2 f_{1_0} \bar{R} \delta_0 + r F_1(\bar{\Box}) r\right).
\ea        
We can simplify the other two terms by recasting the Ricci tensor into the traceless Einstein tensor using \eqref{eq:tracelesseinstein}
, and noting that the $F_3(\Box)$ term depends on the Weyl tensor, both of which vanish in a background de Sitter spacetime. Therefore the only terms which contribute from the $F_2(\Box)$ and $F_3(\Box)$ part of the action are $s^{\mu\nu} F_2(\Box) s_{\mu\nu}$ and $c^{\mu\nu\rho\sigma} F_3(\Box) c_{\mu\nu\rho\sigma}$, where $s^{\mu\nu}$ and $c_{\mu\nu\rho\sigma}$ are the linear variation of the traceless Einstein tensor and the Weyl tensor respectively. As detailed in \cite{Biswas:2016etb}, we can therefore split the second variation of the action up into tensor and scalar
parts as
\ba \label{eq:secondvariationoftheactionformaxsymmapp}
         \hspace{-1mm}\delta^2 S(h^\perp_{\nu\mu})\hspace{-1mm}
        &=& \hspace{-1mm}\frac{1}{2} \int d^4 x \sqrt{-g}\hspace{1.5mm} \widetilde{h}^\perp_{\mu\nu}\left(
\Box  - \frac{\bar{R}}{6} \right)
        \Bigg[ 1 + \frac{2\lambda}{M^2_p} f_{1_0}  \bar{R} \nonumber\\
        &&+ \frac{\lambda}{M^2_p}\Bigg\{\left(\Box - \frac{\bar{R}}{6} \right)
F_2 (\Box)  
        +  2\left( \Box - \frac{\bar{R}}{3}\right) F_3 \left(\Box +  \frac{\bar{R}}{3}\right)
         \Bigg\}\Bigg]\widetilde{h}^{\perp\mu\nu},\non\\
        \delta^2 S(\phi)  &=&- \frac{1}{2} \int d^4 x \sqrt{-g}\hspace{0.4mm}
\widetilde{\phi}\left(\Box+\frac{\bar{R}}{3}\right) \Bigg\{ 1
+2 \frac{\lambda}{M^2_p} f_{1_0} \bar{R} \nonumber\\
        &&  -\frac{\lambda}{M^2_p} \Bigg[ 2F_1 (\Box) \left(3\Box +\bar{R}\right)
+ \frac{1}{2} F_2 \left(
\Box + \frac{2}{3} R \right)\Box    \Bigg]\Bigg\}\widetilde{\phi}.
\ea      
where we have rescaled $h^\perp_{\mu\nu}$ and $\phi$ to $\tilde{h}^\perp_{\mu\nu}$
and $\tilde{\phi}$ by  multiplying by $\frac{M_P}{2}$ and 
$M_P\sqrt{\frac{3}{32}}$ respectively. 

\section{Inverse Fourier transform}
\label{sec:fouriertrans}
For a function $f(k)$,
\ba
        \int \frac{d^3k}{(2\pi)^3} f(k) e^{i\textbf{k}\cdot\textbf{x} }
        &=& \int \frac{d^3k}{(2\pi)^3} f(k) e^{i\textbf{k}\cdot\textbf{x} }\non\\
        &=& \frac{1}{(2\pi)^3} \int^\infty_0 k^2 dk \int^1_{-1} d\cos \theta \int^{2\pi}_0 
        d\phi f(k) e^{ikr\cos\theta}\non\\
        &=& \frac{1}{(2\pi)^3} \int^\infty_0 k^2f(k) dk \int^1_{-1} d\cos \theta e^{ikr\cos\theta}
        \left[\phi\right]^{2\pi}_0 \non\\
        &=& \frac{2\pi}{(2\pi)^3} \int^\infty_0 k^2f(k) dk \int^1_{-1} d\cos \theta e^{ikr\cos\theta}\non\\
        &=& \frac{1}{(2\pi)^2} \int^\infty_0 k^2f(k) dk \left[\frac{ e^{ikr\cos\theta}}{ikr}\right]^1_{-1}\non\\
        &=& \frac{1}{(2\pi)^2} \int^\infty_0 k^2f(k) dk \frac{e^{ikr}- e^{-ikr}}{ikr} \non\\
        &=& \frac{1}{(2\pi)^2r} \int^\infty_0 kf(k) dk \frac{e^{ikr}- e^{-ikr}}{i}  \non\\
        &=& \frac{1}{2\pi^2r} \int^\infty_0 kf(k) dk \sin(kr)
\ea

\newpage        
\section{Raytracing}
\label{sec:raytracing}    
In this section we investigate the effect of IDG on the trajectory of light 
passing a point mass, following the GR calculation given in \cite{taylorwheelerblackholes}.
    
\subsection{Equations of motion}
At large distances, one can use the coordinate transformation 
$r^{'2} =\left(1-\frac{r_*}{r} \text{Erf}\left(\frac{Mr}{2}\right)\right)r^2$ (and then
drop the prime) to write the metric 
\eqref{eq:perturbedmiknowskiscalar} with the potential \eqref{eq:erfpotential} as 
\ba \label{eq:erfradialmetric}
        d\tau^2 =\left(1-\frac{r_*}{r} \text{Erf}\left(\frac{Mr}{2}\right)\right)dt^2-
\left( 1-\frac{r_*}{r} \text{Erf}\left(\frac{Mr}{2} \right)
\right)^{-1} dr^2 -r^2 d\phi^2,
\ea  
where $r_*=2GM$ is the event horizon of the mass in GR. 
We can read off using the law of conservation of energy that 
\ba \label{eq:IDGequationsofmotiontforblackhole}
        dt = \frac{E/m}{1-\frac{r_*}{r} \text{Erf}\left(\frac{Mr}{2}\right)}d\tau,
\ea
and from the law of conservation of angular momentum that
\ba \label{eq:IDGequationsofmotionphiforblackhole}
        d\phi = \frac{L/m}{r^2} d\tau.
\ea
Finally we can substitute these results into the metric \eqref{eq:erfradialmetric}
and solve for $dr$ to find
\ba \label{eq:IDGequationsofmotionrforblackhole}
      dr =\pm \left[ \frac{E^2}{m^2}-\left( 1-\frac{r_*}{r} \text{Erf}\left(\frac{Mr}{2}
\right)\left(1+\frac{L^2}{r^2m^2}\right) \right)
\right]^{1/2} d\tau.         
\ea  
\subsection{Motion of light}
The angular momentum of a particle in flat spacetime is the linear momentum
far from the black hole multiplied by its impact parameter, i.e. $L=b~ p_{far}$.
In flat spacetime, $m^2=E^2 -p^2$ and so $p=(E^2-m^2)^{1/2}$. Therefore
\ba
        b = \frac{L}{p} = \frac{L}{(E^2-m^2)^{1/2}}.
\ea
For the motion of light, we take the limit $m \to 0$ and
so 
\ba
        b_{\text{light}} = \frac{L}{E}.
\ea
Next we multiply the equations of motion for a particle 
(\ref{eq:IDGequationsofmotionphiforblackhole})
and (\ref{eq:IDGequationsofmotionrforblackhole}) by the appropriate power
of $d\tau/dt$ using (\ref{eq:IDGequationsofmotiontforblackhole}).
\ba
       \frac{d\phi}{dt}  = \frac{1-\frac{r_*}{r}
\text{Erf}\left(\frac{Mr}{2}\right)}{r^2} \frac{L}{E},
\ea
\ba
        \left(\frac{dr}{dt}\right)^2 &=&   \left(1-\frac{r_*}{r} \text{Erf}\left(\frac{Mr}{2}\right)\right)^2          \left[ 1-\left( 1-\frac{r_*}{r} \text{Erf}\left(\frac{Mr}{2} \right)\left(\frac{m^2}{E^2}+\frac{1}{r^2}\frac{L^2}{E^2}\right)
\right)\right].~~~~~~~~~~
\ea
Now we take the limit $m\to 0$ to find the equations of motion for light and insert
$b_{\text{light}}=L/E$, yielding
\ba \label{eq:finalequationsnotonshell}
       r \frac{d\phi}{dt} &=& \frac{b_{\text{light}}}{r} \left(1-\frac{r_*}{r}
\text{Erf}\left(\frac{Mr}{2}\right)\right),\nonumber\\
        \frac{dr}{dt}
        &=&  \pm \left(1-\frac{r_*}{r} \text{Erf}\left(\frac{Mr}{2}\right)\right)
         \left[ 1-\frac{b^2_{\text{light}}}{r^2}\left( 1-\frac{r_*}{r} 
         \text{Erf}\left(\frac{Mr}{2}\right) \right)\right]^{1/2}.~~~~~~
\ea       
\subsubsection{Impact parameter}
To calculate the trajectory of a photon, we must first find the impact
parameter $b_{\text{light}}$.
First note that using the time coordinates on a shell of radius $r$ from the origin,
described by 
\ba
        dt_{shell} = dt\left(1-\frac{r_*}{r} \text{Erf}\left(\frac{Mr}{2}\right)
\right)^{1/2}, \nonumber\\
        dr_{shell} = \frac{dr}{\left(1-\frac{r_*}{r} 
        \text{Erf}\left(\frac{Mr}{2}\right)\right)^{1/2}},
\ea        
 we can rewrite (\ref{eq:finalequationsnotonshell}) as  
\ba
       r \frac{d\phi}{dt_{shell}} &=& \frac{b_{\text{light}}}{r} \left(1-\frac{r_*}{r}
\text{Erf}\left(\frac{Mr}{2}\right)\right)^{1/2},\nonumber\\
        \frac{dr_{shell}}{dt_{shell}}
        &=& \left[ 1-\frac{b^2_{\text{light}}}{r^2}\left( 1-\frac{r_*}{r} \text{Erf}\left(\frac{Mr}{2}
\right) \right)\right]^{1/2}.
\ea        
Now a photon coming in from the negative $x$ direction at angle $\phi$ to
the line $y=0$ would give that the speed of light perpendicular to 
the radius (in the +r direction) is equal to
\ba \label{eq:eqnforfindingimpactparam}
        - 1 \cdot \sin(\phi_0) = r_0 \left(\frac{d\phi}{dt_{shell}} \right)
= \frac{b_{\text{light}}}{r_0} 
        \left(1-\frac{r_*}{r_0} \text{Erf}\left(\frac{Mr_0}{2}\right)\right)^{1/2}.
\ea
Then we can simply solve for $b_{\text{light}}$ for each path and then integrate (\ref{eq:finalequationsnotonshell})
to find the path of the photon.
               
\subsubsection{Photon trajectory}
We could also solve for $\frac{dr}{d\phi}$. Using \eqref{eq:finalequationsnotonshell},
\ba
       \frac{dr}{d\phi} &=& \pm\frac{r^2}{b_{\text{light}}}    
       \left[ 1-\frac{b^2_{\text{light}}}{r^2}\left(1-\frac{r_*}{r} 
       \text{Erf}\left(\frac{Mr}{2}\right) \right)\right]^{1/2}.
\ea
We square both sides and define $u\equiv 1/r$ so that $dr=-u^{-2} du$, which means that we have
\ba \label{eq:diffeqnforu}        
        \left(\frac{du}{d\phi}\right)^2 = \frac{1}{b^2} 
        - u^2 \left( 1 -r_*u\text{Erf}\left(\frac{M}{2u}\right)\right).
\ea
We then differentiate both sides with respect to $\phi$ to give
\ba       \label{eq:differentialrayapp}
        u''(\phi) = - u(\phi)+\frac{3}{2}r_*u^2(\phi) \text{Erf}\left(\frac{M}{2u(\phi)}\right) 
        + \frac{r_*u^3(\phi)}{2} \text{Erf}'\left(\frac{M}{2u(\phi)}\right).
\ea   
Now that we have the differential equation describing the path of the light, we simply need to 
find the initial conditions, which requires finding  $b_{\text{light}}$.   
Looking at \eqref{eq:eqnforfindingimpactparam}, we see that to find $b_{\text{light}}$,
 we will take a square with sides 20 (in units of the point mass $m$), and look at photons coming in from the right hand side. 
Consequently, a photon coming in at angle $\phi_0$ to the horizontal will have impact parameter
\ba
        b_{\text{light}}=\frac{r_0 \sin (\phi_0)}{\left[1-\frac{r_*}{r_0} \text{Erf}
         \left(\frac{Mr_0}{2}\right)  \right]^{1/2}}.
\ea       
Finally, if we want a photon coming in at $y=y_i$, then $\tan(\phi_0)
=\frac{y_i}{10}$, i.e. $r_0 \sin(\phi_0)=\frac{10}{\cos(\phi_0)}\sin(\phi_0)
=10 \frac{y_i}{10}=y_i$ gives us       
\ba \label{eq:finalblight}
        b_{\text{light}}=\frac{y_{i}}{\left[1-\frac{r_*}{ \sqrt{100+y_i^2}} \text{Erf}
\left(\frac{ M\sqrt{100+y_i^2}}{2}\right)  \right]^{1/2}}.
\ea       
Combining \eqref{eq:finalblight} and (\ref{eq:diffeqnforu}) with 
$u_0=\frac{1}{r_0} = \frac{1}{\sqrt{100+y_i^2}}$ gives
\ba \label{eq:plotraytracing}
        \frac{\partial u}{\partial \phi}|_{u=u_0}&=&\hspace{-1.3mm} \left[\frac{1-\frac{r_*}{ \sqrt{100+y_i^2}} \text{Erf}
\left(\frac{M \sqrt{100+y_i^2}}{2}\right)  }{y_{i}^2}\right.\non\\
        && - \left.\frac{1}{100+y_i^2} \left( 1 -\frac{r_*}{ \sqrt{100+y_i^2}}\text{Erf}\left(\frac{
M\sqrt{100+y_i^2}}{2}\right)\right)\right]^{1/2}.   ~~~~~ 
\ea        

We use \eqref{eq:plotraytracing} along with \eqref{eq:differentialrayapp}
 to plot the path of photons passing a point mass
in Fig. \ref{raytracing}.

\end{appendices}

\backmatter

\nocite{} 
\renewcommand{\bibname}{References}

\bibliographystyle{unsrt}  
\bibliography{thesisarxiv}        

\end{document}